\newcommand\ha{{H$\alpha$}}
\newcommand\hb{{H$\beta$}}

\newcommand\kms{\:\rm{km\,s^{-1}}}

\newcommand\VEL{\:{\rm km\:s^{-1}}}

\newcommand\OiL{[\ion{O}{1}] $\lambda 6300$}

\newcommand\SiiL{[\ion{S}{2}] $\lambda\lambda 6716, 6731$}
\newcommand\NiiL{[\ion{N}{2}] $\lambda\lambda 6548, 6583$}

\newcommand\FeiiL{[\ion{Fe}{2}] 1.644 $\mu$m}

\newcommand\sii{[\ion{S}{2}]}
\newcommand\nii{[\ion{N}{2}]}
\newcommand\oi{[\ion{O}{1}]}
\newcommand\oii{[\ion{O}{2}]}
\newcommand\oiii{[\ion{O}{3}]}

\newcommand\hii{\ion{H}{2}}

 \newcommand{\SFR}{\mbox{$\:M_{\sun}\;{\rm yr}^{-1}$}}  

\documentclass[modern]{aastex63}
\shorttitle{SNRs in M51}
\shortauthors{Winkler et al.}

\usepackage{natbib}
\usepackage{comment}
\usepackage{todonotes}
\usepackage{graphicx}

\begin{document}

\title{Optical Identification and Spectroscopy of Supernova Remnants in the Galaxy M51\footnote{Based in part on observations with the NASA/ESA Hubble Space Telescope obtained at the Space Telescope Science Institute, which is operated by the Association of Universities for Research in Astronomy, Incorporated, under NASA contract NAS5-26555. Support for program numbers 14638 and 15216 was provided through a grant from the STScI under NASA contract NAS5-26555.}}

\correspondingauthor{P. Frank Winkler}
\email{winkler@middlebury.edu}

\author[0000-0001-6311-277X]{P. Frank Winkler}
\affiliation{Department of Physics, Middlebury College, Middlebury, VT, 05753; 
winkler@middlebury.edu}

\author[0000-0003-3038-8045]{Sadie C. Coffin}
\affiliation{Department of Physics, Middlebury College, Middlebury, VT, 05753}; 
\affiliation{Present affiliation: Southeastern Universities Research Association, Washington, DC 20005; X-ray Astrophysics Laboratory NASA/GSFC, Greenbelt, MD 20771; Center for Research and Exploration in Space Science and Technology, NASA/GSFC, Greenbelt, MD 20771;
sadie.coffin@nasa.gov}

\author[0000-0003-2379-6518]{William P. Blair}
\affiliation{The Henry A. Rowland Department of Physics and Astronomy, 
Johns Hopkins University, 3400 N. Charles Street, Baltimore, MD, 21218, USA; 
wblair@jhu.edu}

\author[0000-0002-4134-864X]{Knox S. Long}
\affil{Space Telescope Science Institute,
3700 San Martin Drive,
Baltimore MD 21218, USA; long@stsci.edu}
\affil{Eureka Scientific, Inc.
2452 Delmer Street, Suite 100,
Oakland, CA 94602-3017}

\author[0000-0001-6654-5378]{Kip D. Kuntz}
\affiliation{The Henry A. Rowland Department of Physics and Astronomy, 
Johns Hopkins University, 3400 N. Charles Street, Baltimore, MD, 21218, USA; 
kkuntz1@jhu.edu}

{\flushleft {\small To be published in {\em The Astrophysical Journal}}}
\vspace{0.2 in}

\begin{abstract}
 
Using a combination of ground-based and HST imaging, we have constructed a catalog of 179 supernova remnants (SNRs) and SNR candidates in the  nearby spiral galaxy M51. Follow-up spectroscopy of 66 of the candidates confirms 61 of these  as SNRs, and suggests that the vast majority of  the unobserved objects are SNRs as well.  A total of 55 of the candidates are coincident with (mostly soft) X-ray sources identified in deep Chandra observations of M51; searching the positions of other soft X-ray sources resulted in several additional possible optical candidates.  There are 16 objects in the catalog coincident with known radio sources.  None of the sources with spectra shows the high velocities ($\gtrsim 500 \kms$) characteristic of young, ejecta-dominated SNRs like Cas A; instead,  most if not all appear to be middle-aged SNRs.  The general properties of the SNRs, size distribution and spectral characteristics, resemble those in other nearby spiral galaxies, notably M33, M83, and NGC\,6946,  where similar samples exist.  However, the spectroscopically observed  \nii:\ha\ ratios appear to be significantly higher than in any of these other galaxies.   Although we have explored various ideas to explain the high ratios in M51, none of the explanations appears to be satisfactory.


\end{abstract}


\section{Introduction} \label{sec:intro}

Following the violent death of a star  as a supernova, material rich in heavy elements is ejected and drives shock waves  into  the surrounding circumstellar and/or interstellar material to form a supernova remnant (SNR).  Supernovae (SNe) are a key part of the cycle that gradually enriches the cosmos; they can trigger new episodes of star formation, and collectively they may influence the evolution of the galaxies in which they take place.  For understanding these  processes, it is most efficacious to study SNR samples in nearby galaxies, especially spirals that are oriented nearly face-on.  Here we typically find large numbers of SNRs, all at effectively the same distance, and in most cases with minimal foreground absorption.  All this contrasts with our own Galaxy, where most of the hundreds of identified SNRs are at  poorly determined distances and suffer from significant extinction---in many cases so much so that they have not even been detected optically.

Relatively nearby, face-on, and with well-defined spiral arms with active star formation, M51 presents an ideal venue for this study of SNRs. M51 (NGC 5194/5) has the distinction of being the first galaxy classified as a spiral \citep{rosse50} and is one of the most well known interacting systems, comprising the grand-design spiral M51A = NGC\,5194 and smaller companion M51B = NGC\,5195\@. M51A is classified as a late-type grand-design Sbc, while its early-type northern companion M51b is classified as a barred S0. At a distance of 8.58 $\pm$ 0.10 Mpc \citep{mcquinn16},  where 1\arcsec\ corresponds to a linear size of 41.6 pc, the Whirlpool Galaxy has been the site of four supernovae since 1945, and is thus expected to host a rich population of  SNRs. 

Most extragalactic SNRs have been first identified optically, through their \SiiL\ lines, where we typically find \sii:\ha\ flux ratios $>0.4$, significantly higher than in  \hii\ regions, where \sii:\ha\ is usually $\lesssim 0.2$ \citep[e.g.,][]{mathewson73,levenson95,long17}.   Most of the bright optical emission from SNRs stems from secondary shocks driven into dense clouds.  The passing shock rapidly heats the material, which then gradually cools in a long-lasting post-shock tail, where we find a variety of low-ionization and even neutral species that radiate following electron collisional excitation, producing optical spectra with strong forbidden lines from (especially) \sii, \nii, \oii, and \oi, in addition to Balmer lines.  By contrast, the emission from \hii\ regions stems from photoionized gas that is kept in predominantly higher ionization states by ongoing UV radiation from hot stars.  

This paper presents the first catalog of SNRs in M51, derived from a combination of HST and ground-based imaging surveys.  We use the traditional \sii:\ha\ ratio criterion to identify likely shock-heated SNR candidates.\footnote{Although our SNR search was limited  to M51A = NGC\,5194, we refer to the galaxy  as M51 for simplicity.}  We also obtain and analyze spectra from many of these, along with a selection of \hii\ regions,  to better measure the  \sii:\ha\ ratios, along with other low-ionization lines that characterize  shock-heated SNRs.  
The paper is organized as follows:  In Sec.\ 2 we describe our imaging observations, candidate selection, and subsequent spectroscopy.  Sec.\ 3 presents the results: our complete catalog of 179 SNR candidates and the spectra we obtained for 66 of these.  In Sec.\ 4 we discuss several aspects of our results, including the unusually strong \nii\ lines that dominate the spectra of many M51 SNRs, overlaps between objects in our search and ones from X-ray and radio surveys, historical SNe in M51, and the population of SNRs in M51 relative to the populations in other similar galaxies.  Finally, Sec.\ 5 gives  a brief summary of our conclusions.

\section{Observations, Data Reduction, and Candidate Selection}

M51 is the most distant galaxy for which an extensive population of SNRs has been identified \citep{long17}.  The most common approach for these studies, which we have followed in our own previous studies of M83 \citep{blair12, blair14, winkler17} and NGC\,6946 \citep{long19, long20}, is to start with a ground-based imaging survey and  follow up with HST\@.
  In the case of M51, the availability of HST data for \ha\ and several broad bands meant that only \sii\ imaging was missing in order to allow an SNR search.  Hence, our observations of M51 began with narrow-band imaging from HST, and then continued to ground-based survey work with the 8.2m Gemini North telescope, where  we first obtained images, and then spectra for over 100 objects, all using the Gemini Multi-Object Spectrograph (GMOS).  Below we describe these data sets in the order they were obtained.   In reality, the order makes little difference; as with our previous studies, the ground- and space-based survey work is largely complementary.

\subsection{HST Imagery}

The HST imagery used in this program is summarized in Table\ 1 
Program GO-10452 (Beckwith, PI) was a Director's Discretionary program that used the ACS/WFC and resulted in the image of M51 that was widely distributed at the  Hubble 15th anniversary.\footnote{See \url{https://hubblesite.org/contents/media/images/2005/12/1677-Image.html?Topic=105-galaxies&keyword=M51&news=true}, and also \url{https://doi.org/10.17909/T9GP4C} for the underlying data.} This imaging covered six overlapping ACS/WFC fields with the data from each field dithered over four exposures, resulting in the total times for each field listed in Table~1.  These observations included three broadband filters plus F658N, which captured \ha\ (plus \NiiL) emission from  nebular regions of active star formation. These data have been used  by \citet{lee11} for a comprehensive study of \hii\ regions in M51, and by many other authors, but  alone they are not sufficient for an SNR search, since a shock diagnostic  such as \SiiL\ is also needed.

\begin{figure}
\plotone{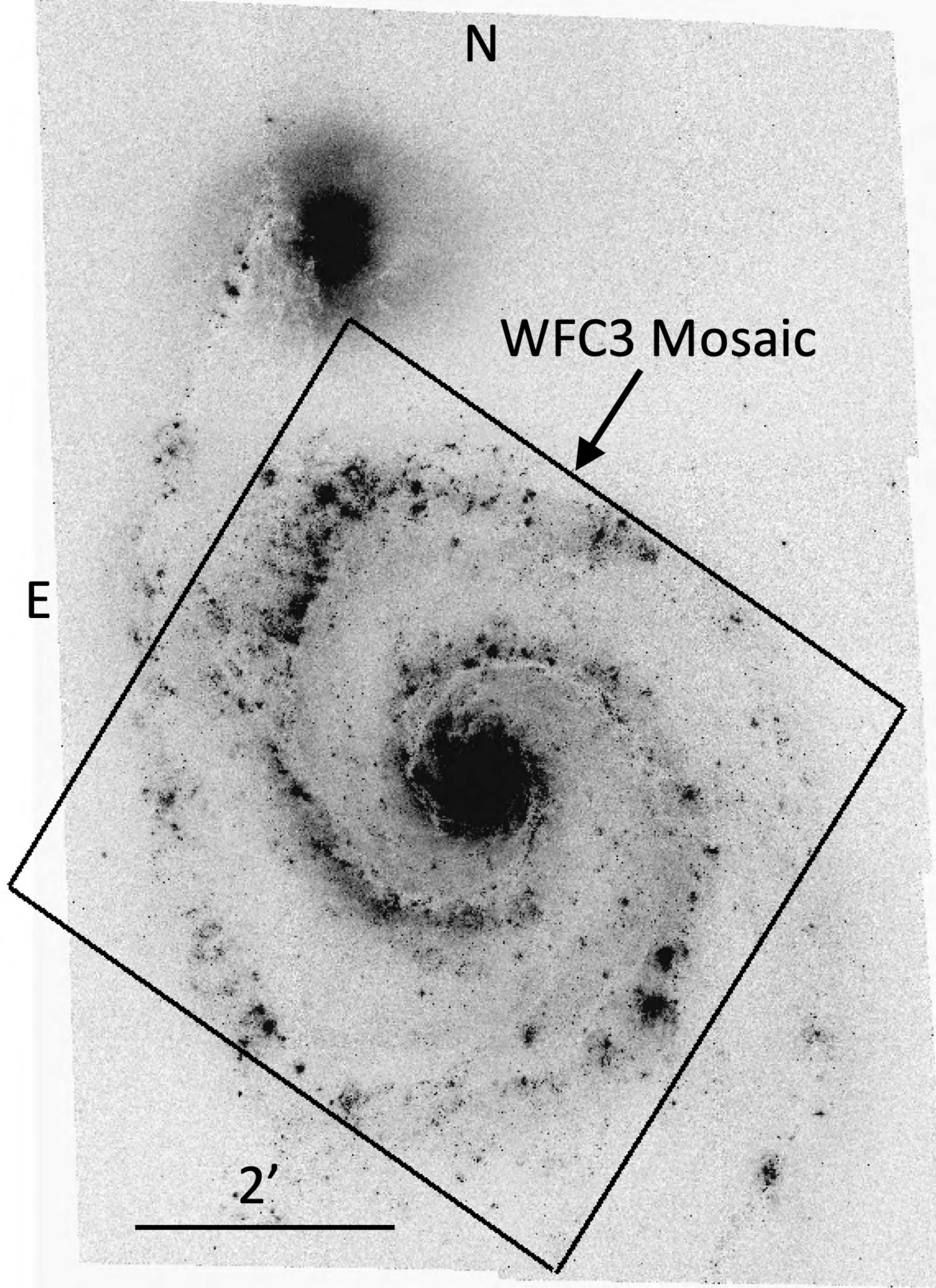}
\caption{Relative field coverage of the HST ACS and WFC3 data are shown.  The background image is the ACS 6-field mosaic  with the F658N filter.  The box shows the footprint of the 4-field WFC3 mosaic for comparison.  WFC3 covered the main body of M51A but missed some of the outermost spiral arms.  It also does not include M51B to the north.
\label{fig_hst_overview}}
\end{figure}

Hence, we leveraged these earlier observations by obtaining WFC3/UVIS imagery in two additional filters, also shown in Table~1.  Program GO-12762 (K. Kuntz, PI) used F673N to capture \SiiL\ emission from four fields, and used F689M to provide a continuum band for subtracting the galaxy background. These data were also dithered, to help mitigate cosmic ray effects and chip gaps in the processed data. 
These data were obtained just prior to the recommendation that a pedestal of electrons (using the FLASH parameter) was advised by STScI, especially for narrow-band imagery, to control charge transfer inefficiency in the  CCDs.  While no FLASH parameters were set, we see no obvious ill effects from not having done so.
Fig.~\ref{fig_hst_overview} shows the relative field coverage of these two data sets after they were aligned and mosaicked.  The newer  \sii\ data cover the main body of M51A but miss some of the outermost spiral arms; they do not cover the companion galaxy M51B.


All of these data were retrieved from the Mikulski Archive for Space Telescopes (MAST) for processing.  The individual dither frames were aligned and combined for each field, then placed into a mosaic.  Since the native pixel scales are different for ACS and WFC3, we used the AstroDrizzle package \citep{fruchter10,gonzaga12} to produce aligned mosaics on a common grid that allow direct comparison.  The procedure closely followed that described in \citet[][see Sec. 2 and the Appendix of that paper]{blair14} for M83 and will not be repeated here. Astrometric alignments relied on centroids of isolated stars in the overlap regions between fields of the mosaic to align all of the frames to a single grid, and then used stars from the UCAC3 astrometric catalog \citep{zacharias10} and 2MASS, selecting objects with small positional uncertainties and checking visually to eliminate close doubles and a few background galaxies.\footnote{This work was done prior to the availability of the Gaia data, and since it is more than accurate enough for our present purposes, the astrometry has not been revisited.}  Finally, the WCS keywords in the FITS file headers were adjusted to place the entire data set on the same absolute astrometric scale, with relative alignment accurate to $< 0\farcs1$. 

Once the images were mosaicked and aligned, the initial search process was straightforward.  We used standard IRAF\footnote{IRAF is distributed by the National Optical Astronomy Observatories,
which are operated by the Association of Universities for Research
in Astronomy, Inc., under cooperative agreement with the National
Science Foundation.} techniques to scale and subtract continuum from the emission-line images.  After some experimentation, we ultimately used the ACS F555W continuum image to scale and subtract from both the \ha\ and \sii\ mosaics. This choice provided a subtraction of the \ha\ image over the entire ACS mosaic area, though our \sii\ coverage was more restricted.  The color differences between the galaxy background in the nuclear region and the outer arms made it difficult to use a single scaling of the continuum image for subtraction, so a separate subtraction (that over-subtracted the outer galaxy) was used specifically for the inner portion of the galaxy.  

\begin{figure}
\plotone{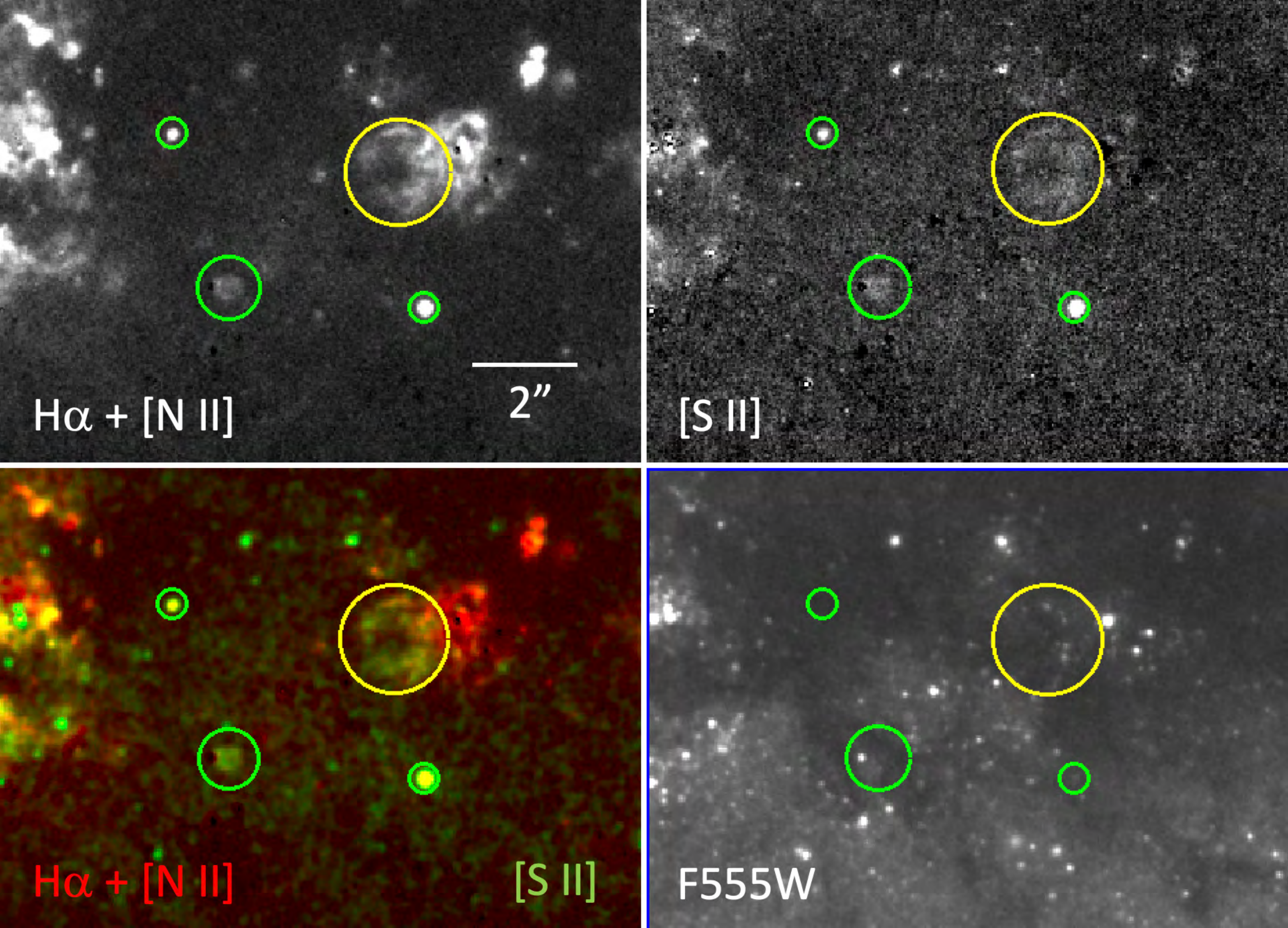}
\caption{HST image data for a small region, located in the first spiral arm south of the nucleus, that contains four SNR candidates. The region is centered at R.A.(J2000) = 13:29:55.93, Decl.(J2000) = +47:10:45.4, and is oriented N-up, E-left.  The green circles indicate three candidates identified in our blind search of the HST images, where we looked for compact nebulae with elevated \sii:\ha\ ratios.  The \ha\ + \nii\ and \sii\ panels have been continuum-subtracted using the F555W data shown at lower right.  The color panel has been scaled to show the SNR candidates as green to yellow, while photoionized regions appear as red.  The yellow circle indicates a large-diameter, low surface-brightness SNR candidate found in the ground-based search that was later corroborated by the HST data when an appropriate display stretch was used.  For scale, the object in the largest green circle is approximately half the size as the Cygnus Loop in our Galaxy.
\label{fig_hst_zoom}}
\end{figure}

The search procedure was to simultaneously display the subtracted \ha\ and \sii\ frames along with a continuum band for reference using SAOimage DS9 \citep{joye03}.   Sub-regions of the WFC3 mosaic were then enlarged to an appropriate scale and inspected visually to locate  compact emission nebulae with enhanced \sii\ emission.  Comparison against the continuum frame prevented us from mistaking stellar residuals for possibly interesting compact nebulae. Cursor readings could be used to estimate the observed ratio of \sii\ to \ha.  Fig.~\ref{fig_hst_zoom} shows an example of a small region containing several SNR candidates as an example, including a color image that also highlights the different relative line intensities for the nebulae in the region shown.

\begin{figure}
\plotone{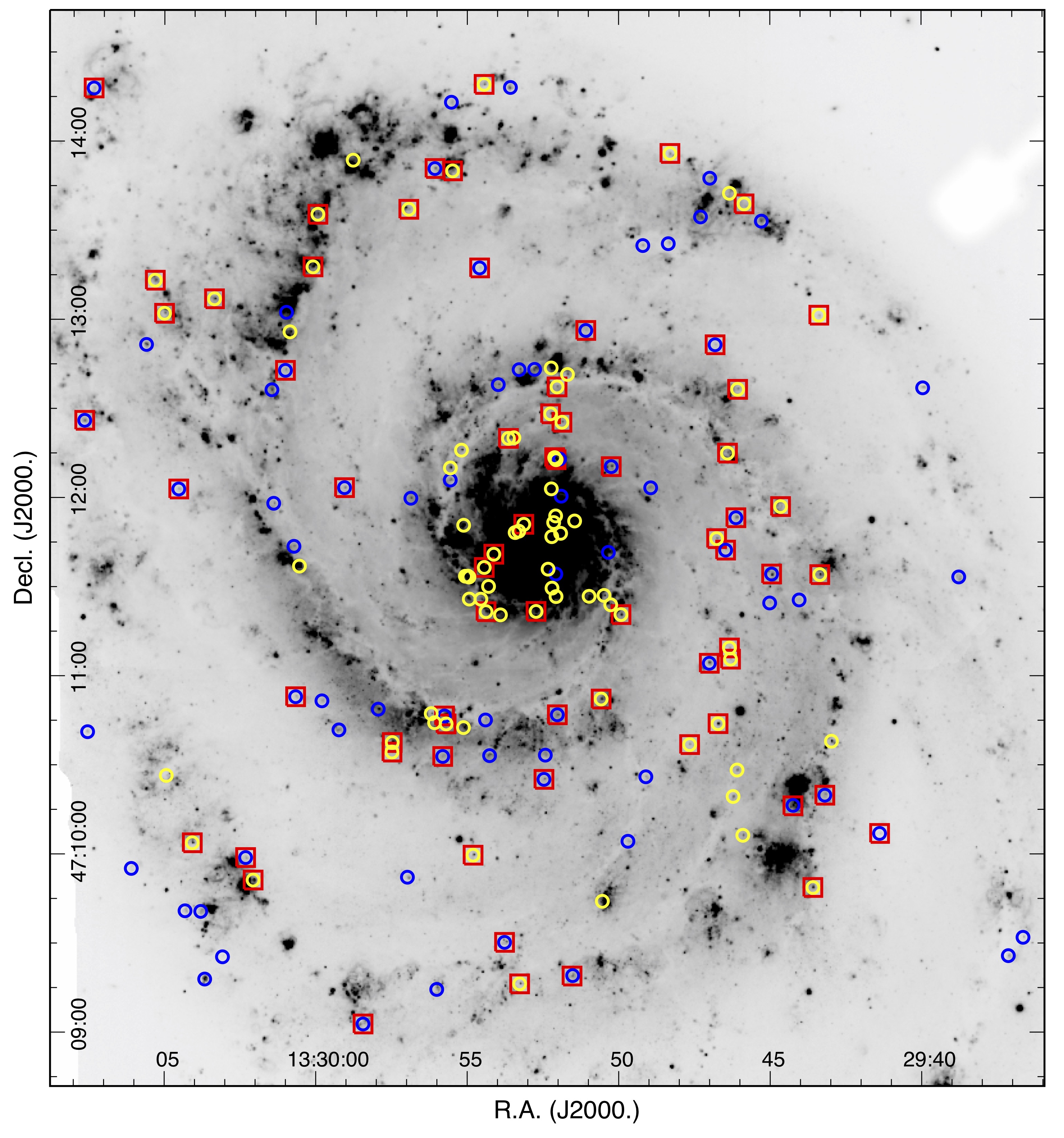}
\caption{\ha\ image of M51 (a mosaic of two GMOS fields).  Yellow circles show the optical SNR candidates identified in HST images; blue circles are ones identified in GMOS images. Candidates for which we obtained spectra are indicated by the slightly  larger red squares. As the figure indicates, candidates with spectra are well distributed around the galaxy.
\label{fig_overview}}
\end{figure}

A significant caveat that made this process less determinative than in previous galaxy searches was that the ACS filter used for \ha\ (F658N) was broad enough to fully include the \nii\ lines that bracket \ha.  \nii\ is both strong and variable in M51, especially in the SNRs (a fact later confirmed with our spectroscopy),
effectively making the images a comparison of the \sii:(\ha\ + \nii) instead of \sii:\ha.  This had the effect of decreasing the contrast between the ratio for SNRs and \hii\ regions that we normally depend on to identify the SNRs.  Nonetheless, with appropriate scaling, a set of compact nebulae that were relatively strong in \sii\ could be identified, as illustrated by the example shown in Fig.~\ref{fig_hst_zoom}.  The overall blind search of the HST data resulted in 80 compact nebulae that were judged to have observed ratios significantly elevated over what was typical for obvious \hii\ regions, but with somewhat less certainty as to their quality as SNR candidates than had been the case for our previous surveys of other galaxies. The spatial distribution of these objects across the region observed is shown by the yellow circles in Fig.~\ref{fig_overview}.

Circular DS9 regions were set on these objects and re-sized to provide a measurement of both the position and the angular size of each object.  Although some objects show some extent and morphology in the {\em HST} data (cf. Fig.~\ref{fig_hst_zoom}),  these objects are almost all  below 1\arcsec\ in size, extending down almost to the resolution limit of the WFC3 data for the smallest nebulae. (Note: One WFC3 pixel = 0\farcs04, or $\sim2$ pc at the distance of M51.)

\subsection{GMOS-N Images}

Although the exquisite resolution of the HST WFC3 images enabled us to identify these small objects of interest, the exposures were not deep enough to see larger, lower surface-brightness (and presumably older) SNRs.  Hence, as a complement to the HST survey, we received approval for imaging M51 with the Gemini Multi-Object Spectrograph (GMOS) on the 8.2m Gemini-North telescope through the ``Fast Turnaround" program GN-2017A-FT-7 (Long, PI). Through this program, we obtained narrow-band images of two overlapping fields covering most of M51 with GMOS in 3 bands: \ha, \sii, and a matched continuum band (HaC filter). The observing details are given in Table~2.

We processed these images using standard IRAF techniques in the {\tt gmos} package, using dozens of stars from 
Gaia DR2 \citep{gaia-collaboration18} to improve the World Coordinate system, and mosaicked the two fields together. 
To enhance the search for faint nebulae, it is useful to use continuum-subtracted images, obtained by scaling and subtracting the  continuum image (separately) from the \ha\ and \sii\ layers.
Finally, we flux calibrated the continuum-subtracted images, using short \ha\ and \sii\ observations of two spectrophotometric standards from the list of \citet{massey88}. 

Unfortunately, this procedure was not efficacious for the \sii\ images of M51,  because at the $600\kms$ redshift of M51, the \nii\ 6584 \AA\ line was shifted well into the bandpass of the HaC filter, with a transmission of almost $ 50\%$.  Since (as we later discovered through our spectroscopy), the \nii\ lines are extremely strong in many M51 SNRs, subtracting the ``continuum" effectively removed much of the nebular emission along with the stars.  We had, in effect, ``thrown out the baby with the bath water!"  To search for SNR candidates, we thus displayed the {\em unsubtracted} flux-calibrated \ha\ and \sii\ images in DS9,  carefully compared small regions, and marked nebulae that appeared to have a relatively strong \sii:\ha\ ratio when compared to obvious \hii\ regions.  While this ersatz procedure was less stringent than using continuum-subtracted images, as we have done for similar searches in other galaxies, we were nevertheless able to identify 107  SNR candidates from the GMOS images.  After eliminating duplicates with candidates already identified from the {\em HST} images, this left 71 new candidates from the GMOS search alone, and a total of 151 from both HST and GMOS.  All of these are listed in Table~3. 
In order to establish priorities to use in designing masks for our subsequent multi-object spectroscopy, we assigned each of these a confidence grade of A, B, or C, based on their \sii:\ha\ ratio as estimated from the images, morphology, and lack of confusion with surrounding emission.

Subsequent to our spectroscopy, we made another careful pass through both the GMOS and the HST images to search for additional candidates that might have been missed earlier.  The GMOS search was aided by the late acquisition of broad $r$-band images of exactly the same fields we had imaged previously using narrow-band filters.  By precisely aligning the $r$-band images with those in \sii, we were at last able to make useful continuum-subtracted \sii\ images.\footnote{The broad $r$ filter passes \ha, \nii, and \sii\ lines, but its far greater bandwidth (1360 \AA, compared with $\sim$\,40-70 \AA\ for the emission-line filters)  makes it relatively more sensitive to stars, and thus effective for continuum subtraction.}  In these improved  images, the candidates we had selected earlier became more obvious, and several new candidates appeared as well.  Our later search of both HST and GMOS images yielded 28 new candidates, which we have also included in Table 3 
for a grand total of 179.


\subsection{GMOS-N Spectroscopy}

We  pursued follow-up multi-object spectroscopy from GMOS-N under program GN-2018A-Q-302 (Winkler, PI)\@.  Based on the spatial distribution of SNR candidates, we determined that masks with slits oriented north-south would give greater efficiency than an east-west orientation, which would have been required had we used our 2017 GMOS images for mask design.  Therefore, we did short dithered exposures in the broad $r$ filter of two overlapping M51 fields with the GMOS chips oriented N-S, which we then used as pre-images to design the MOS masks.  

Three masks were designed and observed, giving preference to higher graded candidates; in total, these included 44 `A' candidates, 18 `B' candidates, and 3 `C' ones, plus one additional SNR that appeared serendipitously on one of the slits:\footnote{Subsequent inspection of the HST images showed that this was clearly a strong candidate that we had overlooked in our initial search; henceforth, we count it among the `A' candidates.} a total of 66 SNR candidates with spectra. The locations of all the SNR candidates with spectra are indicated by the red squares in Fig.~\ref{fig_overview}.  In addition, we explicitly targeted a number of \hii\ regions, and obtained spectra for several others that serendipitously lay along slits with other targets: a total of 44 \hii\ regions spanning a wide range in location, surface brightness, and size.   We list these \hii\ regions in Table~4. 

  Our GMOS-N program used the B600 grating (G5307, 600 lines mm$^{-1}$) and GG455 blocking filter, with the Hamamatsu detector binned $\times 4$ in the dispersion direction, and $\times 2$ spatially, to give a dispersion of 2.06 \AA\,pixel$^{-1}$ and a spatial scale of 0.16 \arcsec\,pixel$^{-1}$.  The spectral coverage varied with the position of the object, but for almost all the objects coverage extended at least from \hb\ through \SiiL, with a resolution of about 5.2 \AA\@.
Spectra were then taken with each of the three masks in May - July 2018.  For each mask, four exposures of 1100 s each were acquired at each of three grating tilts, giving three central wavelength (CWL) settings: 5800\,\AA, 5900\,\AA, and 6000\,\AA, to assure that gaps in spectral coverage (resulting from the chip gaps on the GMOS detector) would be fully covered.\footnote{An exception is  mask 3, where four 1100\,s exposures were obtained at 5800\,\AA\ and 5900\,\AA, but only two 1100 s exposures  at 6000\,\AA.}  Immediately before or after the science frames at each CWL setting, quartz flat and CuAr arc calibration frames were obtained.  


The data were processed using standard procedures  from the {\tt gemini} package in IRAF  for bias subtraction, flat-fielding, wavelength calibration, and combination of spectra with different CWL settings to provide the final 2-dimensional spectra.  
Flux calibration was based on baseline GMOS observations of a few spectrophotometric standard stars, carried out in the same semester as part of standard GMOS operations.  

\begin{figure}
\plotone{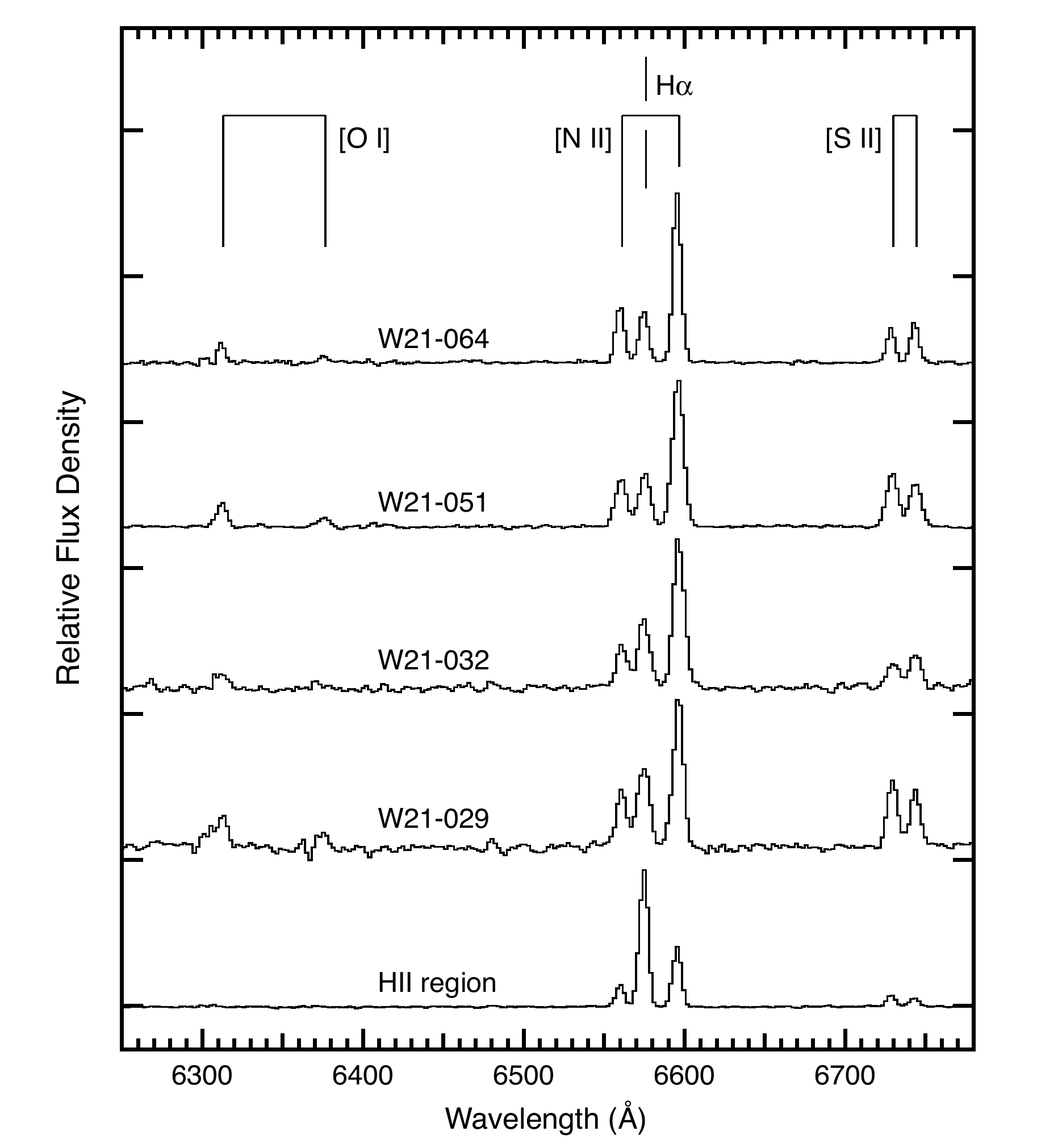}
\caption{Several examples of the SNR spectra we obtained, as well as one \hii\ region spectrum.  The lower three spectra were all objects on the same slit.
\label{fig_examples}}
\end{figure}

During the processing, the 2-D spectra from different slitlets were separated to give individual 2-D spectra from each slitlet. We examined each of these individually to determine the optimum background sky subtraction region.  Many of the objects are located in regions with bright surrounding galactic background (both continuum and emission lines) from M51, so the ability to subtract a representative  local background in the  vicinity of each object is important for obtaining accurate spectra. In most cases this limited the precision of our eventual line flux measurements.  Finally, we  summed rows containing each object to yield our final one-dimensional  spectra.  Several typical examples are shown in Fig.~\ref{fig_examples}.


We then performed  fits to obtain emission-line fluxes  from the spectra, assuming Gaussian profiles, for the following lines and line complexes:  \hb\ alone,  the \oiii\ doublet,  the \oi\ doublet,  the \ha-\nii\ complex, and   the \sii\ doublet. For the fits, we assumed that the background varied linearly with wavelength around each line or complex, and that the FWHM of all lines in each complex was the same.

\section{Results}

Table~5
lists the information we obtained for the SNR candidates for which we obtained spectra.  Specifically we list (1) the source name, (2) the extracted  \ha\ flux, (3-9) the fluxes of primary emission lines, relative to \ha = 300,  and (10) the total \sii:\ha\ ratio.
For doublets where the line ratio is constrained by atomic physics, i.e.\ \oiii, \oi, and \nii, we have listed only the stronger line.  We visually inspected all  the spectra and the fits to them; values which we judged to be more uncertain are indicated by a tilde in the table.  No allowance has been made for additional uncertainties associated with  background sky subtraction.  A few objects were observed with two different masks; in these cases, we used the spectrum that we judged to be of higher quality.  Table~6 lists the same information for the \hii\ region spectra.

\begin{figure}[htb!]
\plottwo{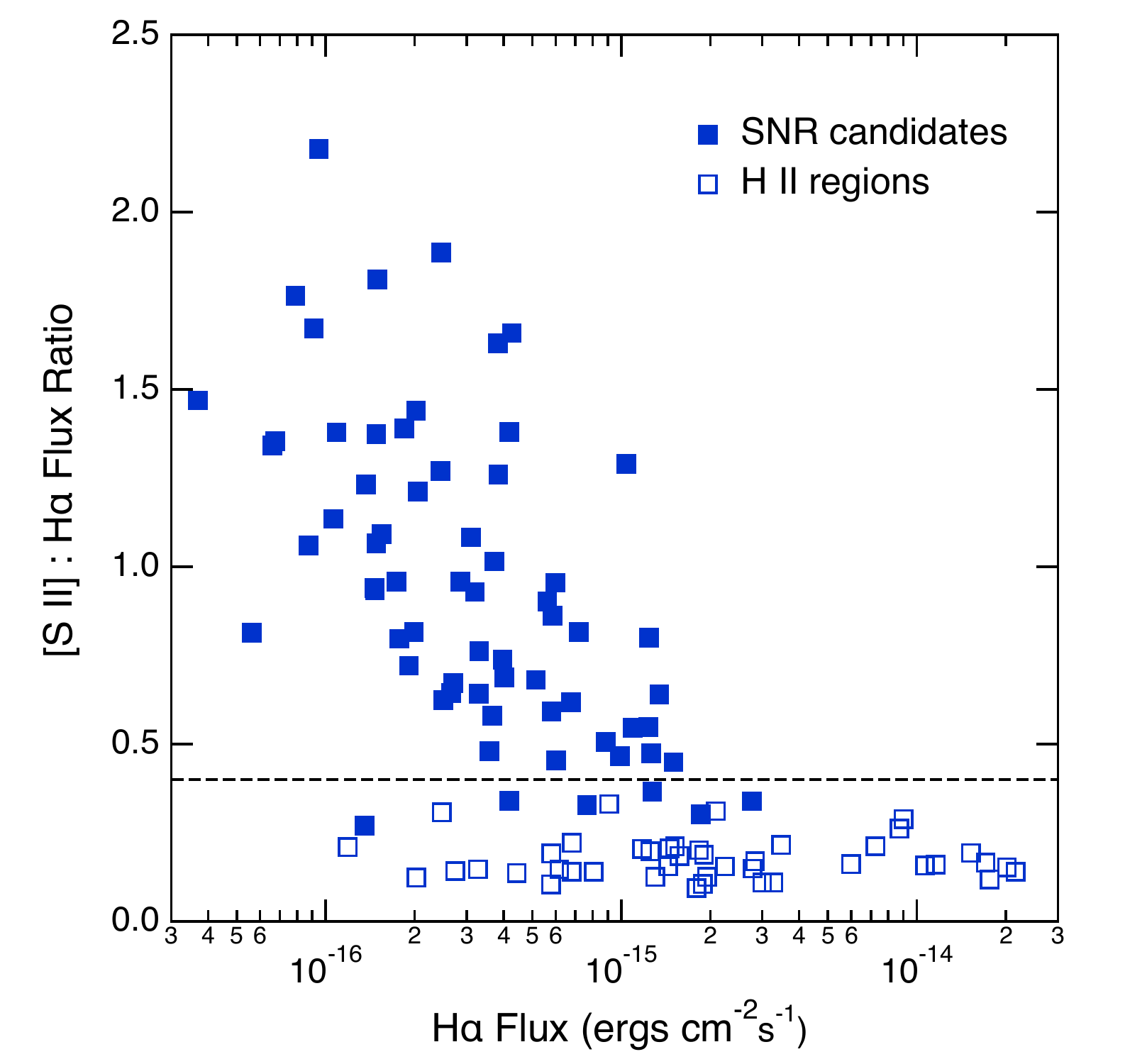}{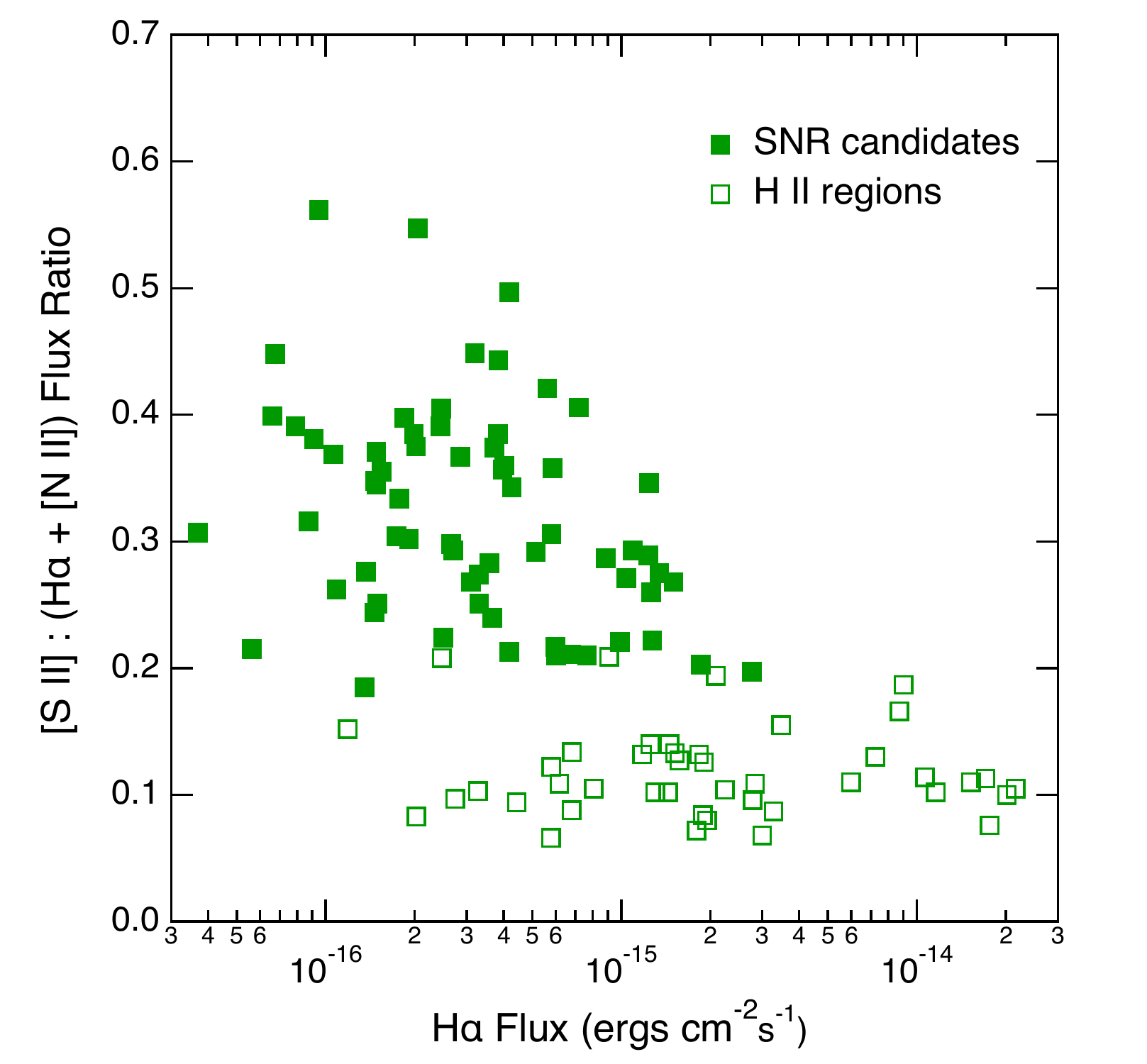}
\caption{({\em left})  \sii:\ha\ ratio for the 66 SNR candidates and 44 \hii\ regions for which we obtained GMOS  spectra. For 60 of the SNR candidates, the \sii:\ha\ ratio measured spectroscopically is $\ge 0.4$ (dashed line), so these  are almost certainly bona fide SNRs.  Meanwhile, the \sii:\ha\ ratios for all the \hii\ regions are well below 0.4.  ({\em right}) Spectroscopic measurements of the \sii:(\ha\ +\ \nii) ratios for the same objects.  This demonstrates the efficacy of selecting candidates using an ``\ha" filter that also passes the \NiiL\ lines, as do both the ACS F658N and the GMOS \ha\ one that we used for candidate selection. Both sets of ratios are plotted as a function of \ha\ flux.}
\label{fig_s2_ha_ratio}
\end{figure}

Fig.~\ref{fig_s2_ha_ratio}, (left panel)  shows a plot of the \sii:\ha\ ratio for all the objects for which we obtained spectra.  Of the 66 SNR candidates with spectra, we measured \sii:\ha\ ratios $>0.4$ for 60 of them, leading us to conclude that these may be considered bona fide SNRs.  (This number includes 43 of the 44 `A' candidates we observed, and 15 of the 18 `B' ones.)   Furthermore, of these 60 with strong \sii:\ha\ ratios, 56 also show clear evidence for \OiL\ emission---another strong indicator of shock excitation.  So too does one of the candidates with a marginal \sii:\ha\ ratio; hence we conclude there are at least 61 bona fide SNRs in our sample.  
 The five candidates not yet confirmed are W21-006, -021, -062, -124, and -179.  While we cannot confirm that these objects are SNRs, neither can we conclude the opposite, since coincident photo-ionized gas could dilute emission from SNR shocks.  Instead, these remain SNR {\em candidates}.

\section{Discussion}

\subsection{Unusually Strong Forbidden Lines}

\begin{figure}
\plottwo{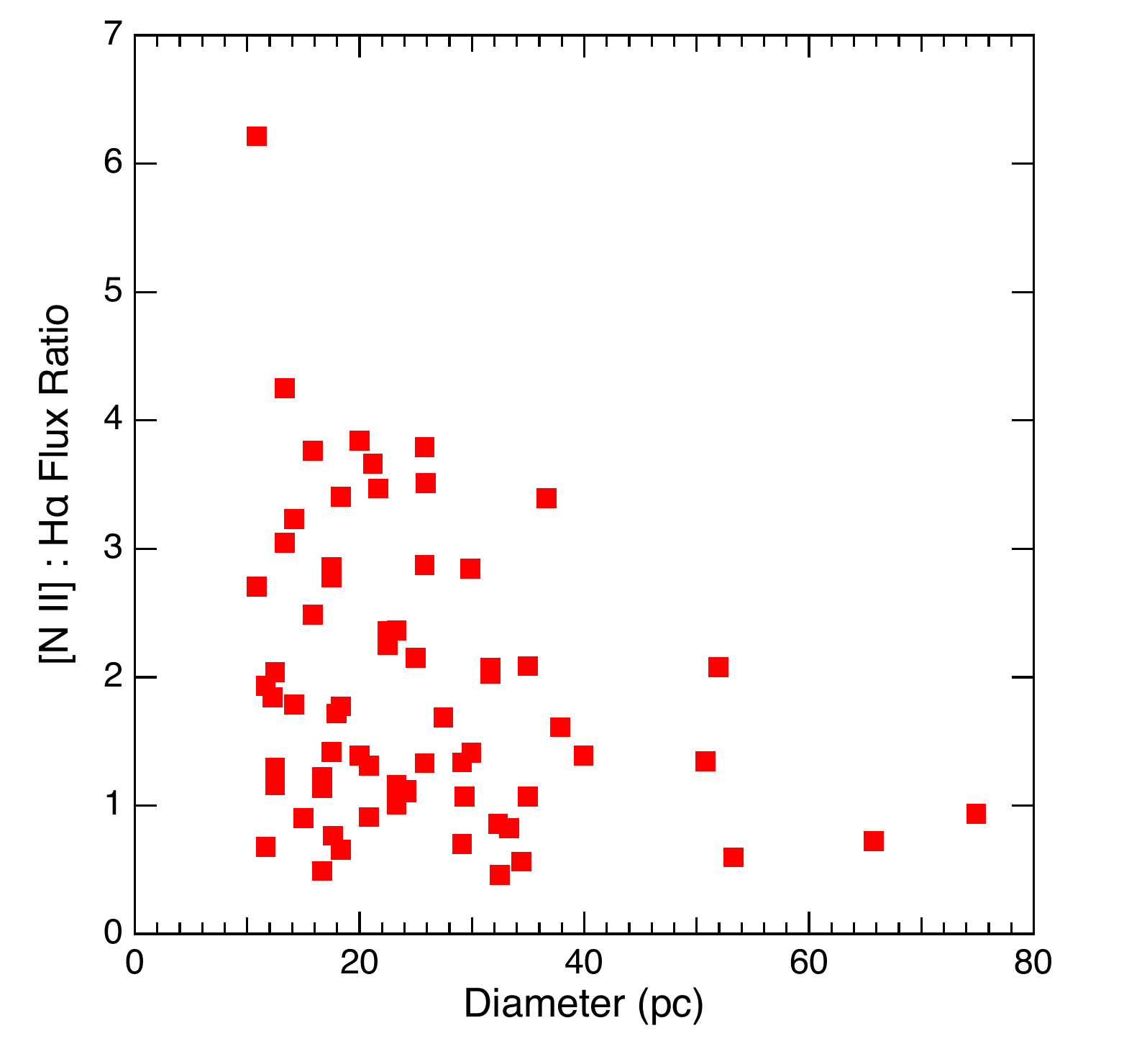}{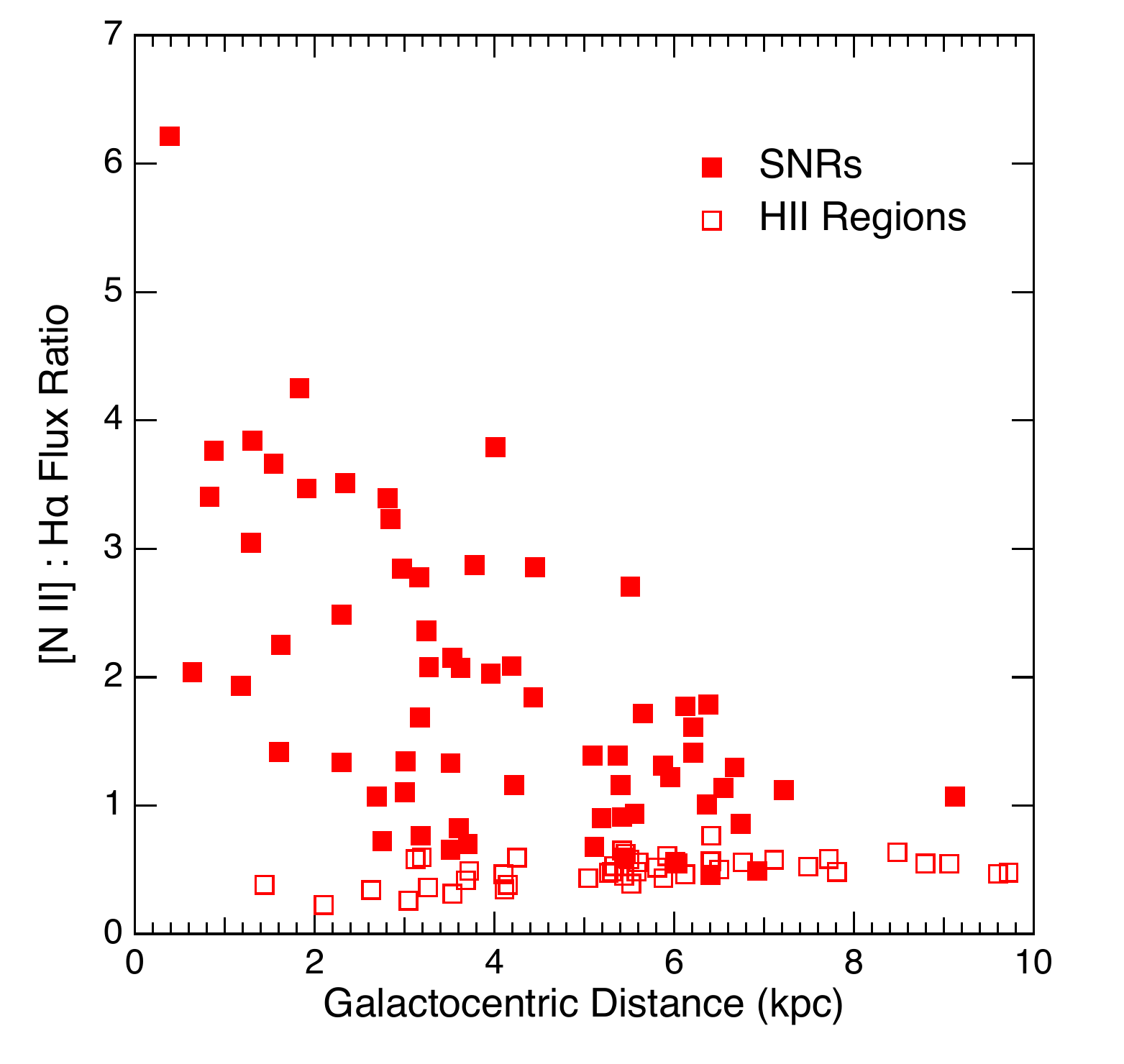}
\caption{({\em left}) Ratio of the nitrogen doublet, \NiiL\ to \ha\ plotted as a function of SNR diameter. ({\em right}) The same  \nii:\ha\ ratio as a function of galactocentric distance (GCD).  The \nii\ lines are extremely strong in the smallest remnants, and they also show a strong gradient with GCD, albeit with large observed dispersion.  \hii\ regions do not show a similar gradient and the ratios are well behaved.
\label{fig_nii_lines}}
\end{figure}

\begin{figure}
\plottwo{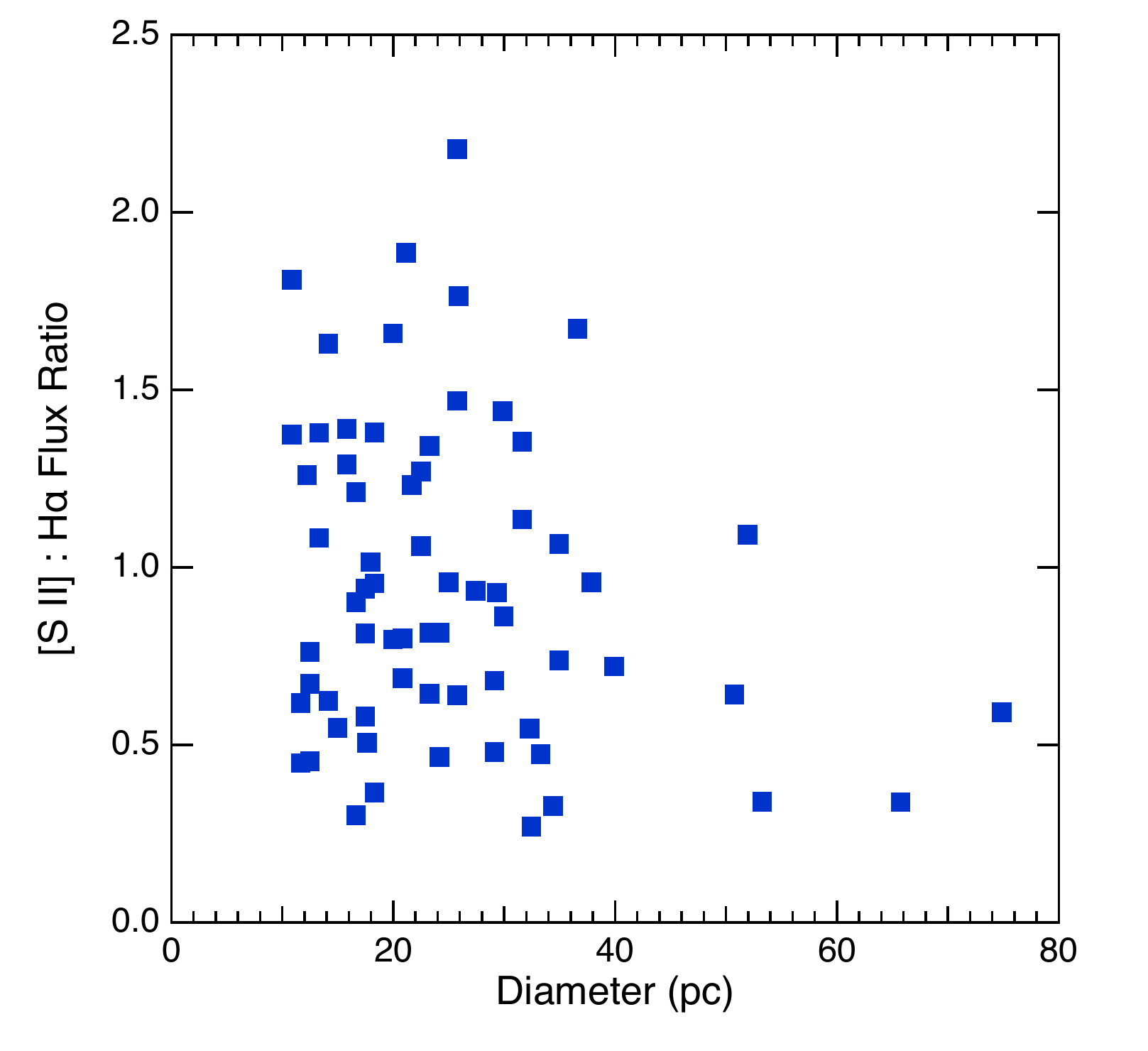}{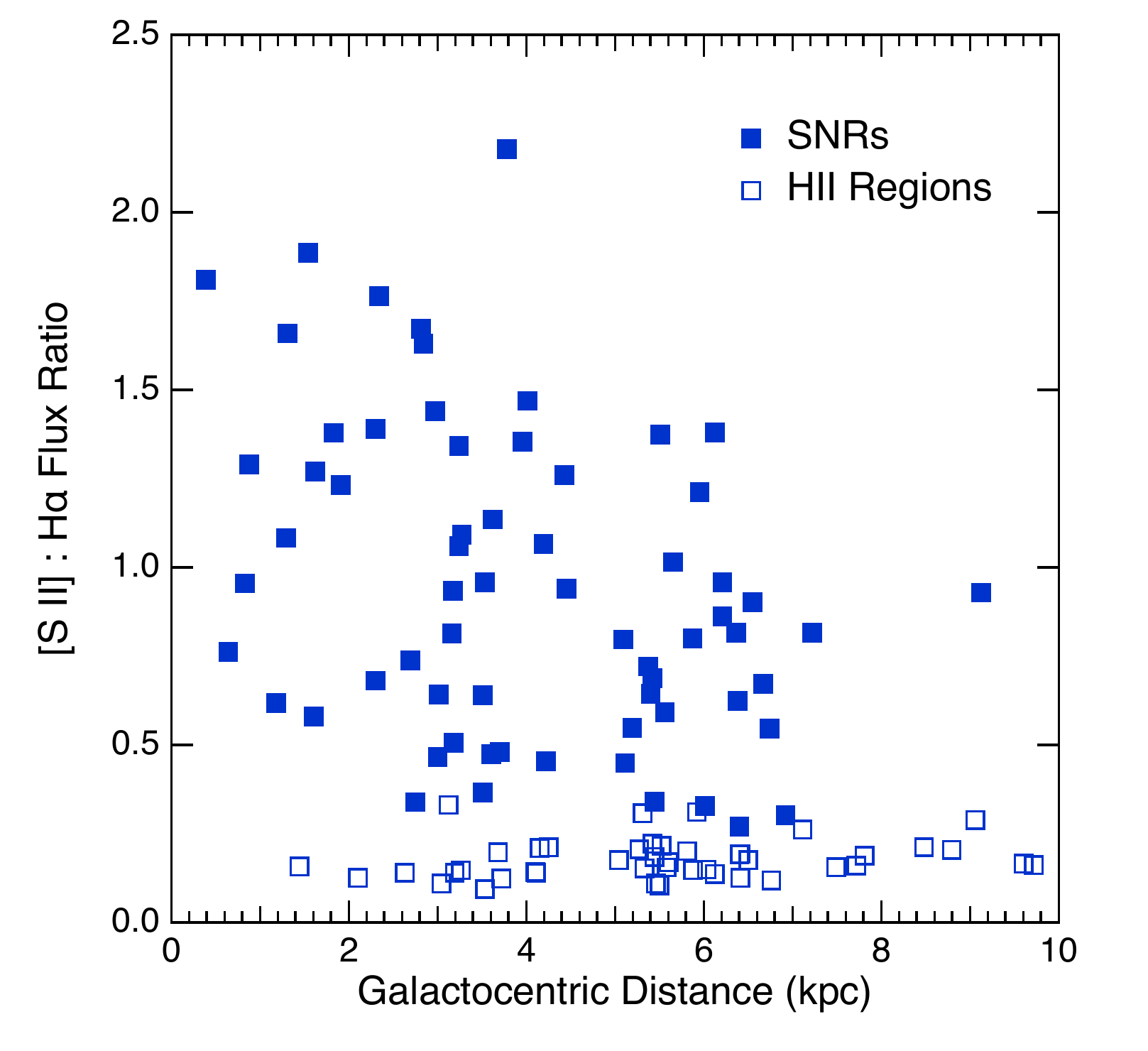}
\caption{Same as Fig.~\ref{fig_nii_lines}, but for the \SiiL:\ha\ ratio.  There is a clear gradient with both diameter and GCD (again with large dispersion), but the gradient is not as strong as for the \nii\ lines.}
\label{fig_sii_ha_ratio}
\end{figure}

The most noticeable feature of our M51 SNR spectra are the extremely strong \NiiL\ lines, and to a lesser extent, the \SiiL\ lines, both with respect to \ha.   Other spiral galaxies with  rich SNR populations, e.g., M33 \citep{long18}, M81 \citep{matonick97, lee15}, M83 \citep{winkler17}, M101 \citep{matonick97}, and NGC~6946 \citep{long19}, show a similar effect, but it is  more extreme in M51\@.  In fact, in a few cases the weaker \nii\ 6548 \AA\ line is stronger than \ha---something we are unaware of in SNRs from any of the other galaxy samples.  As shown in Fig.~\ref{fig_nii_lines}, the \nii\ lines are significantly stronger for smaller remnants, and they also show a strong gradient with increasing galactocentric distance (GCD; listed as R in the tables), albeit with very significant dispersion among objects. Fig.~\ref{fig_sii_ha_ratio} shows a similar but less extreme effect seen in the \sii:\ha\ ratio.  
 The SNRs with the most extreme \nii:\ha\ ratios, all $>3.5$, are W21-088, -064,  -074, -106, -085, and -051.  These are all near the center of M51: five of the six have $R < 2$ kpc, and the sixth, W21-106, is at $R = 4.0$ kpc.  All are relatively small, $D < 26$ pc, and all have \sii:\ha\ ratios $> 1.29$.  Looking instead at the SNRs with the largest \sii:\ha\ ratios, the six with \sii:\ha\  $> 1.65$, three are also among the six with the strongest \nii\ lines: W21-051, -088, and -074.  The others, W21-025, -068,  and -034 also have strong \nii\ lines, \nii:\ha\ $ > 2.8$.  The ones with the strongest \sii\ lines have a wider range of sizes than the \nii-strongest ones, $10 {\rm pc} < d < 37 {\rm pc}$, and they are somewhat less concentrated near the center of M51: only three are at $R < 2$ kpc, though all are at $R < 4$ kpc.

In stark contrast to the SNRs, in the right panels of both Fig.~\ref{fig_nii_lines}
and Fig.~\ref{fig_sii_ha_ratio} we see that the \hii\ regions do {\em not} show a
similar line gradient or scatter with GCD. Indeed, in analyzing the actual
abundances of \hii\ regions in M51, \cite{bresolin04} find quite modest overall
abundance levels near or below solar and almost no abundance gradient in M51, although they did find that nitrogen was somewhat enhanced relative to O:  N/O $\simeq +0.3$ dex, relative to the solar value.\footnote{Abundance analyses in M51 using strong line diagnostics have shown a
range of results.  However,  \cite{bresolin04} obtained
electron temperatures directly and thus improved (over previous work) abundance determinations;
their work is what we reference here.  \citet{croxall15} carried out an analysis similar to that of \cite{bresolin04} for more \hii\ regions in M51 and reached similar conclusions, though they did measure a modest N/O abundance gradient.}
So what is the cause of such elevated and variable forbidden-line strengths
in the SNR population?

The uncertainties for most of the line fluxes were limited by the background sky subtractions. While it is difficult to quantify these, we investigated qualitatively our background subtractions for the most extreme cases. If \ha\ was being over- or under-subtracted significantly on an object-by-object basis, this could impact the ratios to \ha.  Our inspection of the 2-D spectra, however, provides confidence that this is not the case to any significant extent.  Even for the object with the single most extreme \nii:\ha\ ratio (W21-088), we find the background subtraction to have been nominal.  The vast majority of the spectra are well-detected in the red, and low signal-to-noise or poor sky subtraction cannot explain the observed line ratios.  We also note that the \hii\ region sample, processed in the same manner as the SNR sample, shows no such effect.  


Could variations in abundances be responsible for the unusually strong lines?  The smaller SNRs are almost certainly younger than the larger ones, and have consequently swept up less interstellar material.  If many of these smaller/younger remnants encounter circumstellar shells, enriched in N as a result of He-burning and shed during the red-giant and/or asymptotic-giant-branch phases of their progenitors' evolution, this would naturally lead to the strong \nii\ lines we observe. However, any impact from this effect would be expected only for the very smallest/youngest SNRs.  As the left panel in Fig.~\ref{fig_nii_lines} shows, the effect is present for objects as large as 40-50 pc---surely not very young objects.  Furthermore, the \hii\ region sample is well-behaved and shows no hint of any large N-abundance variations from object to object.  While this might have been masked in previous studies that have observed giant \hii\ region complexes, our \hii\ region sample includes many smaller, compact \hii\ regions and ones directly adjacent to many of the SNR candidates themselves.  If large local variations in abundances were present, our \hii\ sample might be expected to show it, yet it does not.

 Less extreme versions of this effect for SNR samples in other galaxies have been attributed to variations in the shock conditions from object to object.   There are  many parameters that can be varied in shock models (e.g., magnetic field, preshock density, precursor emission, shock completeness, in addition to abundances), and perhaps some combination of parameter variations could be responsible for the observed effect.  The more extreme variations seen in M51 SNRs have caused us to revisit this idea in more detail, and to conclude that within the context of currently published models, it appears difficult to account for the observations.

\begin{figure}
\plottwo{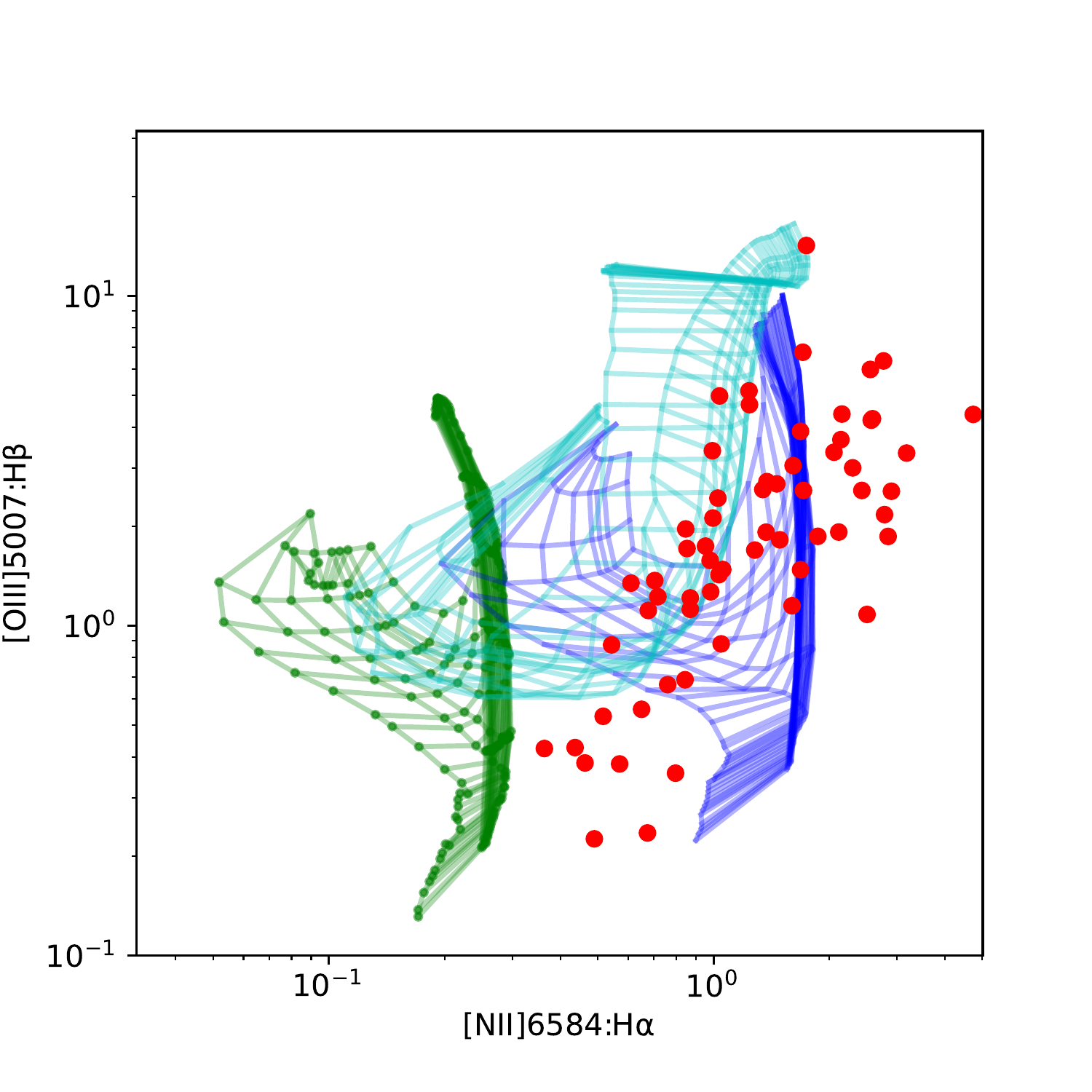}{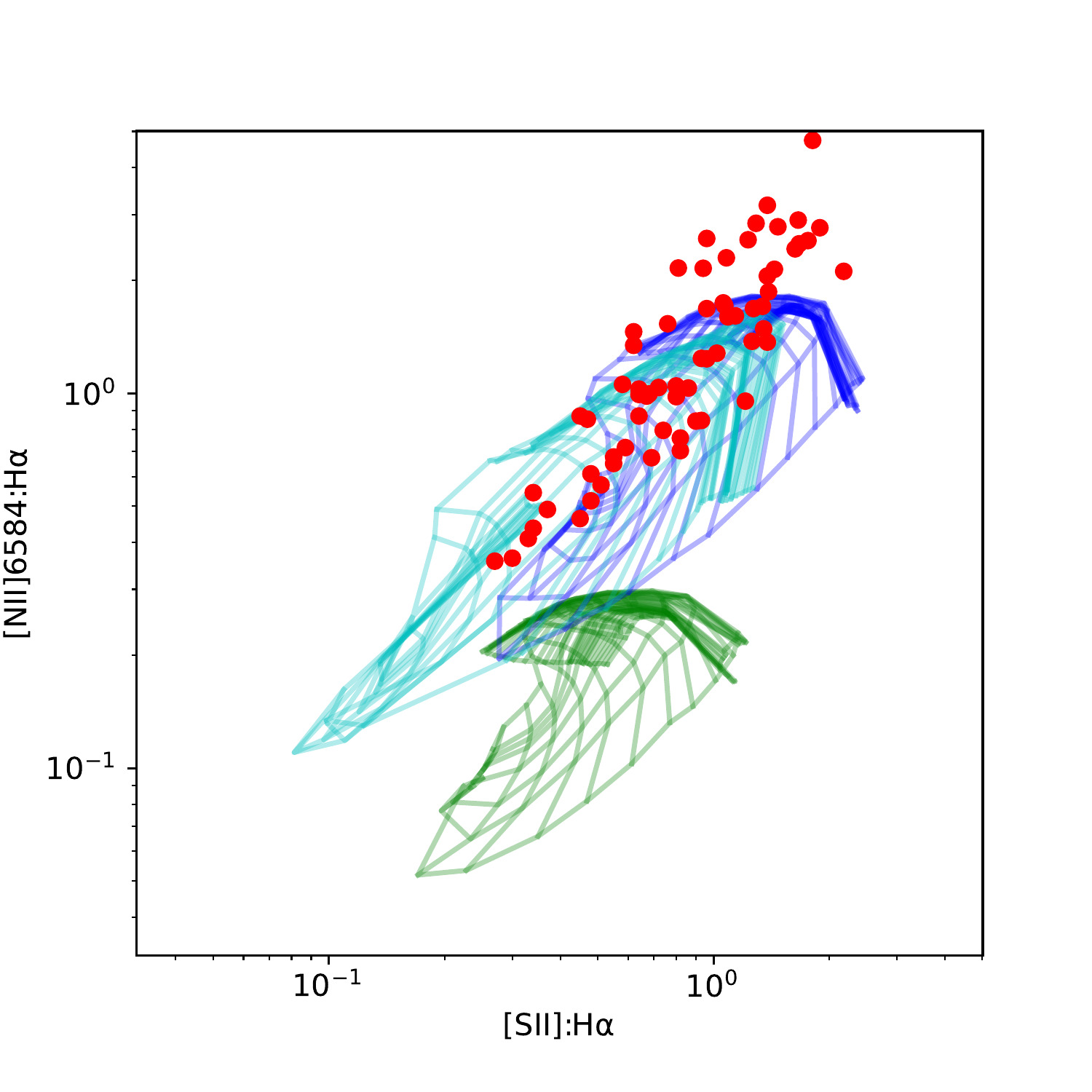}
\caption{Line ratios we have measured in M51, compared to  shock models from \citet{allen08}  with a range of shock velocities and pre-shock magnetic fields, and with metallicities corresponding to the  LMC (green),  solar (blue) and twice solar (cyan).  Both plots suggest a range of metallicities for SNRs in M51, from somewhat below to somewhat above Galactic values. However, the objects with the most extreme \nii:\ha\ ratios are not matched.}
\label{fig_model_compare}
\end{figure}

\begin{figure}[hb!]
\epsscale{.65}
\plotone{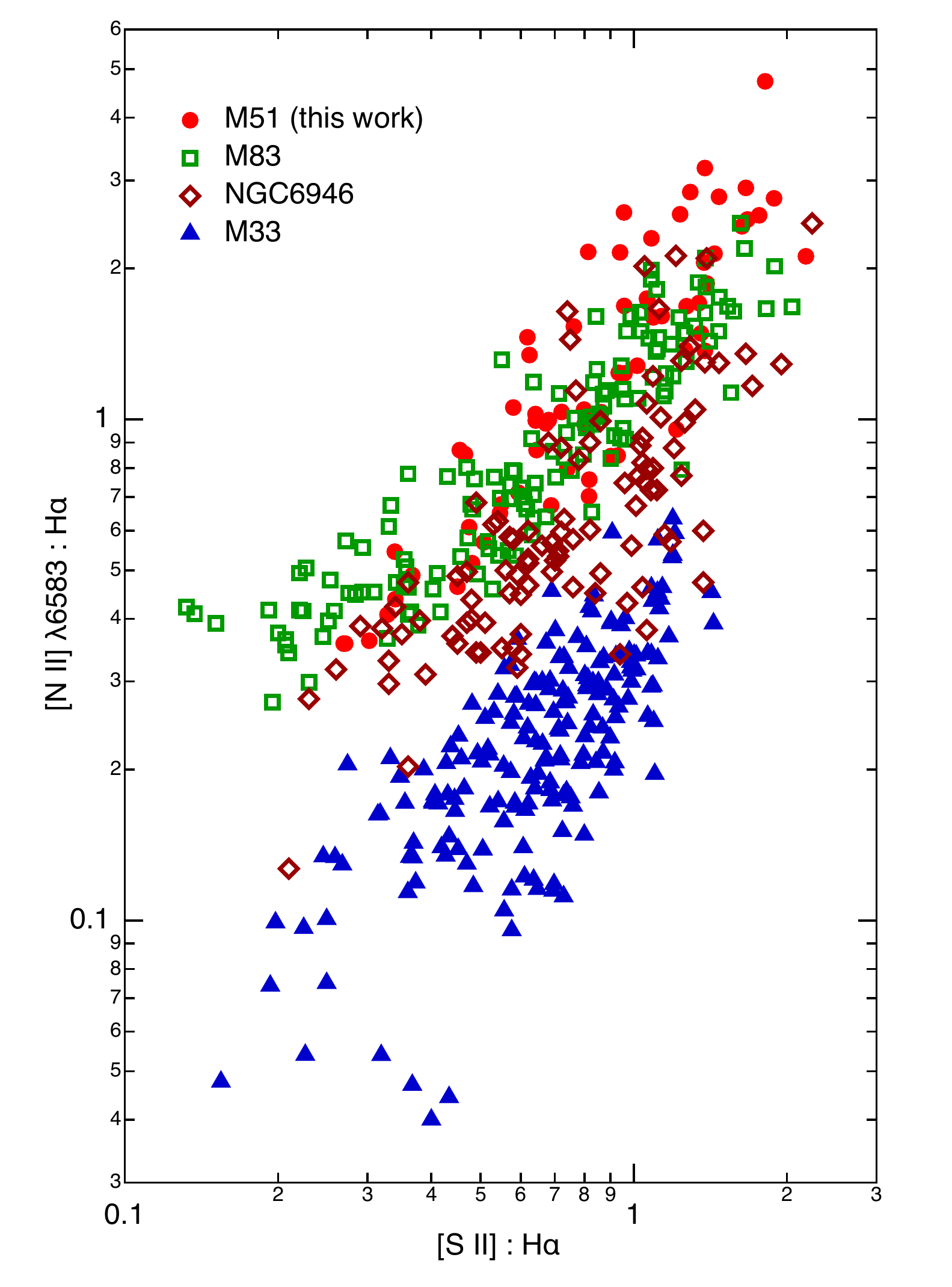}
\caption{\nii$\lambda$\,6583:\ha\ flux ratio plotted against the \SiiL:\ha\ ratio for SNRs in several spiral galaxies.  References: M33: \citet{long18}; M83: \citet{winkler17}; NGC\,6946: \citet{long19}  The same plot for M51, with models overlaid, is shown Fig.~\ref{fig_model_compare}, right.
\label{fig_nii_sii_gals}}
\end{figure}

We have used the extensive grid of shock models from \cite{allen08} to investigate the variations in \nii:\ha\ and \sii:\ha\ for the broad range of the parameter space covered by the grid, as shown in Fig.~\ref{fig_model_compare}.  This grid includes both models  with and without self-consistent pre-ionization, over a velocity range from 100 to 1000 $\kms$, and over a broad range of assumed pre-shock densities and magnetic field strengths.  Additionally, the grid covers three separate abundance sets, characteristic of the LMC, solar, and 2$\times$solar abundances. Our inspections of these models find no cases where the expected \nii:\ha\ ratio exceeds a value of 2, let alone our observed values for many objects up to 4 and above. Indeed, the 2$\times$solar-abundance models actually have somewhat lower ratios than the solar models; apparently, raising the overall abundances affects the cooling carried in various lines, with an overall effect of lowering the \nii:\ha\ ratio.  Fig.~\ref{fig_nii_sii_gals} shows a plot of \nii:\ha\ vs/ \sii:\ha\ for several other galaxies in addition to M51, emphasizing M51's  extreme nature.

The model grids tend to hold one parameter constant while varying other parameters.  While it could be true that varying multiple parameters all in the direction that seems to increase the \nii:\ha\ and \sii:\ha\ somewhat might produce higher values of the ratios than seen in the grid, this remains conjecture until such models are actually calculated. One could also selectively increase the N abundance in the models, but as mentioned above, there is no indication of such an abundance anomaly in the \hii\ sample.


The main difficulty with such elevated ratios is that these singly-ionized species are normally formed in about the same temperature zone behind the shock; hence, driving the ratios to widely variable values is difficult.  It could be that thermal instabilities in combination with magnetic field or other shock parameters are at play,  favoring regions where hydrogen is ionized even as N and S are able to recombine, and thus driving up the observed ratios. However, for these distant SNRs, we are obtaining what are effectively global spectra of their radiative filaments, not localized filaments that might be expected to show such effects more clearly if they were present.  It also would not explain why there is such dramatic variation between different objects.  Hence, at present we are left with a conundrum.


\subsection{Other Emission Line Ratios}

\begin{figure}
\plottwo{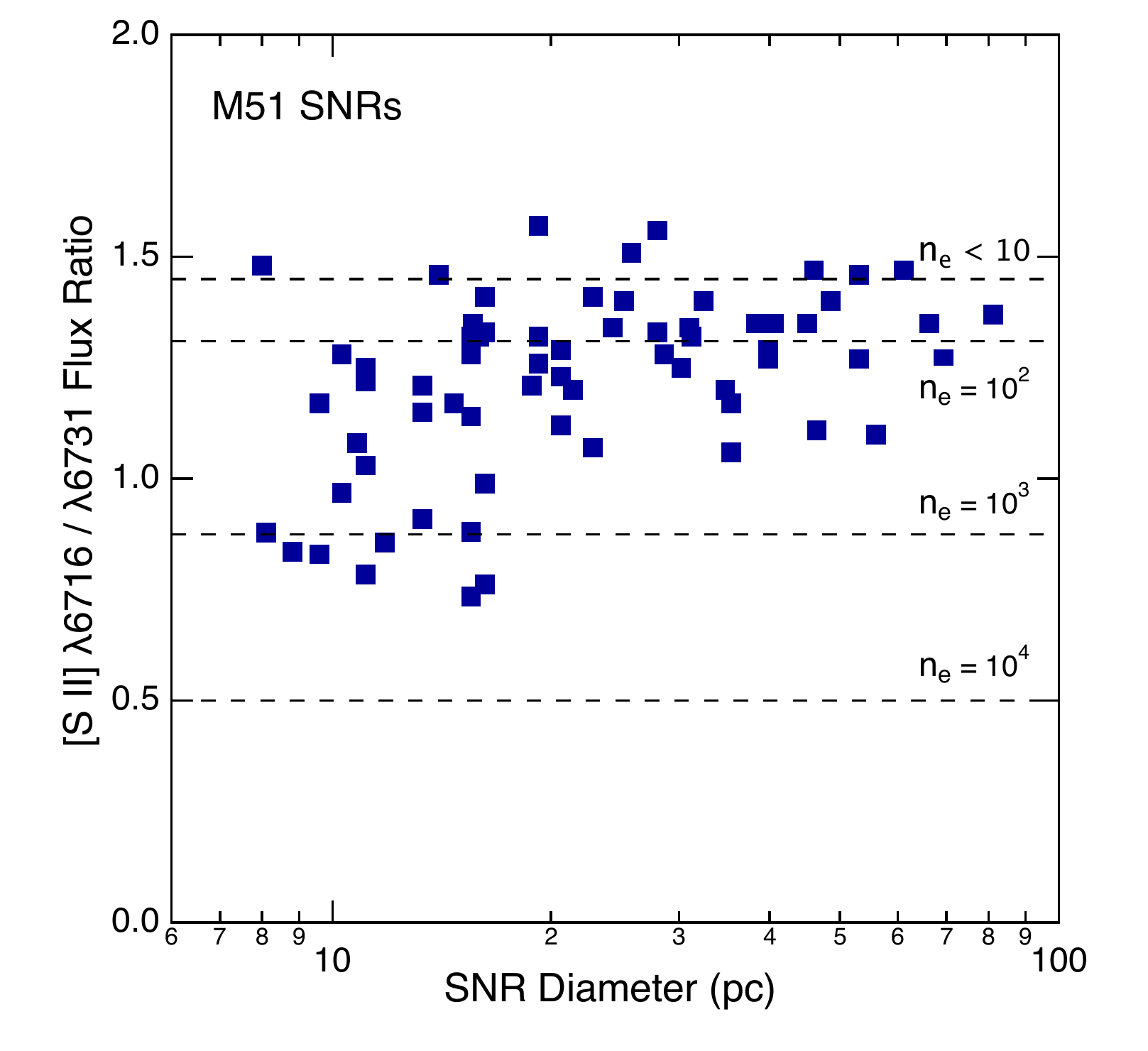}{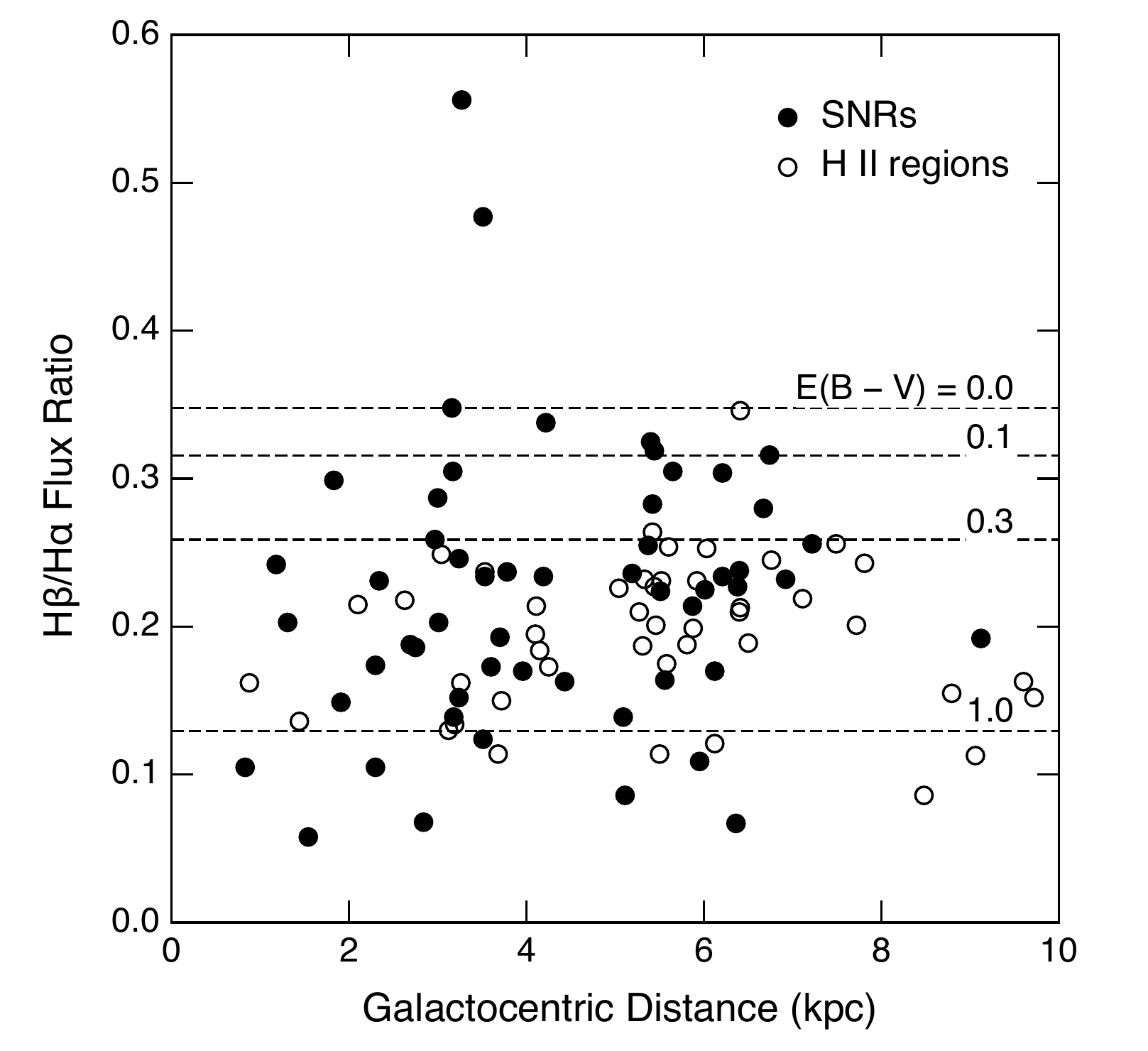}
\caption{({\em left}) Density-sensitive ratio \sii\,$\lambda6717$:$\lambda6731$ flux ratio \citep{osterbrock06}, as a function of the SNR diameter.  Not surprisingly, the smaller (and hence probably younger) objects generally have higher densities.  ({\em right}) The \hb:\ha\ flux ratio as a function of galactocentric distance, for both SNRs and \hii\ regions in M51.  The dashed lines indicate various color excess values.  The relatively high absorption is not surprising, since both SNRs and \hii\ regions are predominantly located in M51's dusty spiral arms.}
\label{fig_hb_ha}
\end{figure}

The \sii\ $\lambda$6717:$\lambda$6731 line ratio is a well-known density diagnostic \citep[e.g.,][]{osterbrock06}.  In SNR spectra, this density refers to the density in the post-shock recombination zone, but it is related to pre-shock density (depending on the details of the model assumptions one uses), which in turn is an indication of the ISM conditions surrounding each SNR\@.  In Fig.\ \ref{fig_hb_ha}, left, we show this ratio, and there is a trend toward lower values of the ratio (higher densities) for smaller diameter (generally younger) objects, but there is significant variation from object to object.  This is a pattern  we have seen in other galaxies as well, and is consistent with an interpretation that varying conditions in the ISM is a significant driver of the evolution of the SNR sample as a whole.

In Fig.\ \ref{fig_hb_ha}, right, we show the \hb:\ha\ Balmer line ratio, which is nominally an indicator of extinction.  Assuming case B, the unreddened ratio would be 0.35, as shown in the figure, with lower values indicating increased extinction.  One caveat is that there is a complication with fainter objects and higher extinctions, both of which can make the \hb\ line weaker and more uncertain, thus affecting the measured ratio. The few values above 0.35 are unphysical and are likely afflicted by this effect. Objects with lower ratios may have weak, poorly determined \hb\ values, and hence uncertain (but high) extinction.  However, for well-measured objects, the figure shows significant variation, with the lowest ratios indicating $E(B-V)$ values of $\sim 1.5$.  Similarly, \citet{calzetti05} found a high dispersion for \hii-emitting knots in M51, especially near the center of the galaxy.
Since most of the SNRs are found within the spiral arms and dust lanes, this variation is not surprising, and is similar to what is observed in other spiral galaxies.

\subsection{Sources Identified at X-ray and Radio Wavelengths}

SNRs emit radiation over a wide wavelength range.  In M33, for example, where 217 SNRs have been identified, 155 were detected at radio wavelengths by \cite{white19}  in a deep survey carried out with the Jansky Very Large Array.  Additionally, 112  were detected (with $>3 \sigma$) at X-ray wavelengths,  either with XMM-Newton \citep{garofali17} or Chandra \citep{long10}, and  98 SNRs were detected in all three  bands.   Similarly, in M83, where  304 SNRs and SNR candidates have been compiled \citep{dopita10,blair12,blair14}, 64 were detected by \cite{russell20} in unconfused regions of a radio image created from data obtained with the Australia Telescope Compact Array.  A total of 87 of the SNRs in M83 have been detected using Chandra \citep{long14}.

M51 has, of course, also been studied at radio and X-ray wavelengths.  \cite{maddox07} compiled a catalog of 107 compact radio sources using the VLA, identifying five as likely SNRs, based on their non-thermal spectral indices and association with resolved shells in \ha\ images obtained with HST\@.  \citet{kilgard05} gave a catalog of 116 X-ray sources from  Chandra observations of M51, and \cite{kuntz16} reported a catalog of 297 X-ray sources within the D25 contour for M51 in a much deeper Chandra image of the galaxy. Both these papers suggested that a substantial number of the sources with soft X-ray hardness ratios were likely SNRs. With our data, we can shed additional light on these.

We have investigated positional coincidences of sources from both the Maddox radio and Kuntz X-ray catalogs with the 179 objects we have identified as SNRs or SNR candidates (Table 3).  There are 16 radio sources that lie within 1\arcsec\ of one of our SNR candidates, compared with three expected by chance.\footnote{We estimate chance coincidences by shifting the source positions of one of the catalogs by several arcsec in various directions and recalculating the the number of coincidences. We also allow for a small systematic offset ($<$0.3\arcsec)  in position  between any two catalogs when carrying out the matching.}  These include four of the five radio sources identified by \citet{maddox07} as probable SNRs.  There are 55 X-ray sources that lie within 1\arcsec\ of a SNR candidate, compared with  four  expected by chance.  Of the 16 optical-radio coincidences, 15 are triple coincidences with an X-ray source as well.  The identification of an X-ray and/or radio source at the position of one of the optical SNR candidates clearly increases the likelihood that the object is a bona fide SNR.

\begin{figure}[htb!]
\epsscale{0.8}
\plotone{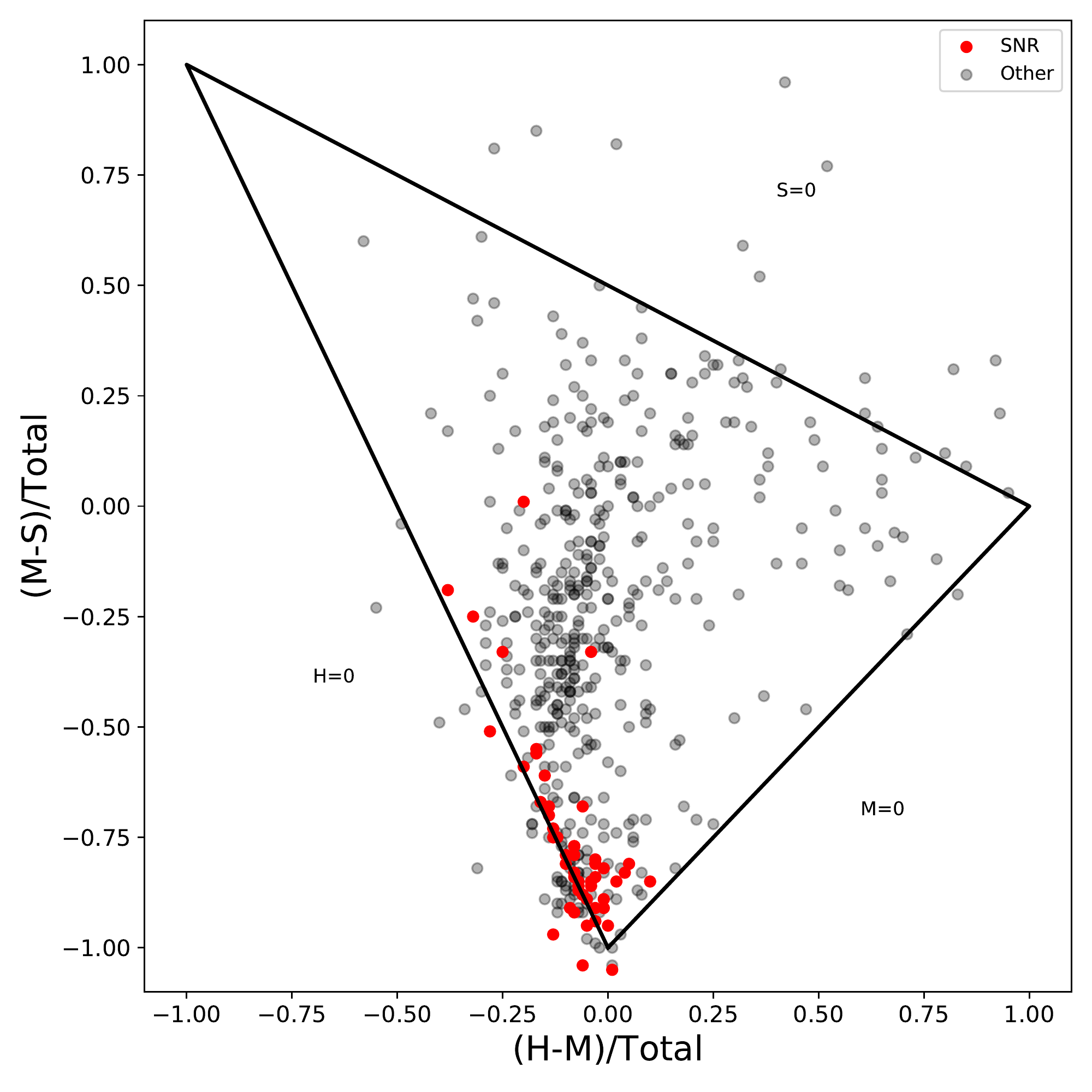}\
\caption{X-ray color-color diagram of X-ray sources in M51, where the soft (S), medium (M), and hard (H) bands correspond to energies of 0.35 - 1.1 keV, 1.1 - 2.6 keV, and 2.6 - 8 keV, respectively, and Total is the sum of all three bands.  X-ray sources lying within 1\arcsec\ of SNRs or SNR candidates from Table 3
are shown in red, while others are in grey.  Not surprisingly, those associated with SNRs are concentrated in the ``soft" region of the diagram (lower vertex of the triangle).}
\label{fig_xray}
\end{figure}

Most X-ray sources in deep X-ray studies of nearby galaxies are  black-hole or neutron-star binaries, or background AGNs---all sources with hard X-ray spectra.  In contrast, SNRs emit X-rays from shock-heated gas and typically have soft X-ray spectra. The two types of objects can be isolated in X-ray hardness-ratio diagrams.  As shown in Fig.\ \ref{fig_xray}, most of the X-ray sources that are spatially coincident with nebulae in our SNR candidate catalog lie in the region where soft X-ray photons dominate, as expected. 

\begin{figure}[htb!]
\epsscale{0.9}
\plotone{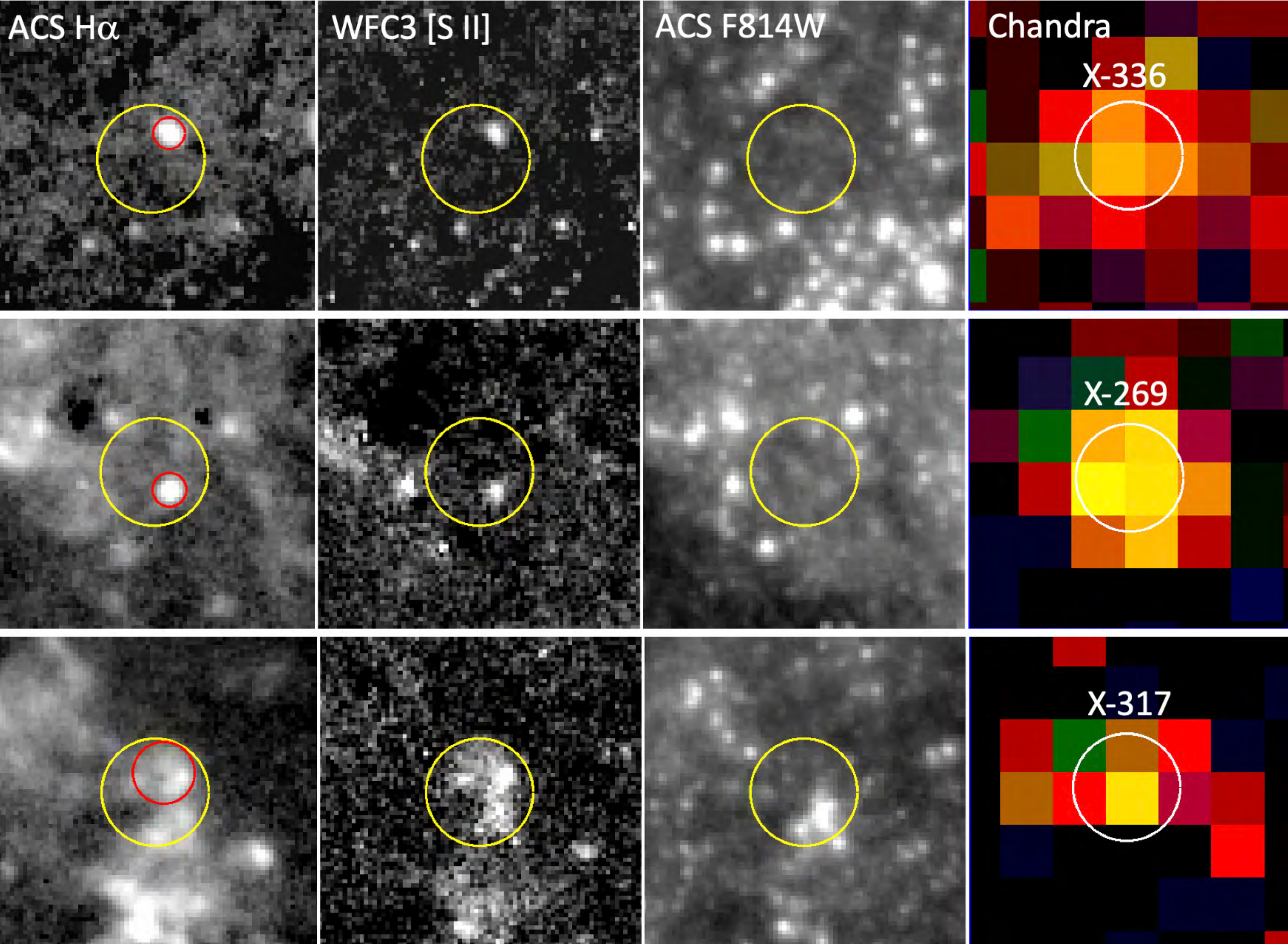}\
\caption{HST and Chandra images of several of the optical objects found to align with soft X-ray sources: X-336, X-269, and X-317. From left to right are the HST \ha, \sii, and F814W (I-band) continuum; the right panels show Chandra soft (0.35 - 1.1 keV) in red and medium (1.1 - 2.6 keV) in green.  The yellow circles are 1\arcsec\ in diameter, and the images are oriented N up, E left; the smaller red circles in the left panels mark the optical counterparts, all of which are relatively free of continuum contamination.
\label{fig_M51XS}}
\end{figure}

We have also turned this technique around and asked what are the other soft X-ray sources in Fig.~\ref{fig_xray}.  Is there optical evidence of possible SNR emission at those locations?  We projected the previously unidentified soft-source positions (those with (M\,--\,S)/Total values  $< -0.5$)  from \cite{kuntz16} onto our HST and GMOS images and performed a visual search for objects of interest.   The majority of these positions were empty or were projected onto general star fields in M51\@.  If these are SNRs, then the SN must have occurred in a sparse region of the ISM with few dense cloudlets, thus producing no bright optical remnant.  This may favor a SN Ia origin, since the time delay between star formation and SN explosion is generally longer for SNe Ia  than for core-collapse SNe \citep[e.g.,][]{totani08}.   Nine additional X-ray sources  do have coincident  emission nebulae and are thus  of possible interest; these  are listed in Table 7.

A selection of these objects is shown in Fig. \ref{fig_M51XS}.  Three of the objects could be bona fide SNRs with elevated \sii:\ha\ ratios, such as X-336, shown in the top panels. Several of the objects found are very small-diameter emission nebulae with relatively low \sii:\ha\ ratios, as in X-269 (middle panels), but the presence of the soft X-ray counterpart certainly makes them intriguing. Two objects are fainter, more extended nebulae that show some clumpiness or other structure, such as X-317 (bottom panels), but it is difficult to measure accurate ratios from the imagery. Finally, one object is just south of the active nuclear region and aligned with a complex and very extended region of emission; while shocks may be involved in this region, it is unlikely to be a SNR in the sense of the other objects we have found.

\begin{figure}[htb!]
\plotone{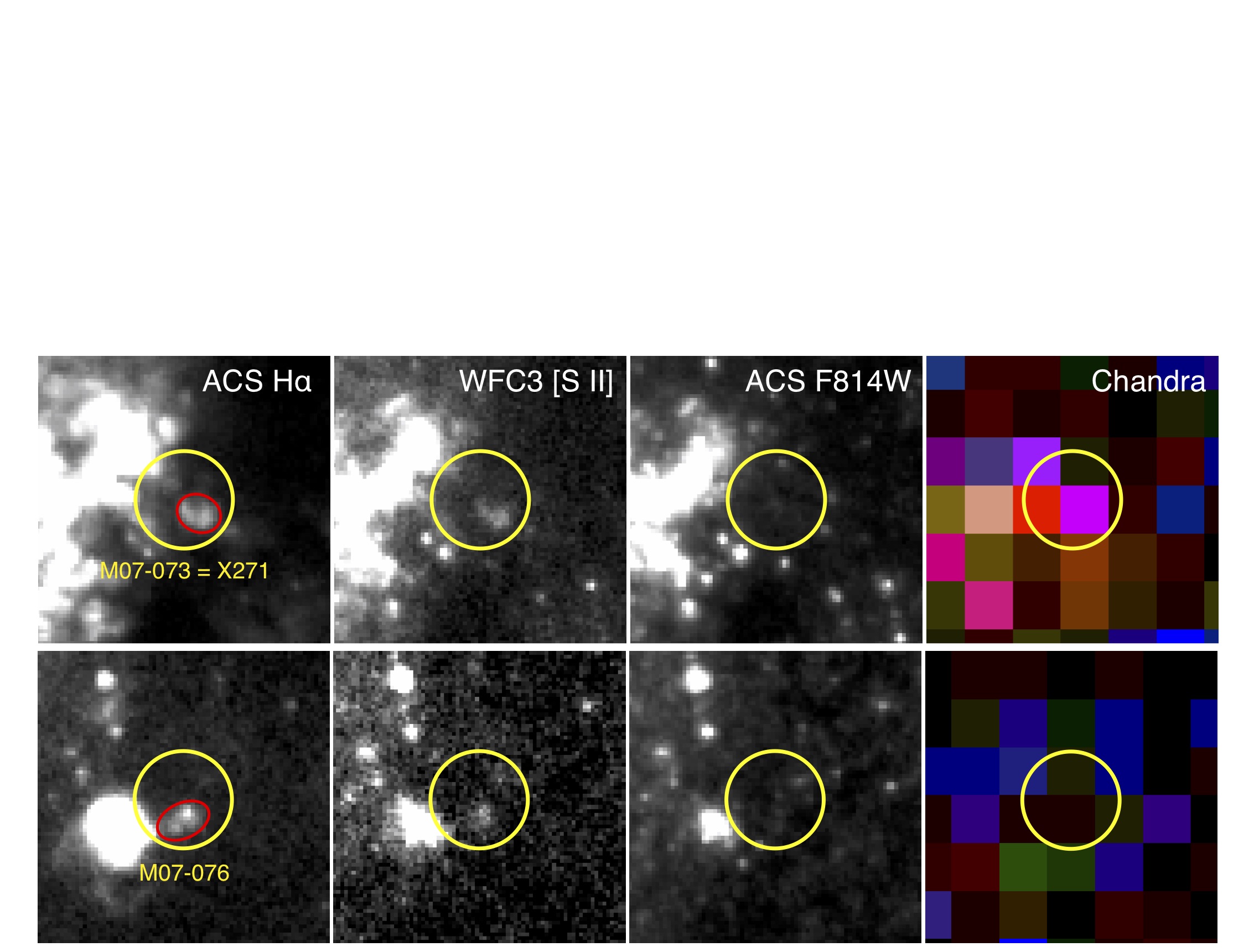}
\caption{HST and Chandra images of  the optical objects found to align with radio sources 73 and 76 from the catalog of M51 sources by \citet{maddox07}. Panels are similar to those in Fig.~\ref{fig_M51XS}:  left to right are the HST \ha, \sii, and F814W (I-band) continuum; the right panels show Chandra  0.35 - 1.1 keV in red,  1.1 - 2.6 keV in green, and  2.6 - 8.0  keV in blue.  The yellow circles are 1\arcsec\ in diameter, and the images are oriented N up, E left; the smaller red ellipses in the left panels mark the optical counterparts, both of which are relatively free of continuum contamination.
\label{fig_m51_radio}}
\end{figure}

We have also checked other sources identified by \citet{maddox07} as possible radio SNRs but that are not included in Table 3.  There are only two of these: M07-073 and M07-076.\footnote{\citet{maddox07} identified M07-076 as a probable SNR based on its radio spectral index and association with an \ha\ nebula.  M07-073 also has a slightly non-thermal spectral index ($-0.19 \pm 0.30$), but it had been identified as a likely SNR in the Chandra survey of M51 by \citet{kilgard05}.}  As shown in Fig.~\ref{fig_m51_radio}, both of these objects have coincident nebulosity, not only in \ha, but in \sii\ as well, though both have nominal \sii:\ha\ (really \sii:(\ha\,+\,\nii) ratios of $\sim$ 0.15 - 0.17, somewhat lower than for most of the SNRs in our sample (Fig.~\ref{fig_s2_ha_ratio}, right).  The source M07-073  is also coincident with the soft X-ray source X271, enhancing the likelihood that it is, indeed, a SNR.  Information about these two sources is also given in Table 7.


At the very least, these comparisons hint at the incompleteness of our current catalog, which is not surprising in the case of M51, since it is the most distant galaxy for which an extensive search for SNRs has so far been carried out.

Finally,  we mention that M51 is host to a number of ultraluminous X-ray sources (ULXs), some of which show eclipses, and others show variability \citep{terashima04,urquhart18,brightman20}.  The numbering of these sources among various authors and the accuracy of the coordinates given are not always clear.  However, projecting those nominal sources onto our data (including the Chandra data), we find two objects of likely interest.  \cite{urquhart18} source ULX-1 corresponds to our source W21-004, a very elongated optical nebula with elevated \sii:\ha.  \cite{urquhart18} show a spectrum and conclude this is a shock-heated jet from the ULX, with some weaker lines possibly indicating some X-ray photoionization as well.  By way of comparison, their ULX-2 is also surrounded by a faint semi-circular nebula, but their spectrum in this case indicates a much lower \sii:\ha\ ratio and normal stellar photoionization.  

From \cite{brightman20}, the source they list as ULX-2 is near our source W21-011, but the coordinate given is significantly off.  However, projecting onto the Chandra images shows the very bright X-ray source (also \cite{terashima04} source 5) to be aligned with W21-011, so the coordinate accuracy in \cite{brightman20} may be suspect.  The optical source in this case is compact and circular, not jet-like. Our spectrum of this source shows \sii:\ha\ = 0.45 and a fairly high density from the $\lambda$6717:$\lambda$6731 ratio of 0.85 ($\sim$1000 $\rm cm^{-3}$).  It is not clear whether this source is a SNR with very bright soft X-ray emission or whether the source is more similar to \cite{urquhart18} ULX-1 mentioned above but with a different geometry.

Of the remaining ULXs listed by \cite{brightman20}, ULX-9 is not within our field of view, but none of the others aligns even approximately with any of our sources.  Although we retain W21-004 and W21-011 in our catalog, the exceedingly strong X-ray emission from these two sources makes it clear they are not normal SNRs per se, but the \sii:\ha\ ratio criterion found them anyway.  This is directly analogous to the situation in M83, where the jet-like microquasar MQ1 was initially associated with a SNR candidate \citep{soria14}, and where a second nebula with elevated \sii:\ha\  may align with another microquasar \citep{soria20}.

\subsection{Historical Supernovae}

\begin{figure}[t!]
\epsscale{0.7}
\plotone{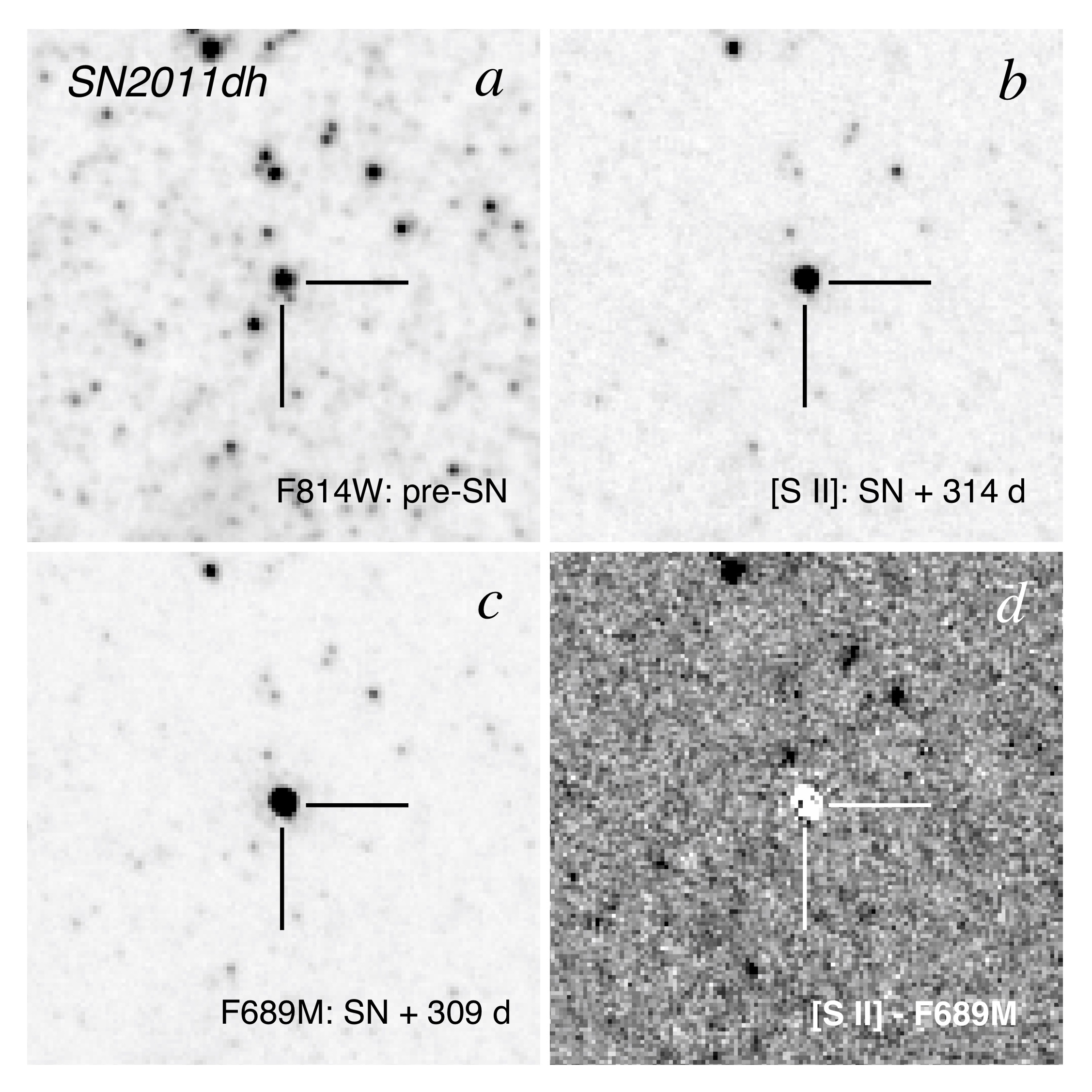}\
\caption{HST images showing the position of SN\,2011dh,  ({\em a}) red continuum (F814W) about 6 yr prior to the SN explosion \citep[this is the same data shown in][Fig.\ 1]{van-dyk11}; ({\em b}) Image in \sii\ (F673N) about  10 months post-explosion; ({\em c}) F689M band about  10 months post-explosion; ({\em d}) Difference image between \sii\ and F689M.  The SN is somewhat over-subtracted, yet residuals remain for several stars, indicating the absence of \sii\ nebular emission from SN2011dh.  The field size is 5\arcsec, oriented N up, E left; the tick marks indicate the SN position and are 1\arcsec\ in length.
\label{fig_sn20011dh}}
\end{figure}

 While M51's four historical SNe within the past 75 years (SN\,1945A, SN\,1994I, SN\,2005cs, and SN\,2011dh) may not be remarkable in an absolute sense, it does place M51 among the top five galaxies within 10 Mpc in SN productivity.
The only one of M51's historical SNe detected in the images discussed here is the Type IIb SN\,2011dh, located in the galaxy's outer spiral arm, $\sim 2\farcm6$ SE of the nucleus.  The declining SN is shown in the \sii\ and F689M images from 2012, over 10 months after the explosion (Fig.\ \ref{fig_sn20011dh}, panels b and c).  
The progenitor candidate was first identified by \citet{li11}, and is discussed by \citet{van-dyk11}.  Fig.\ \ref{fig_sn20011dh}a shows the progenitor \citep[or more likely its brighter companion, see][]{van-dyk11} in the 2005 data.
By taking a difference image between the \sii\ image and the F689M image from our 2012 data (Fig.\ \ref{fig_sn20011dh}d), we see that the SN itself appears somewhat over-subtracted, while residuals from several stars remain visible.  This indicates that no nebular emission (at least in the \sii\ lines) had developed at the time these images were taken, 10 months post-explosion.  The late-time evolution of SN\,2011dh is discussed in detail by \citet{maund19}.

\begin{figure}[ht!]
\epsscale{0.66}
\plotone{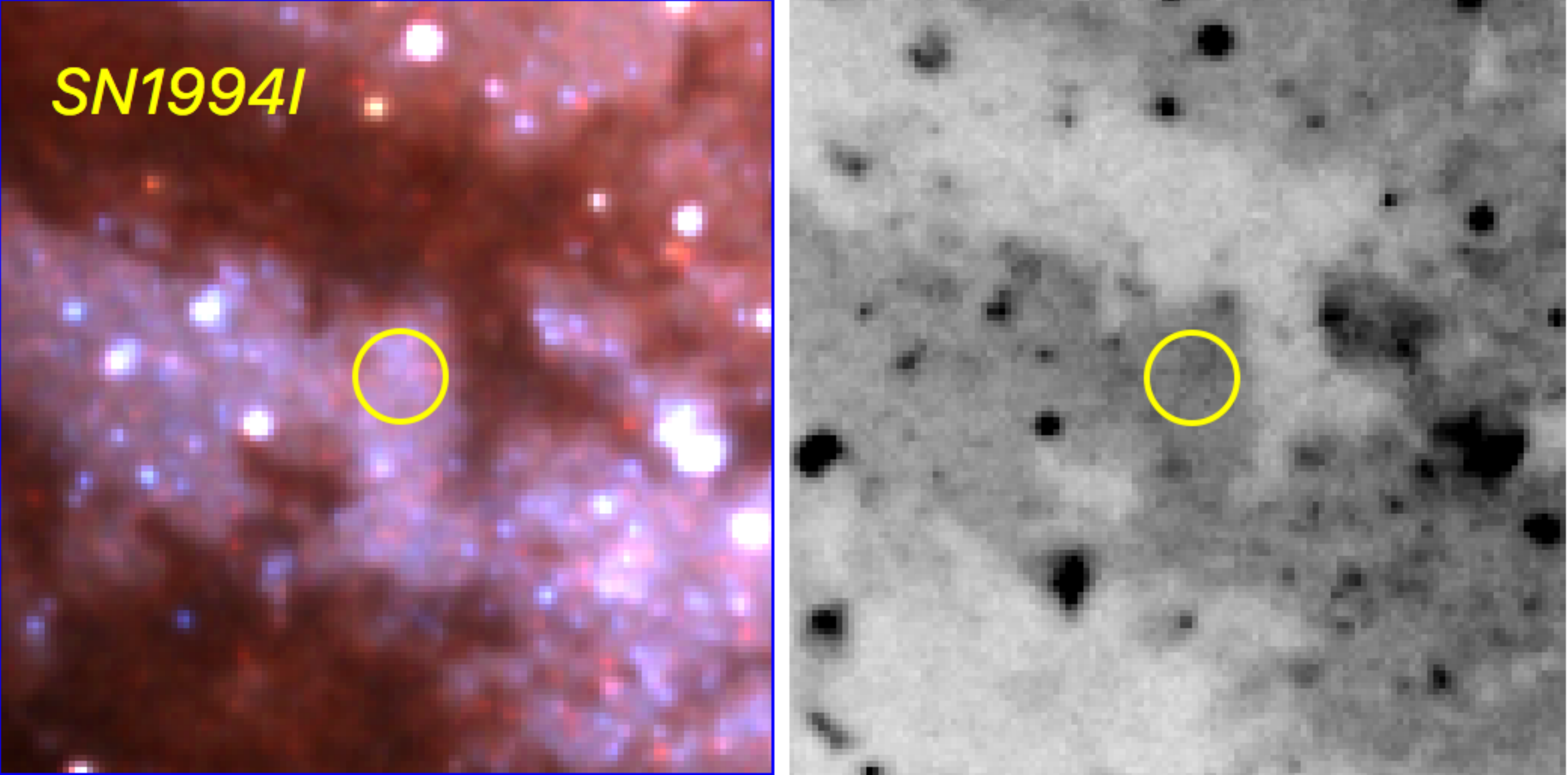}
\caption{HST ACS images from 2005 showing the position of SN\,1994i:  ({\em left})  color image where R = F814W, G = F555W, B = F435W; ({\em right}) F658N (\ha + \nii).  The field is 5\arcmin\ square, oriented N up, E left.  The circle marks the SN location and is 0\farcs6 in diameter, indicating the $3\sigma$ uncertainty in the SN position.  No nebular remnant is apparent at this position.
\label{fig_sn1994i}}
\end{figure}

 We have also examined the positions of the other three historical SNe.  The Type I SN\,1945A took place in  M51B = NGC\,5195, at a location  only about $\sim 7\arcsec$ southwest of the bright nucleus, with a large uncertainty in its position.  Nothing stands out in the HST ACS images (the only ones to cover this crowded field).    SN\,1994I (Type Ic) was located $18 \arcsec$ SE from the nucleus of M51A, also in a crowded field.  Fig.~\ref{fig_sn1994i} shows the ACS images of this field; there is no evidence for nebulosity that might indicate a remnant from SN\,1994I . 

\begin{figure}[hb!]
\epsscale{0.99}
\plotone{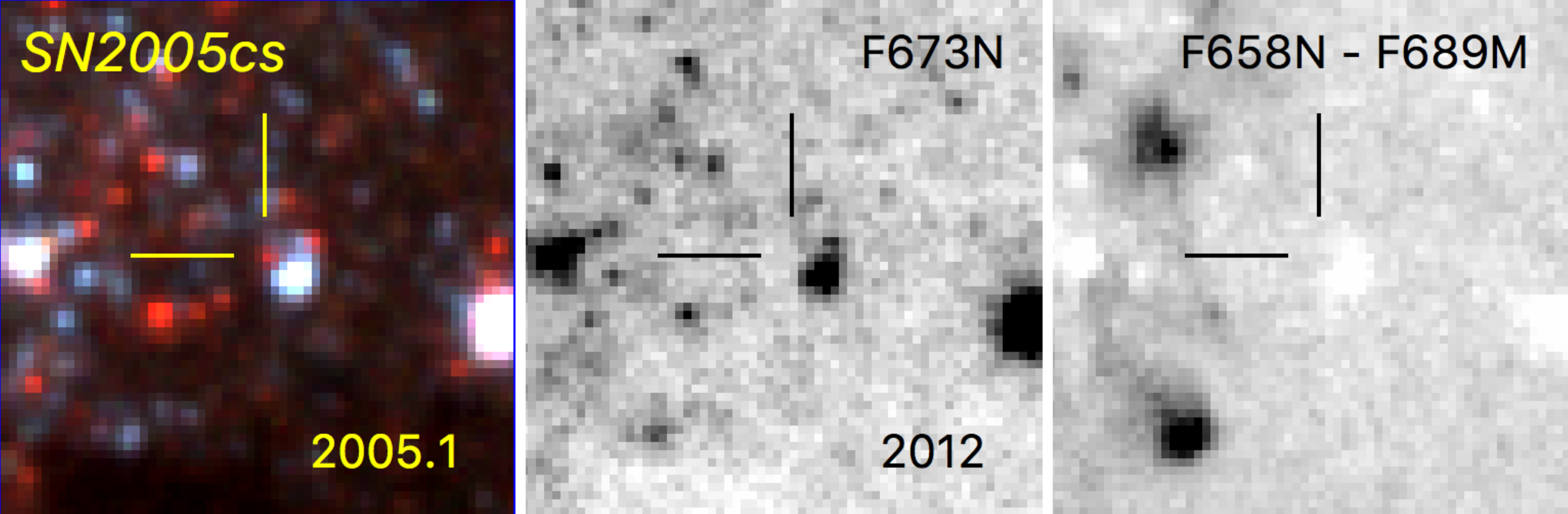}
\caption{HST images of a 3\arcsec\ square field centered on the location of SN\,2005cs: ({\em left}) color image where R = F814W, G = F555W, B = F435W, taken 5 months prior to the SN event.  The position of the red supergiant identified by \citet{li06} as the progenitor star is marked, immediately NE of a bright object that is probably a compact star cluster (the tick marks are 0\farcs 5 in length).
Li et al.\ show these same images individually, but with an even smaller field, in their Fig.~3.
({\em center}) WFC3 F673N image from 2012, 7 yr after the event.  The progenitor star is no longer visible, nor is there any resulting \sii\ nebulosity visible.  ({\em right}) Difference image between ACS F658N (\ha\ + \nii) and the WFC3 F689M continuum, clearly showing two small nebulae, unrelated to the SN, about 1\arcsec\ NE and SE from the SN position.   
\label{fig_sn2005cs}}
\end{figure}

SN\,2005cs (Type IIP) exploded on 2005 June 26, about five months after the HST ACS images in program 10452 (Table 1) were taken.  In Fig.~\ref{fig_sn2005cs}, the left panel shows a 3-color image from 5 months prior to the event in which the red supergiant progenitor identified by \citet{li06} is marked.  This star is apparently missing in the central panel, showing the  \sii\ (F673N)image from 2012.  The right panel shows a continuum-subtracted \ha\ image (ACS F658N -- WFC3 F689M), which shows two small nebulae just NE and SE of the SN position (also faintly visible in \sii\ in the central panel).  These are present both before and after the SN event; there is no apparent remnant from the SN itself.


\subsection{Comparison with Other Spiral Galaxies}

 In recent years, large populations of SNRs have been identified in numerous galaxies, including M31 \citep{lee14a}, 
M33 \citep{lee14a, long18},  M81 \citep{lee15}, M83 \citep{winkler17}, and NGC\,6946 \citep{long19}. 
How does the  population of SNRs in M51 compare with ones in many of these other galaxies?



\begin{figure}
\plotone{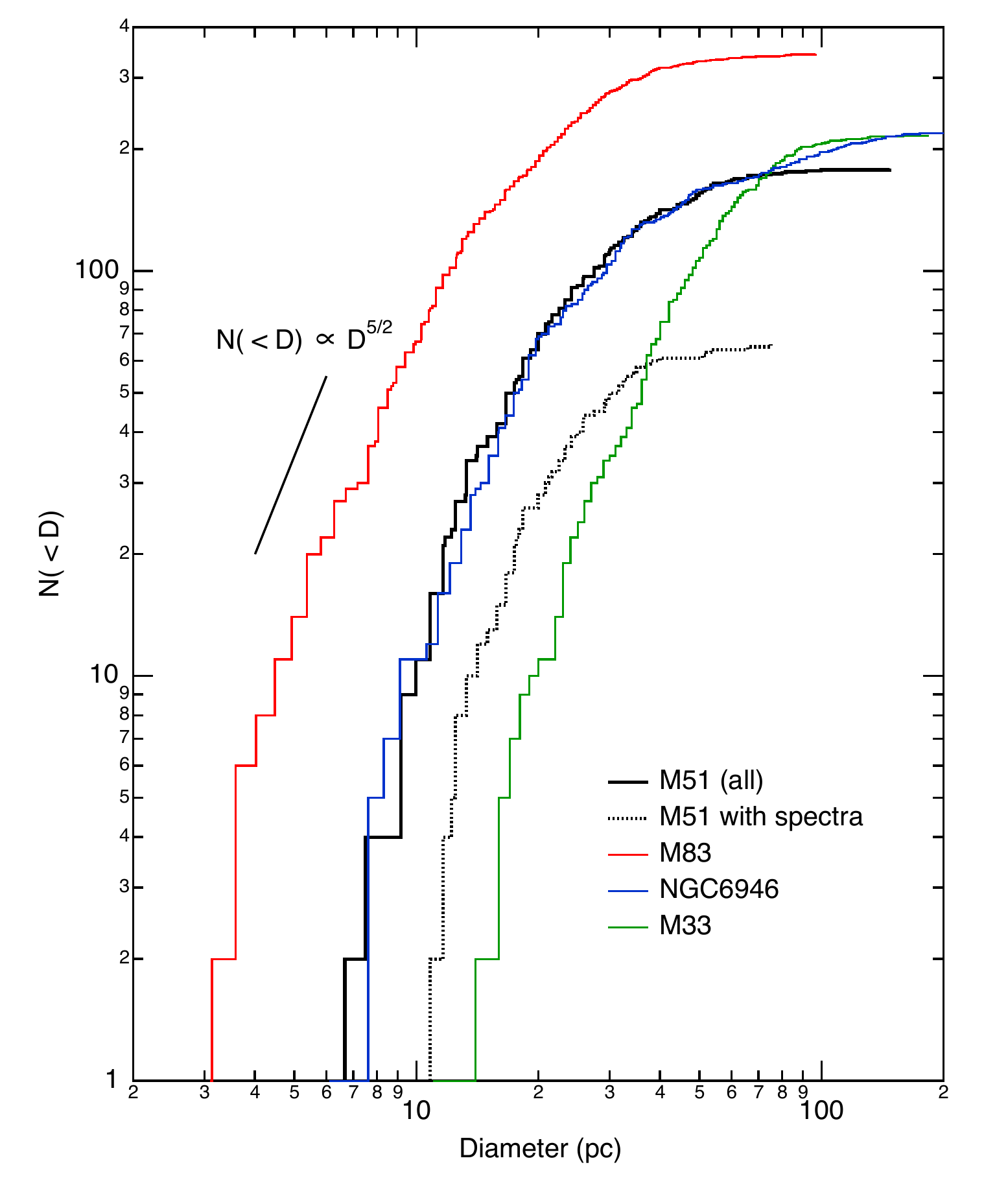}
\caption{Cumulative number of SNR candidates smaller than a given diameter for M51 and for several other spiral galaxies.  (For M51 we show both the complete sample (Table 3)
and those for which we have obtained spectra, to demonstrate that those with spectra represent a fair subset of the entire sample.)  For each of the galaxies, the heart of the distribution, where there are a few dozen SNRs, has a slope consistent with Sedov expansion ($D(t) \propto t^{2/5}$, so $N(<D) \propto D^{5/2}$ for a uniform SN rate).  References: M83: \citet{blair12, blair14}; NGC\,6946: \citet{long20}; M33: \citet{lee14b,long18}. 
\label{fig_n_vs_d}}
\end{figure}

We show a plot of the cumulative size distribution, $N(<D)$ as a function of the diameter, $D$, for SNRs in M51, and for three other spiral galaxies with extensive SNR samples, in Fig.~\ref{fig_n_vs_d}.  (Data sources are in the figure caption.)  For M51, as for M83 and NGC\,6946, all the diameters have been measured from HST images.\footnote{For NGC\,6946, data are from Tables~1 and 2 of \citet{long20}. Virtually all diameters were measured from HST images---\ha\ where available, but some  objects were detected only in the \FeiiL\ line so were measured on those images. A very few objects lay outside the HST footprint, so those diameters were estimated from ground-based images.}  
 The number-diameter distributions for SNRs in other galaxies are similar.  For example, \citet{long17} shows  number-diameter plots for several galaxies, including the SMC and LMC \citep[based on data from][]{badenes10} and M31 \citep[based on data from][]{lee14a} that resemble those for the galaxies shown here.

It is interesting  to note that  for all four galaxies in Fig.~\ref{fig_n_vs_d}, that portion of the distributions with a few to a few tens of objects have slopes approximately consistent with  Sedov expansion and a uniform average supernova rate, $N(<D) \propto D^{5/2}$; in contrast, one expects $N(<D) \propto D$ for free expansion.  In actuality, this result is most likely due to a combination of selection effects rather than something physical.   The optical emission that we observe from (most) SNRs arises from relatively low velocity (100\,-\,300 $\VEL$) radiative shocks propelled by the primary shock into dense cloudlets in the ISM\@. Because the surface area of the primary shock increases with time, we generally expect the optical luminosity to increase with time as well, until the primary shock velocity drops to the point where it can no longer drive secondary shocks into these cloudlets. We then expect the SNR (now in the so-called radiative phase) to fade.  

The maximum luminosity a SNR reaches---and importantly the radius at which it reaches maximum---depends on a variety of factors, including the average density of the ISM, the number of cloudlets, and the energy of the SNR explosion.  SNRs expanding into dense media brighten, and eventually  fade, at smaller radii than those expanding into less dense material.   At some level, the samples of SNRs that we have observed in all of these galaxies is luminosity-limited at small diameters, and surface-brightness-limited at large diameters.  At the small-diameter end of the distribution, we are unable to measure diameters of unresolved objects, which effectively sets a lower limit at the point-spread function (PSF) of the detector used.  Furthermore, SNRs with sizes on the  order of the PSF are  hard to pick out against stars (because it is difficult to remove stars completely from  narrow-band images).  These two effects limit the low end of the distribution, and probably explain why the limits for M51 and NGC\,6946 are about twice that for the similar galaxy M83---since the former two galaxies are at almost twice M83's distance.

The number of SNRs in a galaxy should be proportional to the star formation rate (SFR), since about 75\% of the SNe observed in external galaxies arise from the core-collapse explosion of massive stars whose main sequence lifetimes are short \citep{li11a}, and since the ages of stellar populations that produce Type I SNe are also typically less than a Gyr \citep{maoz10}.  The SFRs of the galaxies shown in Fig.~\ref{fig_n_vs_d} vary.  According to \cite{jarrett13}, the SFRs for M83,  NGC6946, and M51 are all similar: estimates range from 0.9-3.2 \SFR, 1.4-3.2 \SFR, and 1.3-3.5 \SFR, respectively, depending on the wavelength band used to estimate the SFR.   

On the other hand, there have been four historical SNe in  M51, whereas M83 and NGC~6946 have hosted six and ten, respectively, suggesting that the SFRs in the latter two galaxies are somewhat higher than in M51\@.  For M33,\footnote{For M33, the SFR  is variously estimated by \cite{williams18} to be 0.17$\pm$0.06 \SFR  from a multi-wavelength analysis to 0.25$^{+0.10}_{-0.07}$ 
 \SFR\ using FUV and 24$\mu$ imaging to 0.33$\pm$0.10 \SFR\ using SED fitting, whereas \cite{verley09} suggest 0.45 $\pm$ 0.10 \SFR\ from a multi-wavelength analysis.} the SFR is between 0.2-0.5  \SFR\@.    No SN has been observed historically in M33\@.  Therefore, if all the samples had equal luminosity sensitivity, we might expect there to be roughly equal numbers of SNRs in M83, NGC6946, and M51, and far fewer in M33\@.  In point of fact, however, there are 300 SNRs (and candidates) in M83, 225 in NGC~6946, 179 in M51, and 217 in M33\@.  The relatively large number of SNRs in M33 is clearly due to its proximity; SNRs exhibit a large range of luminosities at any diameter; and the various SNR samples are mainly flux-, not luminosity-limited.  M51 is the most distant galaxy in the sample (8.58 Mpc), compared to M83  at 4.6 Mpc \citep{saha06} and NGC~6946  at 7.8 Mpc \citep{murphy18,anand18}.  As a result one would expect M51 to have somewhat fewer SNRs in a flux-limited sample.
 
 The biggest difference in the samples is in the number of small-diameter SNRs.  M51 and NGC~6946 both have about 10 SNRs with diameters of 10 pc or less, whereas M83 has almost 100, and M33 has none.  With a couple of exceptions, all of the small-diameter SNRs in these galaxies were identified on the basis of elevated \sii:\ha\ ratios, not from properties that one would expect from a young, ejecta-dominated SNR like Cas A. Also, the \sii:\ha\ technique cannot find young Balmer-dominated SNRs that arise from SNIa.  Therefore the most likely reason for this difference is that the small-diameter SNRs that we do detect are ones expanding into locally dense regions of the ISM.  M83 is known to have very substantial diffuse soft X-ray emission, indicative of a hot, high pressure ISM throughout its spiral arms, which may account for the larger number of small diameter SNRs.

\section{Summary}

We have used a combination of ground-based and space-based imagery to construct the first catalog of SNRs for M51, based on elevated \sii:\ha\ line ratios compared to \hii\  regions in the same images.   Our list of candidates totals 179 objects.  We obtained GMOS spectra of  66 of the candidates (along with a number of \hii\ regions); 60 of the SNR candidates have measured \sii:\ha\ ratios that exceed 0.4,  the standard value for declaring an emission nebula to be a SNR.  Moreover, 51 of the candidates show intrinsic \oi$\lambda$6300, another line associated with radiative shocks, including one object with a  \sii:\ha\ ratio slightly below the 0.4 limit. This suggests not only that a high percentage of our SNR candidates (61 of 66) are actually SNRs, but also that the vast majority of other SNR candidates in our catalog are also likely to be bona fide SNRs.  

The SNRs/candidates in the sample are mostly distributed along the very prominent spiral arms of M51. Nearly a third (55 of 179) of the  SNRs/candidates are coincident with X-ray sources identified with Chandra  \citep{kuntz16}, and most of these X-ray sources have the soft spectra normally seen in SNRs. A search at the positions of other soft X-ray sources turned up another handful of possibly interesting candidates.  Only 16 of the SNRs/candidates  are associated with with VLA radio sources cataloged by \cite{maddox07}, and a search at the positions of additional objects they identified as likely SNRs shows interesting optical emission from two of those.   The median diameter of the SNRs in the sample is 24 pc; none of the SNRs for which we have spectra shows evidence of the type of peculiar abundances or line broadening expected from a young SNR in the free expansion phase such as Cas A in our Galaxy; most are likely in the Sedov (or possibly radiative phase) of their evolution.

We have detected nebular emission from none of the four historical SNe that have occurred in M51.  But we do detect fading continuum from SN2011dh in HST images taken ten months after that event.

The most surprising feature of the spectra of the SNRs in M51 is the behavior of the \nii\ lines relative to \ha.  We see very high \nii:\ha\ line ratios in many of them which, from an observational perspective, explains why the imaging \sii:\ha\ line ratios for the SNR candidates were lower than typically seen in other galaxies. (That is,  the \ha\ filters used passed significant \nii\ emission as well, lowering the observed ratio.)  Additionally, the dispersion in \nii:\ha\ between objects of similar size or similar galactocentric distance is very large. While both of these effects have been seen in other galaxies, they are more extreme (and hence more obvious) in M51 and seem to be pointing toward something more fundamental.  

The cause for such high ratios is hard to explain astrophysically.  The  highest \nii:\ha\ ratios tend to appear in smaller diameter SNRs, or alternatively in objects at relatively small galactocentric radius, but the large dispersion in values seems to indicate that neither of these parameters is a dominant effect.  
As far as the abundance of N is concerned, the \nii:\ha\ ratios in M51 are higher on average than in galaxies like M83, where the {\em overall} metallicity is significantly higher. Existing shock models with elevated abundances do not reproduce the observed \nii:\ha\ ratios we see in M51 SNRs; furthermore, the \nii:\ha\ ratios in M51 \hii\ regions are not unusual and are nearly constant with galactocentric radius, as expected from the overall metallicity. Hence, abundance variations do not provide an obvious explanation.  Further spectroscopic studies and modeling of the SNRs in M51 are clearly warranted to understand this conundrum.


\acknowledgments

We thank Derek Hammer (previously of STScI), Heather Greenfield (STScI) and Jennifer Mack (STScI) for producing the aligned, mosaicked HST images and properly applying the CTE corrections to the  HST data used in this paper. We are also grateful for valuable comments from John Raymond.  Partial support for the analysis of the data was provided by NASA through grant numbers HST-GO-12762 and HST-GO-15216 from the Space Telescope Science Institute, which is operated by AURA, Inc., under NASA contract NAS 5-26555. PFW acknowledges additional support from the NSF through grant AST-1714281.  WPB acknowledges partial support from the JHU Center for Astrophysical Sciences.  

\vspace{5mm}
\facilities{HST(WFC3), HST(ACS), Gemini:North (GMOS)}

\software{astropy \citep{astropy},  AstroDrizzle \citep{AstroDrizzle}, matplotlib\citep{hunter07}, SAOImageDS9 \cite{joye03}}

\pagebreak

\bibliographystyle{aasjournal}


\begin{thebibliography}{}
\expandafter\ifx\csname natexlab\endcsname\relax\def\natexlab#1{#1}\fi
\providecommand{\url}[1]{\href{#1}{#1}}
\providecommand{\dodoi}[1]{doi:~\href{http://doi.org/#1}{\nolinkurl{#1}}}
\providecommand{\doeprint}[1]{\href{http://ascl.net/#1}{\nolinkurl{http://ascl.net/#1}}}
\providecommand{\doarXiv}[1]{\href{https://arxiv.org/abs/#1}{\nolinkurl{https://arxiv.org/abs/#1}}}

\bibitem[{{Allen} {et~al.}(2008){Allen}, {Groves}, {Dopita}, {Sutherland}, \&
  {Kewley}}]{allen08}
{Allen}, M.~G., {Groves}, B.~A., {Dopita}, M.~A., {Sutherland}, R.~S., \&
  {Kewley}, L.~J. 2008, \apjs, 178, 20, \dodoi{10.1086/589652}

\bibitem[{{Anand} {et~al.}(2018){Anand}, {Rizzi}, \& {Tully}}]{anand18}
{Anand}, G.~S., {Rizzi}, L., \& {Tully}, R.~B. 2018, \aj, 156, 105,
  \dodoi{10.3847/1538-3881/aad3b2}

\bibitem[{{Astropy Collaboration} {et~al.}(2013){Astropy Collaboration},
  {Robitaille}, {Tollerud}, {Greenfield}, {Droettboom}, {Bray}, {Aldcroft},
  {Davis}, {Ginsburg}, {Price-Whelan}, {Kerzendorf}, {Conley}, {Crighton},
  {Barbary}, {Muna}, {Ferguson}, {Grollier}, {Parikh}, {Nair}, {Unther},
  {Deil}, {Woillez}, {Conseil}, {Kramer}, {Turner}, {Singer}, {Fox}, {Weaver},
  {Zabalza}, {Edwards}, {Azalee Bostroem}, {Burke}, {Casey}, {Crawford},
  {Dencheva}, {Ely}, {Jenness}, {Labrie}, {Lim}, {Pierfederici}, {Pontzen},
  {Ptak}, {Refsdal}, {Servillat}, \& {Streicher}}]{astropy}
{Astropy Collaboration}, {Robitaille}, T.~P., {Tollerud}, E.~J., {et~al.} 2013,
  \aap, 558, A33, \dodoi{10.1051/0004-6361/201322068}

\bibitem[{{Badenes} {et~al.}(2010){Badenes}, {Maoz}, \& {Draine}}]{badenes10}
{Badenes}, C., {Maoz}, D., \& {Draine}, B.~T. 2010, \mnras, 407, 1301,
  \dodoi{10.1111/j.1365-2966.2010.17023.x}

\bibitem[{{Blair} {et~al.}(2012){Blair}, {Winkler}, \& {Long}}]{blair12}
{Blair}, W.~P., {Winkler}, P.~F., \& {Long}, K.~S. 2012, \apjs, 203, 8,
  \dodoi{10.1088/0067-0049/203/1/8}

\bibitem[{{Blair} {et~al.}(2014){Blair}, {Chandar}, {Dopita}, {Ghavamian},
  {Hammer}, {Kuntz}, {Long}, {Soria}, {Whitmore}, \& {Winkler}}]{blair14}
{Blair}, W.~P., {Chandar}, R., {Dopita}, M.~A., {et~al.} 2014, \apj, 788, 55,
  \dodoi{10.1088/0004-637X/788/1/55}

\bibitem[{{Bresolin} {et~al.}(2004){Bresolin}, {Garnett}, \&
  {Kennicutt}}]{bresolin04}
{Bresolin}, F., {Garnett}, D.~R., \& {Kennicutt}, Robert~C., J. 2004, \apj,
  615, 228, \dodoi{10.1086/424377}

\bibitem[{{Brightman} {et~al.}(2020){Brightman}, {Earnshaw}, {F{\"u}rst},
  {Harrison}, {Heida}, {Israel}, {Pike}, {Stern}, \& {Walton}}]{brightman20}
{Brightman}, M., {Earnshaw}, H., {F{\"u}rst}, F., {et~al.} 2020, \apj, 895,
  127, \dodoi{10.3847/1538-4357/ab7e2a}

\bibitem[{{Calzetti} {et~al.}(2005){Calzetti}, {Kennicutt}, {Bianchi},
  {Thilker}, {Dale}, {Engelbracht}, {Leitherer}, {Meyer}, {Sosey}, {Mutchler},
  {Regan}, {Thornley}, {Armus}, {Bendo}, {Boissier}, {Boselli}, {Draine},
  {Gordon}, {Helou}, {Hollenbach}, {Kewley}, {Madore}, {Martin}, {Murphy},
  {Rieke}, {Rieke}, {Roussel}, {Sheth}, {Smith}, {Walter}, {White}, {Yi},
  {Scoville}, {Polletta}, \& {Lindler}}]{calzetti05}
{Calzetti}, D., {Kennicutt}, R.~C., J., {Bianchi}, L., {et~al.} 2005, \apj,
  633, 871, \dodoi{10.1086/466518}

\bibitem[{{Croxall} {et~al.}(2015){Croxall}, {Pogge}, {Berg}, {Skillman}, \&
  {Moustakas}}]{croxall15}
{Croxall}, K.~V., {Pogge}, R.~W., {Berg}, D.~A., {Skillman}, E.~D., \&
  {Moustakas}, J. 2015, \apj, 808, 42, \dodoi{10.1088/0004-637X/808/1/42}

\bibitem[{{Dopita} {et~al.}(2010){Dopita}, {Blair}, {Long}, {Mutchler},
  {Whitmore}, {Kuntz}, {Balick}, {Bond}, {Calzetti}, {Carollo}, {Disney},
  {Frogel}, {O'Connell}, {Hall}, {Holtzman}, {Kimble}, {MacKenty}, {McCarthy},
  {Paresce}, {Saha}, {Silk}, {Sirianni}, {Trauger}, {Walker}, {Windhorst}, \&
  {Young}}]{dopita10}
{Dopita}, M.~A., {Blair}, W.~P., {Long}, K.~S., {et~al.} 2010, \apj, 710, 964,
  \dodoi{10.1088/0004-637X/710/2/964}

\bibitem[{{Fruchter}(2010)}]{fruchter10}
{Fruchter}, A.~S.~e. 2010, in 2010 Space Telescope Science Institute
  Calibration Workshop, 382--387

\bibitem[{{Gaia Collaboration} {et~al.}(2018){Gaia Collaboration}, {Brown},
  {Vallenari}, {Prusti}, {de Bruijne}, {Babusiaux}, {Bailer-Jones}, {Biermann},
  {Evans}, {Eyer}, {Jansen}, {Jordi}, {Klioner}, {Lammers}, {Lindegren},
  {Luri}, {Mignard}, {Panem}, {Pourbaix}, {Randich}, {Sartoretti}, {Siddiqui},
  {Soubiran}, {van Leeuwen}, {Walton}, {Arenou}, {Bastian}, {Cropper},
  {Drimmel}, {Katz}, {Lattanzi}, {Bakker}, {Cacciari}, {Casta{\~n}eda},
  {Chaoul}, {Cheek}, {De Angeli}, {Fabricius}, {Guerra}, {Holl}, {Masana},
  {Messineo}, {Mowlavi}, {Nienartowicz}, {Panuzzo}, {Portell}, {Riello},
  {Seabroke}, {Tanga}, {Th{\'e}venin}, {Gracia-Abril}, {Comoretto},
  {Garcia-Reinaldos}, {Teyssier}, {Altmann}, {Andrae}, {Audard},
  {Bellas-Velidis}, {Benson}, {Berthier}, {Blomme}, {Burgess}, {Busso},
  {Carry}, {Cellino}, {Clementini}, {Clotet}, {Creevey}, {Davidson}, {De
  Ridder}, {Delchambre}, {Dell'Oro}, {Ducourant},
  {Fern{\'a}ndez-Hern{\'a}ndez}, {Fouesneau}, {Fr{\'e}mat}, {Galluccio},
  {Garc{\'\i}a-Torres}, {Gonz{\'a}lez-N{\'u}{\~n}ez}, {Gonz{\'a}lez-Vidal},
  {Gosset}, {Guy}, {Halbwachs}, {Hambly}, {Harrison}, {Hern{\'a}ndez},
  {Hestroffer}, {Hodgkin}, {Hutton}, {Jasniewicz}, {Jean-Antoine-Piccolo},
  {Jordan}, {Korn}, {Krone-Martins}, {Lanzafame}, {Lebzelter}, {L{\"o}ffler},
  {Manteiga}, {Marrese}, {Mart{\'\i}n-Fleitas}, {Moitinho}, {Mora}, {Muinonen},
  {Osinde}, {Pancino}, {Pauwels}, {Petit}, {Recio-Blanco}, {Richards},
  {Rimoldini}, {Robin}, {Sarro}, {Siopis}, {Smith}, {Sozzetti}, {S{\"u}veges},
  {Torra}, {van Reeven}, {Abbas}, {Abreu Aramburu}, {Accart}, {Aerts},
  {Altavilla}, {{\'A}lvarez}, {Alvarez}, {Alves}, {Anderson}, {Andrei},
  {Anglada Varela}, {Antiche}, {Antoja}, {Arcay}, {Astraatmadja}, {Bach},
  {Baker}, {Balaguer-N{\'u}{\~n}ez}, {Balm}, {Barache}, {Barata}, {Barbato},
  {Barblan}, {Barklem}, {Barrado}, {Barros}, {Barstow}, {Bartholom{\'e}
  Mu{\~n}oz}, {Bassilana}, {Becciani}, {Bellazzini}, {Berihuete}, {Bertone},
  {Bianchi}, {Bienaym{\'e}}, {Blanco-Cuaresma}, {Boch}, {Boeche}, {Bombrun},
  {Borrachero}, {Bossini}, {Bouquillon}, {Bourda}, {Bragaglia}, {Bramante},
  {Breddels}, {Bressan}, {Brouillet}, {Br{\"u}semeister}, {Brugaletta},
  {Bucciarelli}, {Burlacu}, {Busonero}, {Butkevich}, {Buzzi}, {Caffau},
  {Cancelliere}, {Cannizzaro}, {Cantat-Gaudin}, {Carballo}, {Carlucci},
  {Carrasco}, {Casamiquela}, {Castellani}, {Castro-Ginard}, {Charlot},
  {Chemin}, {Chiavassa}, {Cocozza}, {Costigan}, {Cowell}, {Crifo}, {Crosta},
  {Crowley}, {Cuypers}, {Dafonte}, {Damerdji}, {Dapergolas}, {David}, {David},
  {de Laverny}, {De Luise}, {De March}, {de Martino}, {de Souza}, {de Torres},
  {Debosscher}, {del Pozo}, {Delbo}, {Delgado}, {Delgado}, {Di Matteo},
  {Diakite}, {Diener}, {Distefano}, {Dolding}, {Drazinos}, {Dur{\'a}n},
  {Edvardsson}, {Enke}, {Eriksson}, {Esquej}, {Eynard Bontemps}, {Fabre},
  {Fabrizio}, {Faigler}, {Falc{\~a}o}, {Farr{\`a}s Casas}, {Federici},
  {Fedorets}, {Fernique}, {Figueras}, {Filippi}, {Findeisen}, {Fonti},
  {Fraile}, {Fraser}, {Fr{\'e}zouls}, {Gai}, {Galleti}, {Garabato},
  {Garc{\'\i}a-Sedano}, {Garofalo}, {Garralda}, {Gavel}, {Gavras}, {Gerssen},
  {Geyer}, {Giacobbe}, {Gilmore}, {Girona}, {Giuffrida}, {Glass}, {Gomes},
  {Granvik}, {Gueguen}, {Guerrier}, {Guiraud}, {Guti{\'e}rrez-S{\'a}nchez},
  {Haigron}, {Hatzidimitriou}, {Hauser}, {Haywood}, {Heiter}, {Helmi}, {Heu},
  {Hilger}, {Hobbs}, {Hofmann}, {Holland}, {Huckle}, {Hypki}, {Icardi},
  {Jan{\ss}en}, {Jevardat de Fombelle}, {Jonker}, {Juh{\'a}sz}, {Julbe},
  {Karampelas}, {Kewley}, {Klar}, {Kochoska}, {Kohley}, {Kolenberg},
  {Kontizas}, {Kontizas}, {Koposov}, {Kordopatis}, {Kostrzewa-Rutkowska},
  {Koubsky}, {Lambert}, {Lanza}, {Lasne}, {Lavigne}, {Le Fustec}, {Le
  Poncin-Lafitte}, {Lebreton}, {Leccia}, {Leclerc}, {Lecoeur-Taibi},
  {Lenhardt}, {Leroux}, {Liao}, {Licata}, {Lindstr{\o}m}, {Lister}, {Livanou},
  {Lobel}, {L{\'o}pez}, {Managau}, {Mann}, {Mantelet}, {Marchal}, {Marchant},
  {Marconi}, {Marinoni}, {Marschalk{\'o}}, {Marshall}, {Martino}, {Marton},
  {Mary}, {Massari}, {Matijevi{\v{c}}}, {Mazeh}, {McMillan}, {Messina},
  {Michalik}, {Millar}, {Molina}, {Molinaro}, {Moln{\'a}r}, {Montegriffo},
  {Mor}, {Morbidelli}, {Morel}, {Morris}, {Mulone}, {Muraveva}, {Musella},
  {Nelemans}, {Nicastro}, {Noval}, {O'Mullane}, {Ord{\'e}novic},
  {Ord{\'o}{\~n}ez-Blanco}, {Osborne}, {Pagani}, {Pagano}, {Pailler},
  {Palacin}, {Palaversa}, {Panahi}, {Pawlak}, {Piersimoni}, {Pineau}, {Plachy},
  {Plum}, {Poggio}, {Poujoulet}, {Pr{\v{s}}a}, {Pulone}, {Racero}, {Ragaini},
  {Rambaux}, {Ramos-Lerate}, {Regibo}, {Reyl{\'e}}, {Riclet}, {Ripepi}, {Riva},
  {Rivard}, {Rixon}, {Roegiers}, {Roelens}, {Romero-G{\'o}mez}, {Rowell},
  {Royer}, {Ruiz-Dern}, {Sadowski}, {Sagrist{\`a} Sell{\'e}s}, {Sahlmann},
  {Salgado}, {Salguero}, {Sanna}, {Santana-Ros}, {Sarasso}, {Savietto},
  {Schultheis}, {Sciacca}, {Segol}, {Segovia}, {S{\'e}gransan}, {Shih},
  {Siltala}, {Silva}, {Smart}, {Smith}, {Solano}, {Solitro}, {Sordo}, {Soria
  Nieto}, {Souchay}, {Spagna}, {Spoto}, {Stampa}, {Steele},
  {Steidelm{\"u}ller}, {Stephenson}, {Stoev}, {Suess}, {Surdej}, {Szabados},
  {Szegedi-Elek}, {Tapiador}, {Taris}, {Tauran}, {Taylor}, {Teixeira},
  {Terrett}, {Teyssand ier}, {Thuillot}, {Titarenko}, {Torra Clotet}, {Turon},
  {Ulla}, {Utrilla}, {Uzzi}, {Vaillant}, {Valentini}, {Valette}, {van Elteren},
  {Van Hemelryck}, {van Leeuwen}, {Vaschetto}, {Vecchiato}, {Veljanoski},
  {Viala}, {Vicente}, {Vogt}, {von Essen}, {Voss}, {Votruba}, {Voutsinas},
  {Walmsley}, {Weiler}, {Wertz}, {Wevers}, {Wyrzykowski}, {Yoldas},
  {{\v{Z}}erjal}, {Ziaeepour}, {Zorec}, {Zschocke}, {Zucker}, {Zurbach}, \&
  {Zwitter}}]{gaia-collaboration18}
{Gaia Collaboration}, {Brown}, A.~G.~A., {Vallenari}, A., {et~al.} 2018, \aap,
  616, A1, \dodoi{10.1051/0004-6361/201833051}

\bibitem[{{Garofali} {et~al.}(2017){Garofali}, {Williams}, {Plucinsky},
  {Gaetz}, {Wold}, {Haberl}, {Long}, {Blair}, {Pannuti}, {Winkler}, \&
  {Gross}}]{garofali17}
{Garofali}, K., {Williams}, B.~F., {Plucinsky}, P.~P., {et~al.} 2017, \mnras,
  472, 308, \dodoi{10.1093/mnras/stx1905}

\bibitem[{Gonsaga {et~al.}(2012)Gonsaga, Fruchter, \& Mack}]{AstroDrizzle}
Gonsaga, S., Fruchter, A., \& Mack, J., eds. 2012, The DrizzlePack Handbook
  (STScI)

\bibitem[{{Gonzaga} \& {et al.}(2012)}]{gonzaga12}
{Gonzaga}, S., \& {et al.} 2012, {The DrizzlePac Handbook}

\bibitem[{Hunter(2007)}]{hunter07}
Hunter, J.~D. 2007, Computing in Science \& Engineering, 9, 90,
  \dodoi{10.1109/MCSE.2007.55}

\bibitem[{{Jarrett} {et~al.}(2013){Jarrett}, {Masci}, {Tsai}, {Petty},
  {Cluver}, {Assef}, {Benford}, {Blain}, {Bridge}, {Donoso}, {Eisenhardt},
  {Koribalski}, {Lake}, {Neill}, {Seibert}, {Sheth}, {Stanford}, \&
  {Wright}}]{jarrett13}
{Jarrett}, T.~H., {Masci}, F., {Tsai}, C.~W., {et~al.} 2013, \aj, 145, 6,
  \dodoi{10.1088/0004-6256/145/1/6}

\bibitem[{{Joye} \& {Mandel}(2003)}]{joye03}
{Joye}, W.~A., \& {Mandel}, E. 2003, in Astronomical Society of the Pacific
  Conference Series, Vol. 295, Astronomical Data Analysis Software and Systems
  XII, ed. H.~E. {Payne}, R.~I. {Jedrzejewski}, \& R.~N. {Hook}, 489

\bibitem[{{Kilgard} {et~al.}(2005){Kilgard}, {Cowan}, {Garcia}, {Kaaret},
  {Krauss}, {McDowell}, {Prestwich}, {Primini}, {Stockdale}, {Trinchieri},
  {Ward}, \& {Zezas}}]{kilgard05}
{Kilgard}, R.~E., {Cowan}, J.~J., {Garcia}, M.~R., {et~al.} 2005, \apjs, 159,
  214, \dodoi{10.1086/430443}

\bibitem[{{Kuntz} {et~al.}(2016){Kuntz}, {Long}, \& {Kilgard}}]{kuntz16}
{Kuntz}, K.~D., {Long}, K.~S., \& {Kilgard}, R.~E. 2016, \apj, 827, 46,
  \dodoi{10.3847/0004-637X/827/1/46}

\bibitem[{{Lee} {et~al.}(2011){Lee}, {Hwang}, \& {Lee}}]{lee11}
{Lee}, J.~H., {Hwang}, N., \& {Lee}, M.~G. 2011, \apj, 735, 75,
  \dodoi{10.1088/0004-637X/735/2/75}

\bibitem[{{Lee} \& {Lee}(2014{\natexlab{a}})}]{lee14a}
{Lee}, J.~H., \& {Lee}, M.~G. 2014{\natexlab{a}}, \apj, 786, 130,
  \dodoi{10.1088/0004-637X/786/2/130}

\bibitem[{{Lee} \& {Lee}(2014{\natexlab{b}})}]{lee14b}
---. 2014{\natexlab{b}}, \apj, 793, 134, \dodoi{10.1088/0004-637X/793/2/134}

\bibitem[{{Lee} {et~al.}(2015){Lee}, {Sohn}, {Lee}, {Lim}, {Jang}, {Ko}, {Koo},
  {Hwang}, {Kim}, \& {Park}}]{lee15}
{Lee}, M.~G., {Sohn}, J., {Lee}, J.~H., {et~al.} 2015, \apj, 804, 63,
  \dodoi{10.1088/0004-637X/804/1/63}

\bibitem[{{Levenson} {et~al.}(1995){Levenson}, {Kirshner}, {Blair}, \&
  {Winkler}}]{levenson95}
{Levenson}, N.~A., {Kirshner}, R.~P., {Blair}, W.~P., \& {Winkler}, P.~F. 1995,
  \aj, 110, 739, \dodoi{10.1086/117558}

\bibitem[{{Li} {et~al.}(2011{\natexlab{a}}){Li}, {Chornock}, {Leaman},
  {Filippenko}, {Poznanski}, {Wang}, {Ganeshalingam}, \& {Mannucci}}]{li11a}
{Li}, W., {Chornock}, R., {Leaman}, J., {et~al.} 2011{\natexlab{a}}, \mnras,
  412, 1473, \dodoi{10.1111/j.1365-2966.2011.18162.x}

\bibitem[{{Li} {et~al.}(2011{\natexlab{b}}){Li}, {Filippenko}, \& {van
  Dyk}}]{li11}
{Li}, W., {Filippenko}, A.~V., \& {van Dyk}, S.~D. 2011{\natexlab{b}}, The
  Astronomer's Telegram, 3401, 1

\bibitem[{{Li} {et~al.}(2006){Li}, {Van Dyk}, {Filippenko}, {Cuillandre},
  {Jha}, {Bloom}, {Riess}, \& {Livio}}]{li06}
{Li}, W., {Van Dyk}, S.~D., {Filippenko}, A.~V., {et~al.} 2006, \apj, 641,
  1060, \dodoi{10.1086/499916}

\bibitem[{{Long}(2017)}]{long17}
{Long}, K.~S. 2017, {Galactic and Extragalactic Samples of Supernova Remnants:
  How They Are Identified and What They Tell Us} (Springer), 2005,
  \dodoi{10.1007/978-3-319-21846-5_90}

\bibitem[{{Long} {et~al.}(2018){Long}, {Blair}, {Milisavljevic}, {Raymond}, \&
  {Winkler}}]{long18}
{Long}, K.~S., {Blair}, W.~P., {Milisavljevic}, D., {Raymond}, J.~C., \&
  {Winkler}, P.~F. 2018, \apj, 855, 140, \dodoi{10.3847/1538-4357/aaac7e}

\bibitem[{{Long} {et~al.}(2020){Long}, {Blair}, {Winkler}, \& {Lacey}}]{long20}
{Long}, K.~S., {Blair}, W.~P., {Winkler}, P.~F., \& {Lacey}, C.~K. 2020, arXiv
  e-prints, arXiv:2007.01415.
\newblock \doarXiv{2007.01415}

\bibitem[{{Long} {et~al.}(2014){Long}, {Kuntz}, {Blair}, {Godfrey},
  {Plucinsky}, {Soria}, {Stockdale}, \& {Winkler}}]{long14}
{Long}, K.~S., {Kuntz}, K.~D., {Blair}, W.~P., {et~al.} 2014, \apjs, 212, 21,
  \dodoi{10.1088/0067-0049/212/2/21}

\bibitem[{{Long} {et~al.}(2019){Long}, {Winkler}, \& {Blair}}]{long19}
{Long}, K.~S., {Winkler}, P.~F., \& {Blair}, W.~P. 2019, \apj, 875, 85,
  \dodoi{10.3847/1538-4357/ab0d94}

\bibitem[{{Long} {et~al.}(2010){Long}, {Blair}, {Winkler}, {Becker}, {Gaetz},
  {Ghavamian}, {Helfand}, {Hughes}, {Kirshner}, {Kuntz}, {McNeil}, {Pannuti},
  {Plucinsky}, {Saul}, {T{\"u}llmann}, \& {Williams}}]{long10}
{Long}, K.~S., {Blair}, W.~P., {Winkler}, P.~F., {et~al.} 2010, \apjs, 187,
  495, \dodoi{10.1088/0067-0049/187/2/495}

\bibitem[{{Maddox} {et~al.}(2007){Maddox}, {Cowan}, {Kilgard}, {Schinnerer}, \&
  {Stockdale}}]{maddox07}
{Maddox}, L.~A., {Cowan}, J.~J., {Kilgard}, R.~E., {Schinnerer}, E., \&
  {Stockdale}, C.~J. 2007, \aj, 133, 2559, \dodoi{10.1086/515573}

\bibitem[{{Maoz} {et~al.}(2010){Maoz}, {Sharon}, \& {Gal-Yam}}]{maoz10}
{Maoz}, D., {Sharon}, K., \& {Gal-Yam}, A. 2010, \apj, 722, 1879,
  \dodoi{10.1088/0004-637X/722/2/1879}

\bibitem[{{Massey} {et~al.}(1988){Massey}, {Strobel}, {Barnes}, \&
  {Anderson}}]{massey88}
{Massey}, P., {Strobel}, K., {Barnes}, J.~V., \& {Anderson}, E. 1988, \apj,
  328, 315, \dodoi{10.1086/166294}

\bibitem[{{Mathewson} \& {Clarke}(1973)}]{mathewson73}
{Mathewson}, D.~S., \& {Clarke}, J.~N. 1973, \apj, 180, 725,
  \dodoi{10.1086/152002}

\bibitem[{{Matonick} \& {Fesen}(1997)}]{matonick97}
{Matonick}, D.~M., \& {Fesen}, R.~A. 1997, \apjs, 112, 49,
  \dodoi{10.1086/313034}

\bibitem[{{Maund}(2019)}]{maund19}
{Maund}, J.~R. 2019, \apj, 883, 86, \dodoi{10.3847/1538-4357/ab2386}

\bibitem[{{McQuinn} {et~al.}(2016){McQuinn}, {Skillman}, {Dolphin}, {Berg}, \&
  {Kennicutt}}]{mcquinn16}
{McQuinn}, K. B.~W., {Skillman}, E.~D., {Dolphin}, A.~E., {Berg}, D., \&
  {Kennicutt}, R. 2016, \apj, 826, 21, \dodoi{10.3847/0004-637X/826/1/21}

\bibitem[{{Murphy} {et~al.}(2018){Murphy}, {Khan}, {Williams}, {Dolphin},
  {Dalcanton}, \& {D{\'\i}az-Rodr{\'\i}guez}}]{murphy18}
{Murphy}, J.~W., {Khan}, R., {Williams}, B., {et~al.} 2018, \apj, 860, 117,
  \dodoi{10.3847/1538-4357/aac2be}

\bibitem[{{Osterbrock} \& {Ferland}(2006)}]{osterbrock06}
{Osterbrock}, D.~E., \& {Ferland}, G.~J. 2006, {Astrophysics of Gaseous Nebulae
  and Active Galactic Nuclei, second edition} (Sausalito, CA: University
  Science Books), 121--127

\bibitem[{{Rosse}(1850)}]{rosse50}
{Rosse}, T. E.~O. 1850, Philosophical Transactions of the Royal Society of
  London Series I, 140, 499

\bibitem[{{Russell} {et~al.}(2020){Russell}, {White}, {Long}, {Blair}, {Soria},
  \& {Winkler}}]{russell20}
{Russell}, T.~D., {White}, R.~L., {Long}, K.~S., {et~al.} 2020, \mnras, 495,
  479, \dodoi{10.1093/mnras/staa1177}

\bibitem[{{Saha} {et~al.}(2006){Saha}, {Thim}, {Tammann}, {Reindl}, \&
  {Sandage}}]{saha06}
{Saha}, A., {Thim}, F., {Tammann}, G.~A., {Reindl}, B., \& {Sandage}, A. 2006,
  \apjs, 165, 108, \dodoi{10.1086/503800}

\bibitem[{{Soria} {et~al.}(2020){Soria}, {Blair}, {Long}, {Russell}, \&
  {Winkler}}]{soria20}
{Soria}, R., {Blair}, W.~P., {Long}, K.~S., {Russell}, T.~D., \& {Winkler},
  P.~F. 2020, \apj, 888, 103, \dodoi{10.3847/1538-4357/ab5b0c}

\bibitem[{{Soria} {et~al.}(2014){Soria}, {Long}, {Blair}, {Godfrey}, {Kuntz},
  {Lenc}, {Stockdale}, \& {Winkler}}]{soria14}
{Soria}, R., {Long}, K.~S., {Blair}, W.~P., {et~al.} 2014, Science, 343, 1330,
  \dodoi{10.1126/science.1248759}

\bibitem[{{Terashima} \& {Wilson}(2004)}]{terashima04}
{Terashima}, Y., \& {Wilson}, A.~S. 2004, \apj, 601, 735,
  \dodoi{10.1086/380505}

\bibitem[{{Totani} {et~al.}(2008){Totani}, {Morokuma}, {Oda}, {Doi}, \&
  {Yasuda}}]{totani08}
{Totani}, T., {Morokuma}, T., {Oda}, T., {Doi}, M., \& {Yasuda}, N. 2008,
  \pasj, 60, 1327, \dodoi{10.1093/pasj/60.6.1327}

\bibitem[{{Urquhart} {et~al.}(2018){Urquhart}, {Soria}, {Johnston}, {Pakull},
  {Motch}, {Schwope}, {Miller-Jones}, \& {Anderson}}]{urquhart18}
{Urquhart}, R., {Soria}, R., {Johnston}, H.~M., {et~al.} 2018, \mnras, 475,
  3561, \dodoi{10.1093/mnras/sty014}

\bibitem[{{Van Dyk} {et~al.}(2011){Van Dyk}, {Li}, {Cenko}, {Kasliwal},
  {Horesh}, {Ofek}, {Kraus}, {Silverman}, {Arcavi}, {Filippenko}, {Gal-Yam},
  {Quimby}, {Kulkarni}, {Yaron}, \& {Polishook}}]{van-dyk11}
{Van Dyk}, S.~D., {Li}, W., {Cenko}, S.~B., {et~al.} 2011, \apjl, 741, L28,
  \dodoi{10.1088/2041-8205/741/2/L28}

\bibitem[{{Verley} {et~al.}(2009){Verley}, {Corbelli}, {Giovanardi}, \&
  {Hunt}}]{verley09}
{Verley}, S., {Corbelli}, E., {Giovanardi}, C., \& {Hunt}, L.~K. 2009, \aap,
  493, 453, \dodoi{10.1051/0004-6361:200810566}

\bibitem[{{White} {et~al.}(2019){White}, {Long}, {Becker}, {Blair}, {Helfand},
  \& {Winkler}}]{white19}
{White}, R.~L., {Long}, K.~S., {Becker}, R.~H., {et~al.} 2019, \apj, submitted

\bibitem[{{Williams} {et~al.}(2018){Williams}, {Gear}, \& {Smith}}]{williams18}
{Williams}, T.~G., {Gear}, W.~K., \& {Smith}, M. W.~L. 2018, \mnras, 479, 297,
  \dodoi{10.1093/mnras/sty1476}

\bibitem[{{Winkler} {et~al.}(2017){Winkler}, {Blair}, \& {Long}}]{winkler17}
{Winkler}, P.~F., {Blair}, W.~P., \& {Long}, K.~S. 2017, \apj, 839, 83,
  \dodoi{10.3847/1538-4357/aa683d}

\bibitem[{{Zacharias} {et~al.}(2010){Zacharias}, {Finch}, {Girard}, {Hambly},
  {Wycoff}, {Zacharias}, {Castillo}, {Corbin}, {DiVittorio}, {Dutta}, {Gaume},
  {Gauss}, {Germain}, {Hall}, {Hartkopf}, {Hsu}, {Holdenried}, {Makarov},
  {Martinez}, {Mason}, {Monet}, {Rafferty}, {Rhodes}, {Siemers}, {Smith},
  {Tilleman}, {Urban}, {Wieder}, {Winter}, \& {Young}}]{zacharias10}
{Zacharias}, N., {Finch}, C., {Girard}, T., {et~al.} 2010, \aj, 139, 2184,
  \dodoi{10.1088/0004-6256/139/6/2184}

\end{thebibliography}


\clearpage

\begin{deluxetable}{lccc}
\tablewidth{6.5in}

\tablecaption{HST Imaging Data for M51\tablenotemark{a}\label{table_hst_obsns}}

\tablehead{
\colhead{Camera} & \colhead{Filter} & \colhead{Dithers} & \colhead{Exp(s)\tablenotemark{b}} }
\startdata
{\bf Prog. 10452:\tablenotemark{c}} &    &     &   \\
ACS/WFC &   F658N &  4x680  & 2720 \\
ACS/WFC &   F435W &  4x680  & 2720 \\
ACS/WFC &   F555W &  4x340  & 1360 \\
ACS/WFC &   F814W &  4x340  & 1360 \\
{\bf Prog. 12762:}      &         &     &   \\
WFC3/UVIS &  F673N &  6x900  & 5400 \\
WFC3/UVIS &  F689M &  2x500  & 1000 \\
\enddata
\tablenotetext{a}{Prog.\ 10452 used six fields of ACS/WFC, and were obtained in 2005 January. 
Prog.\ 12762 obtained four fields of WFC3/UVIS in 2012 April; 
refer to Figure 1 for relative field coverage.} 
\tablenotetext{b}{Exposure time for each field including all dithers.}
\tablenotetext{c}{A MAST High Level Science Program for these data can be
found at URL \url{https://archive.stsci.edu/prepds/m51/}.}

\end{deluxetable}

\begin{deluxetable}{lcccccrc}
\tablewidth{0pt}
\tablecaption{GMOS Imaging Observations of M51\label{table2_imaging_obsns}}
\label{table2_imaging_obsns}

\tablehead{
\colhead {} & \colhead{R.A.} & \colhead{Decl.}& & & \multicolumn{2}{c}{Filter} & \colhead {Exposure}\\ 
\cline{2-3}   \cline{6-7}
\colhead{Field} & \multicolumn{2}{c}{(J2000.)} & Date &\colhead{Designation} &
\colhead{$\rm \lambda_{c}$(\AA)} &
\colhead{$\Delta \lambda$(\AA)\tablenotemark{a}} &
\colhead {(s)} 
}

\startdata
 & & & & H$\alpha$  & 6573 & 72\phn    & $6\times350$ \\
M51 North &  13:29:53.0 & 47:13:40.8 & 2017 Mar 27 & HaC  & 6642& 69\phn & $6\times440$ \\
 & & & & \sii & 6718 & 43\phn & $6\times500$   \\
 & & & 2018 Jun 2 & r & 6300 & 1360\phn & $6\times 60$ \\
 \\
 & & & & H$\alpha$  & 6573 & 72\phn    & $6\times350$ \\
M51 South &  13:29:51.0 & 47:10:10.0 & 2017 May 21 & HaC  & 6642& 69\phn &$6\times440$ \\
 & & & & \sii & 6718 & 43\phn & $6\times500$   \\
 & & & 2018 Jun 2 & r & 6300 & 1360\phn & $6\times 60$ \\
\enddata

\tablenotetext{a}{Full width at half maximum.}
\end{deluxetable}

\startlongtable
\begin{deluxetable}{lrrrrrrrcc}
\tablecaption{SNR Candidates in M51 \label{table3_snr}}
\tablehead{
\colhead{Name} &
\colhead{R.A.} &
\colhead{Decl.} &
\colhead{Diam.} &
\colhead{R} &
\colhead{X-ray} &
\colhead{Radio} &
\colhead{Spec.} &
\colhead{\sii:H$\alpha>0.4$}
\\
\colhead{} &
\colhead{(J2000)} &
\colhead{(J2000)} &
\colhead{(pc)} &
\colhead{(kpc)} &
\colhead{} &
\colhead{} &
\colhead{} &
\colhead{}
}
\startdata
W21-001 & 13:29:37.05 & 47:09:25.5 & 27 & 9.0 & -- & -- & n & -- \\ 
W21-002 & 13:29:37.34 & 47:10:07.6 & 54 & 7.9 & -- & -- & n & -- \\ 
W21-003 & 13:29:38.77 & 47:11:33.3 & 100 & 6.3 & -- & -- & n & -- \\ 
W21-004\tablenotemark{a} & 13:29:39.96 & 47:12:37.1 & 45 & 6.2 & X107 & -- & n & -- \\ 
W21-005 & 13:29:41.24 & 47:08:13.4 & 50 & 10.0 & -- & -- & n & -- \\ 
W21-006 & 13:29:41.39 & 47:10:06.9 & 32 & 6.4 & -- & -- & y & n \\ 
W21-007 & 13:29:41.69 & 47:07:54.1 & 47 & 10.6 & -- & -- & n & -- \\ 
W21-008 & 13:29:42.42 & 47:07:42.5 & 92 & 10.9 & -- & -- & n & -- \\ 
W21-009 & 13:29:42.97 & 47:10:38.0 & 21 & 5.1 & X123 & -- & n & -- \\ 
W21-010 & 13:29:43.19 & 47:10:19.8 & 53 & 5.4 & -- & -- & y & n \\ 
W21-011\tablenotemark{b} & 13:29:43.36 & 47:11:34.1 & 12 & 4.2 & X124 & -- & y & y \\ 
W21-012 & 13:29:43.38 & 47:13:01.4 & 23 & 5.4 & -- & -- & y & y \\ 
W21-013 & 13:29:43.59 & 47:09:48.7 & 30 & 6.2 & -- & -- & y & y \\ 
W21-014 & 13:29:43.98 & 47:12:23.2 & 34 & 4.3 & -- & -- & n & -- \\ 
W21-015 & 13:29:44.01 & 47:11:51.0 & 20 & 4.0 & -- & -- & n & -- \\ 
W21-016 & 13:29:44.04 & 47:11:25.5 & 44 & 4.0 & -- & -- & n & -- \\ 
W21-017 & 13:29:44.25 & 47:10:16.2 & 15 & 5.2 & -- & -- & y & y \\ 
W21-018 & 13:29:44.40 & 47:11:40.1 & 32 & 3.7 & -- & -- & n & -- \\ 
W21-019 & 13:29:44.56 & 47:11:40.3 & 44 & 3.7 & -- & -- & n & -- \\ 
W21-020 & 13:29:44.65 & 47:11:57.0 & 29 & 3.7 & -- & -- & y & y \\ 
W21-021 & 13:29:44.95 & 47:11:34.3 & 18 & 3.5 & -- & -- & y & n \\ 
W21-022 & 13:29:45.01 & 47:11:24.5 & 35 & 3.5 & -- & -- & n & -- \\ 
W21-023 & 13:29:45.28 & 47:13:33.1 & 27 & 5.8 & -- & -- & n & -- \\ 
W21-024 & 13:29:45.90 & 47:10:06.3 & 27 & 5.0 & -- & -- & n & -- \\ 
W21-025 & 13:29:46.07 & 47:12:36.5 & 26 & 3.8 & -- & -- & y & y \\ 
W21-026 & 13:29:46.09 & 47:10:28.3 & 30 & 4.2 & -- & -- & n & -- \\ 
W21-027 & 13:29:46.13 & 47:11:53.2 & 51 & 3.0 & -- & -- & y & y \\ 
W21-028 & 13:29:46.22 & 47:10:19.4 & 29 & 4.5 & -- & -- & n & -- \\ 
W21-029 & 13:29:46.30 & 47:11:05.6 & 23 & 3.2 & -- & -- & y & y \\ 
W21-030 & 13:29:46.32 & 47:08:38.4 & 82 & 8.2 & -- & -- & n & -- \\ 
W21-031 & 13:29:46.34 & 47:13:42.5 & 7 & 5.8 & -- & -- & n & -- \\ 
W21-032 & 13:29:46.34 & 47:11:09.4 & 17 & 3.2 & X144 & -- & y & y \\ 
W21-033 & 13:29:46.40 & 47:12:15.0 & 18 & 3.2 & -- & -- & y & y \\ 
W21-034 & 13:29:46.47 & 47:11:42.3 & 37 & 2.8 & -- & -- & y & y \\ 
W21-035 & 13:29:46.72 & 47:10:43.9 & 33 & 3.6 & -- & -- & y & y \\ 
W21-036 & 13:29:46.72 & 47:08:36.8 & 49 & 8.2 & -- & -- & n & -- \\ 
W21-037 & 13:29:46.76 & 47:11:46.2 & 35 & 2.7 & -- & -- & y & y \\ 
W21-038 & 13:29:46.81 & 47:12:51.5 & 32 & 4.0 & -- & -- & y & y \\ 
W21-039 & 13:29:46.99 & 47:13:47.5 & 39 & 5.9 & -- & -- & n & -- \\ 
W21-040 & 13:29:47.01 & 47:11:04.2 & 24 & 3.0 & X149 & -- & y & y \\ 
W21-041 & 13:29:47.29 & 47:13:34.5 & 31 & 5.3 & -- & -- & n & -- \\ 
W21-042 & 13:29:47.66 & 47:10:36.9 & 25 & 3.5 & -- & -- & y & y \\ 
W21-043 & 13:29:48.10 & 47:09:29.3 & 43 & 5.9 & -- & -- & n & -- \\ 
W21-044 & 13:29:48.30 & 47:13:55.9 & 17 & 6.0 & -- & -- & y & y \\ 
W21-045 & 13:29:48.36 & 47:13:25.5 & 24 & 4.8 & -- & -- & n & -- \\ 
W21-046 & 13:29:48.36 & 47:09:42.7 & 53 & 5.3 & -- & -- & n & -- \\ 
W21-047 & 13:29:48.94 & 47:12:03.3 & 66 & 1.9 & -- & -- & n & -- \\ 
W21-048 & 13:29:49.10 & 47:10:26.0 & 29 & 3.5 & -- & -- & n & -- \\ 
W21-049 & 13:29:49.20 & 47:13:24.9 & 26 & 4.6 & -- & -- & n & -- \\ 
W21-050 & 13:29:49.69 & 47:10:04.3 & 37 & 4.3 & -- & -- & n & -- \\ 
W21-051 & 13:29:49.92 & 47:11:20.6 & 21 & 1.5 & X161 & M07-017 & y & y \\ 
W21-052 & 13:29:50.24 & 47:12:10.5 & 22 & 1.6 & X166 & -- & y & y \\ 
W21-053 & 13:29:50.26 & 47:11:23.9 & 7 & 1.3 & -- & -- & n & -- \\ 
W21-054 & 13:29:50.34 & 47:11:41.6 & 36 & 1.1 & X169 & -- & n & -- \\ 
W21-055 & 13:29:50.48 & 47:11:27.1 & 9 & 1.2 & X171 & M07-031 & n & -- \\ 
W21-056 & 13:29:50.54 & 47:09:44.1 & 19 & 5.0 & -- & -- & n & -- \\ 
W21-057 & 13:29:50.54 & 47:11:44.6 & 24 & 1.0 & X170 & -- & n & -- \\ 
W21-058 & 13:29:50.58 & 47:10:52.2 & 29 & 2.3 & X172 & -- & y & y \\ 
W21-059 & 13:29:50.98 & 47:11:26.8 & 12 & 1.0 & X177 & -- & n & -- \\ 
W21-060 & 13:29:51.08 & 47:12:56.3 & 27 & 3.2 & -- & -- & y & y \\ 
W21-061 & 13:29:51.45 & 47:11:52.1 & 13 & 0.7 & X180 & -- & n & -- \\ 
W21-062 & 13:29:51.54 & 47:09:18.9 & 34 & 6.0 & -- & -- & y & n \\ 
W21-063 & 13:29:51.69 & 47:12:41.5 & 19 & 2.5 & -- & -- & n & -- \\ 
W21-064 & 13:29:51.87 & 47:12:25.4 & 13 & 1.8 & X187 & -- & y & y \\ 
W21-065 & 13:29:51.90 & 47:12:00.5 & 21 & 0.8 & -- & -- & n & -- \\ 
W21-066 & 13:29:51.92 & 47:11:48.0 & 13 & 0.4 & -- & -- & n & -- \\ 
W21-067 & 13:29:51.95 & 47:12:13.0 & 18 & 1.3 & -- & -- & n & -- \\ 
W21-068 & 13:29:52.02 & 47:10:46.9 & 26 & 2.3 & X189 & -- & y & y \\ 
W21-069 & 13:29:52.03 & 47:12:37.3 & 16 & 2.3 & -- & -- & y & y \\ 
W21-070 & 13:29:52.07 & 47:12:12.8 & 13 & 1.3 & X192 & -- & y & y \\ 
W21-071 & 13:29:52.07 & 47:11:26.7 & 17 & 0.7 & X191 & M07-047 & n & -- \\ 
W21-072 & 13:29:52.07 & 47:11:34.3 & 59 & 0.5 & -- & -- & n & -- \\ 
W21-073 & 13:29:52.08 & 47:11:53.9 & 16 & 0.6 & -- & -- & n & -- \\ 
W21-074 & 13:29:52.11 & 47:12:13.5 & 20 & 1.3 & X192 & -- & y & y \\ 
W21-075 & 13:29:52.15 & 47:11:51.6 & 10 & 0.5 & -- & -- & n & -- \\ 
W21-076 & 13:29:52.20 & 47:11:29.6 & 12 & 0.6 & X196 & M07-049 & n & -- \\ 
W21-077 & 13:29:52.21 & 47:11:46.9 & 13 & 0.3 & -- & -- & n & -- \\ 
W21-078 & 13:29:52.22 & 47:12:03.0 & 14 & 0.9 & X194 & -- & n & -- \\ 
W21-079 & 13:29:52.23 & 47:12:43.7 & 9 & 2.5 & -- & -- & n & -- \\ 
W21-080 & 13:29:52.25 & 47:12:28.2 & 22 & 1.9 & X199 & -- & y & y \\ 
W21-081 & 13:29:52.29 & 47:11:59.1 & 10 & 0.7 & -- & -- & n & -- \\ 
W21-082 & 13:29:52.33 & 47:11:35.9 & 13 & 0.3 & -- & M07-050 & n & -- \\ 
W21-083 & 13:29:52.42 & 47:10:33.3 & 39 & 2.9 & -- & -- & n & -- \\ 
W21-084 & 13:29:52.47 & 47:10:25.2 & 22 & 3.2 & -- & -- & y & y \\ 
W21-085 & 13:29:52.73 & 47:11:21.7 & 16 & 0.9 & X208 & M07-052 & y & y \\ 
W21-086 & 13:29:52.78 & 47:12:43.2 & 66 & 2.5 & -- & -- & n & -- \\ 
W21-087 & 13:29:53.00 & 47:11:42.3 & 7 & 0.1 & -- & -- & n & -- \\ 
W21-088 & 13:29:53.13 & 47:11:51.0 & 11 & 0.4 & -- & -- & y & y \\ 
W21-089 & 13:29:53.27 & 47:09:16.3 & 18 & 6.1 & -- & -- & y & y \\ 
W21-090 & 13:29:53.29 & 47:12:43.1 & 23 & 2.5 & -- & -- & n & -- \\ 
W21-091 & 13:29:53.42 & 47:11:48.3 & 29 & 0.4 & -- & -- & n & -- \\ 
W21-092 & 13:29:53.46 & 47:12:20.2 & 19 & 1.6 & -- & -- & n & -- \\ 
W21-093 & 13:29:53.58 & 47:14:18.1 & 34 & 6.5 & -- & -- & n & -- \\ 
W21-094 & 13:29:53.64 & 47:12:20.0 & 17 & 1.6 & -- & -- & y & y \\ 
W21-095 & 13:29:53.77 & 47:09:30.3 & 75 & 5.6 & -- & -- & y & y \\ 
W21-096 & 13:29:53.96 & 47:09:23.5 & 12 & 5.8 & X235 & -- & n & -- \\ 
W21-097 & 13:29:53.98 & 47:12:38.1 & 31 & 2.4 & -- & -- & n & -- \\ 
W21-098 & 13:29:54.12 & 47:11:41.0 & 12 & 0.6 & -- & -- & y & y \\ 
W21-099 & 13:29:54.26 & 47:10:33.1 & 49 & 3.0 & -- & -- & n & -- \\ 
W21-100 & 13:29:54.30 & 47:11:30.1 & 11 & 0.9 & X242 & M07-061 & n & -- \\ 
W21-101 & 13:29:54.38 & 47:11:21.6 & 12 & 1.2 & X243 & -- & y & y \\ 
W21-102 & 13:29:54.40 & 47:10:45.2 & 36 & 2.5 & -- & -- & n & -- \\ 
W21-103 & 13:29:54.44 & 47:11:36.5 & 18 & 0.8 & X245 & -- & y & y \\ 
W21-104 & 13:29:54.44 & 47:14:19.1 & 17 & 6.5 & X246 & -- & y & y \\ 
W21-105 & 13:29:54.55 & 47:11:25.8 & 11 & 1.1 & -- & -- & n & -- \\ 
W21-106 & 13:29:54.59 & 47:13:17.3 & 26 & 4.0 & X249 & -- & y & y \\ 
W21-107 & 13:29:54.66 & 47:09:35.4 & 40 & 5.4 & -- & -- & n & -- \\ 
W21-108 & 13:29:54.81 & 47:09:59.6 & 12 & 4.4 & X252 & -- & y & y \\ 
W21-109 & 13:29:54.93 & 47:11:25.9 & 12 & 1.2 & X256 & -- & n & -- \\ 
W21-110 & 13:29:54.94 & 47:11:33.3 & 15 & 1.1 & X259 & M07-064 & n & -- \\ 
W21-111 & 13:29:54.98 & 47:11:33.6 & 9 & 1.1 & X259 & M07-064 & n & -- \\ 
W21-112 & 13:29:55.06 & 47:11:33.6 & 9 & 1.1 & X259 & -- & n & -- \\ 
W21-113 & 13:29:55.12 & 47:11:50.7 & 22 & 1.1 & X261 & -- & n & -- \\ 
W21-114 & 13:29:55.12 & 47:10:42.6 & 9 & 2.8 & X262 & -- & n & -- \\ 
W21-115 & 13:29:55.19 & 47:12:16.0 & 22 & 1.8 & X264 & -- & n & -- \\ 
W21-116 & 13:29:55.21 & 47:09:36.1 & 48 & 5.4 & -- & -- & n & -- \\ 
W21-117 & 13:29:55.24 & 47:10:46.5 & 18 & 2.6 & X265 & M07-068 & n & -- \\ 
W21-118 & 13:29:55.49 & 47:13:50.0 & 21 & 5.4 & -- & -- & y & y \\ 
W21-119 & 13:29:55.53 & 47:14:13.1 & 39 & 6.4 & -- & -- & n & -- \\ 
W21-120 & 13:29:55.56 & 47:12:10.0 & 22 & 1.7 & X274 & M07-074 & n & -- \\ 
W21-121 & 13:29:55.56 & 47:12:05.9 & 37 & 1.6 & -- & -- & n & -- \\ 
W21-122 & 13:29:55.68 & 47:09:28.2 & 47 & 5.8 & -- & -- & n & -- \\ 
W21-123 & 13:29:55.70 & 47:10:43.8 & 14 & 2.8 & X277 & -- & y & y \\ 
W21-124 & 13:29:55.75 & 47:10:46.5 & 66 & 2.8 & X278 & -- & y & n \\ 
W21-125 & 13:29:55.81 & 47:10:32.9 & 52 & 3.3 & -- & -- & y & y \\ 
W21-126 & 13:29:55.86 & 47:11:44.7 & 17 & 1.4 & X279 & M07-079 & n & -- \\ 
W21-127 & 13:29:56.01 & 47:09:14.4 & 50 & 6.4 & -- & -- & n & -- \\ 
W21-128 & 13:29:56.07 & 47:13:50.8 & 11 & 5.5 & X281 & -- & y & y \\ 
W21-129 & 13:29:56.08 & 47:10:44.2 & 20 & 2.9 & -- & -- & n & -- \\ 
W21-130 & 13:29:56.19 & 47:10:47.3 & 11 & 2.8 & X283 & M07-083 & n & -- \\ 
W21-131 & 13:29:56.86 & 47:11:59.8 & 13 & 2.0 & -- & -- & n & -- \\ 
W21-132 & 13:29:56.93 & 47:13:37.1 & 20 & 5.1 & -- & -- & y & y \\ 
W21-133 & 13:29:56.97 & 47:09:52.2 & 47 & 5.0 & -- & -- & n & -- \\ 
W21-134 & 13:29:57.48 & 47:10:37.6 & 26 & 3.5 & X293 & M07-084 & y & y \\ 
W21-135 & 13:29:57.49 & 47:10:34.3 & 32 & 3.6 & -- & -- & y & y \\ 
W21-136 & 13:29:57.72 & 47:09:06.6 & 61 & 7.0 & -- & -- & n & -- \\ 
W21-137 & 13:29:57.94 & 47:10:48.8 & 29 & 3.3 & -- & -- & n & -- \\ 
W21-138 & 13:29:58.26 & 47:09:05.7 & 24 & 7.1 & -- & -- & n & -- \\ 
W21-139 & 13:29:58.44 & 47:09:02.8 & 24 & 7.2 & -- & -- & y & y \\ 
W21-140 & 13:29:58.44 & 47:14:03.8 & 7 & 6.4 & -- & -- & n & -- \\ 
W21-141 & 13:29:58.77 & 47:13:53.6 & 20 & 6.0 & -- & -- & n & -- \\ 
W21-142 & 13:29:59.05 & 47:12:03.4 & 30 & 3.0 & X313 & -- & y & y \\ 
W21-143 & 13:29:59.24 & 47:10:41.8 & 51 & 3.9 & -- & -- & n & -- \\ 
W21-144 & 13:29:59.34 & 47:12:51.0 & 147 & 4.1 & -- & -- & n & -- \\ 
W21-145 & 13:29:59.73 & 47:13:02.5 & 53 & 4.5 & -- & -- & n & -- \\ 
W21-146 & 13:29:59.80 & 47:10:51.6 & 49 & 3.9 & -- & -- & n & -- \\ 
W21-147 & 13:29:59.94 & 47:13:35.4 & 18 & 5.7 & -- & -- & y & y \\ 
W21-148 & 13:30:00.10 & 47:13:17.7 & 12 & 5.1 & -- & -- & y & y \\ 
W21-149 & 13:30:00.21 & 47:12:38.7 & 36 & 4.1 & -- & -- & n & -- \\ 
W21-150 & 13:30:00.44 & 47:09:15.2 & 52 & 7.2 & -- & -- & n & -- \\ 
W21-151 & 13:30:00.54 & 47:11:36.9 & 20 & 3.5 & X326 & -- & n & -- \\ 
W21-152 & 13:30:00.67 & 47:10:53.0 & 35 & 4.2 & -- & -- & y & y \\ 
W21-153 & 13:30:00.72 & 47:11:43.6 & 30 & 3.6 & -- & -- & n & -- \\ 
W21-154 & 13:30:00.86 & 47:12:55.8 & 17 & 4.7 & -- & -- & n & -- \\ 
W21-155 & 13:30:01.01 & 47:12:42.9 & 17 & 4.5 & X335 & -- & y & y \\ 
W21-156 & 13:30:01.39 & 47:11:58.1 & 12 & 4.0 & X339 & M07-096 & n & -- \\ 
W21-157 & 13:30:01.40 & 47:12:01.5 & 28 & 4.0 & -- & -- & n & -- \\ 
W21-158 & 13:30:01.45 & 47:12:36.3 & 54 & 4.5 & -- & -- & n & -- \\ 
W21-159 & 13:30:01.52 & 47:12:41.1 & 17 & 4.6 & -- & -- & n & -- \\ 
W21-160 & 13:30:01.99 & 47:10:31.2 & 24 & 5.2 & -- & -- & n & -- \\ 
W21-161 & 13:30:02.07 & 47:09:51.2 & 23 & 6.4 & X347 & M07-099 & y & y \\ 
W21-162 & 13:30:02.32 & 47:09:58.8 & 38 & 6.2 & X349 & -- & y & y \\ 
W21-163 & 13:30:03.08 & 47:09:25.4 & 27 & 7.5 & -- & -- & n & -- \\ 
W21-164 & 13:30:03.35 & 47:13:06.9 & 21 & 5.9 & X352 & -- & y & y \\ 
W21-165 & 13:30:03.66 & 47:09:17.9 & 71 & 7.9 & X354 & -- & n & -- \\ 
W21-166 & 13:30:03.80 & 47:09:40.7 & 58 & 7.2 & -- & -- & n & -- \\ 
W21-167 & 13:30:04.08 & 47:10:03.8 & 12 & 6.7 & X357 & -- & y & y \\ 
W21-168 & 13:30:04.32 & 47:09:40.9 & 60 & 7.4 & -- & -- & n & -- \\ 
W21-169 & 13:30:04.53 & 47:12:02.9 & 40 & 5.4 & X363 & -- & y & y \\ 
W21-170 & 13:30:04.77 & 47:13:01.2 & 18 & 6.3 & -- & -- & n & -- \\ 
W21-171 & 13:30:04.94 & 47:10:26.4 & 34 & 6.4 & -- & -- & n & -- \\ 
W21-172 & 13:30:05.00 & 47:13:02.0 & 14 & 6.4 & X366 & -- & y & y \\ 
W21-173 & 13:30:05.31 & 47:13:13.2 & 32 & 6.7 & -- & -- & y & y \\ 
W21-174 & 13:30:05.65 & 47:12:51.8 & 17 & 6.4 & -- & -- & n & -- \\ 
W21-175 & 13:30:06.09 & 47:09:55.1 & 80 & 7.6 & -- & -- & n & -- \\ 
W21-176 & 13:30:06.79 & 47:12:10.3 & 40 & 6.4 & -- & -- & n & -- \\ 
W21-177 & 13:30:07.34 & 47:14:17.9 & 29 & 9.1 & -- & -- & y & y \\ 
W21-178 & 13:30:07.54 & 47:10:41.1 & 44 & 7.2 & -- & -- & n & -- \\ 
W21-179 & 13:30:07.64 & 47:12:26.0 & 17 & 6.9 & -- & -- & y & n \\ 
\enddata
\tablenotetext{a}{This object corresponds with ULX-1 reported by \cite{urquhart18}; the object is distinctly elongated and is likely a jet-like structure.  The listed diameter corresponds to the long dimension of the observed nebula.}
\tablenotetext{b}{This object corresponds with source 5 in \cite{terashima04} and is likely also source ULX-2 reported by \cite{brightman20}; the listed coordinate in the latter reference is off, but inspection of Chandra data shows the strong X-ray source near this position actually aligns with W21-011.}
\end{deluxetable}

\begin{deluxetable}{rrrrrr}
\tablecaption{\hii\ regions in M51\label{table_h2}}
\tablehead{
\colhead{Name} &
\colhead{R.A.} &
\colhead{Decl.} &
\colhead{R} &
\colhead{X-ray} &
\colhead{Radio}
\\
\colhead{} &
\colhead{(J2000)} &
\colhead{(J2000)} &
\colhead{(kpc)} &
\colhead{} &
\colhead{}
}
\startdata
HII-01 & 13:29:39.31 & 47:08:40.4 & 9.6 & -- & M07-005 \\ 
HII-02 & 13:29:39.38 & 47:08:35.9 & 9.7 & -- & -- \\ 
HII-03 & 13:29:43.14 & 47:10:21.4 & 5.4 & -- & -- \\ 
HII-04 & 13:29:43.18 & 47:10:24.3 & 5.3 & -- & -- \\ 
HII-05 & 13:29:43.23 & 47:10:26.0 & 5.3 & -- & -- \\ 
HII-06 & 13:29:43.66 & 47:09:51.0 & 6.1 & -- & -- \\ 
HII-07 & 13:29:43.73 & 47:13:08.3 & 5.5 & -- & -- \\ 
HII-08 & 13:29:43.77 & 47:13:10.1 & 5.5 & -- & -- \\ 
HII-09 & 13:29:44.12 & 47:10:22.9 & 5.0 & X130 & M07-007 \\ 
HII-10 & 13:29:44.46 & 47:10:58.6 & 4.1 & -- & -- \\ 
HII-11 & 13:29:44.50 & 47:10:55.6 & 4.2 & -- & -- \\ 
HII-12 & 13:29:44.65 & 47:11:54.9 & 3.7 & -- & -- \\ 
HII-13 & 13:29:44.91 & 47:11:32.2 & 3.5 & -- & -- \\ 
HII-14 & 13:29:45.86 & 47:13:41.8 & 5.9 & -- & -- \\ 
HII-15 & 13:29:45.86 & 47:13:32.3 & 5.6 & -- & -- \\ 
HII-16 & 13:29:46.10 & 47:12:34.3 & 3.7 & -- & -- \\ 
HII-17 & 13:29:46.31 & 47:12:17.6 & 3.3 & -- & -- \\ 
HII-18 & 13:29:46.33 & 47:11:07.8 & 3.2 & -- & -- \\ 
HII-19 & 13:29:46.39 & 47:12:11.8 & 3.1 & -- & -- \\ 
HII-20 & 13:29:47.15 & 47:08:52.1 & 7.5 & -- & -- \\ 
HII-21 & 13:29:48.02 & 47:10:17.2 & 4.1 & -- & -- \\ 
HII-22 & 13:29:49.27 & 47:09:25.9 & 5.9 & -- & -- \\ 
HII-23 & 13:29:49.95 & 47:11:24.4 & 1.4 & X162 & -- \\ 
HII-24 & 13:29:51.07 & 47:12:53.1 & 3.0 & -- & -- \\ 
HII-25 & 13:29:52.02 & 47:12:32.5 & 2.1 & -- & -- \\ 
HII-26 & 13:29:52.08 & 47:12:45.2 & 2.6 & X193 & -- \\ 
HII-27 & 13:29:52.79 & 47:11:23.6 & 0.9 & -- & -- \\ 
HII-28 & 13:29:52.80 & 47:14:07.3 & 6.0 & -- & -- \\ 
HII-29 & 13:29:53.24 & 47:09:32.6 & 5.4 & -- & -- \\ 
HII-30 & 13:29:54.87 & 47:10:04.2 & 4.2 & -- & -- \\ 
HII-31 & 13:29:55.51 & 47:13:47.2 & 5.3 & -- & -- \\ 
HII-32 & 13:29:55.59 & 47:13:52.3 & 5.5 & -- & -- \\ 
HII-33 & 13:29:57.26 & 47:09:18.7 & 6.4 & -- & -- \\ 
HII-34 & 13:29:59.26 & 47:13:44.6 & 5.8 & -- & -- \\ 
HII-35 & 13:29:59.70 & 47:13:58.6 & 6.4 & X319 & -- \\ 
HII-36 & 13:30:00.93 & 47:09:29.6 & 6.8 & -- & M07-094 \\ 
HII-37 & 13:30:01.43 & 47:09:03.5 & 7.8 & -- & -- \\ 
HII-38 & 13:30:02.36 & 47:09:49.6 & 6.5 & -- & M07-100 \\ 
HII-39 & 13:30:03.41 & 47:12:53.8 & 5.6 & -- & -- \\ 
HII-40 & 13:30:03.47 & 47:09:41.2 & 7.1 & -- & M07-102 \\ 
HII-41 & 13:30:05.00 & 47:13:03.7 & 6.4 & -- & -- \\ 
HII-42 & 13:30:07.12 & 47:13:57.8 & 8.5 & -- & -- \\ 
HII-43 & 13:30:07.25 & 47:14:07.3 & 8.8 & -- & -- \\ 
HII-44 & 13:30:07.41 & 47:13:21.9 & 7.7 & -- & M07-105 \\ 
HII-45 & 13:30:07.45 & 47:14:14.8 & 9.1 & -- & -- \\ 
\enddata
\label{table6_h2}
\end{deluxetable}


\begin{longrotatetable}
\begin{deluxetable}{rrrrrrrrrr}
\tablecaption{Spectra of SNR Candidates \label{table6_snr_spectra}}
\tablehead{
\colhead{Name} &
\colhead{H$\alpha$ flux$^{a}$} &
\colhead{H$\beta$} &
\colhead{\oiii\,$\lambda$5007} &
\colhead{H$\alpha$} &
\colhead{\nii\,$\lambda$6584} &
\colhead{\oi\,$\lambda$6300} &
\colhead{\sii\,$\lambda$6716} &
\colhead{\sii\,$\lambda$6731} &
\colhead{\sii:H$\alpha$}
}
\startdata
W21-006 & 13 & 71 & ... & 300 & 107 & ... & 47 & 34 & 0.27 \\ 
W21-010 & 41 & 96 & ... & 300 & 131 & ... & 59 & 43 & 0.34 \\ 
W21-011 & 60 & 84 & 102 & 300 & 261 & 51 & 62 & 75 & 0.45 \\ 
W21-012 & 26 & 74 & 83 & 300 & 261 & 40 & 113 & 80 & 0.64 \\ 
W21-013 & 58 & 70 & 100 & 300 & 310 & $\sim$61 & 145 & 114 & 0.86 \\ 
W21-017 & 123 & 71 & 79 & 300 & 203 & 30 & 95 & 70 & 0.55 \\ 
W21-020 & 35 & 32 & ... & 300 & 155 & $\sim$60 & 83 & 61 & 0.48 \\ 
W21-021 & 127 & 31 & ... & 300 & 147 & ... & 62 & 48 & 0.37 \\ 
W21-025 & 9 & 143 & 275 & 300 & 635 & 124 & 382 & 271 & 2.18 \\ 
W21-027 & 32 & 61 & 207 & 300 & 298 & 32 & 114 & 78 & 0.64 \\ 
W21-029 & 6 & 20 & 135 & 300 & 512 & 152 & 222 & 181 & 1.34 \\ 
W21-032 & 5 & ... & 196 & 300 & 648 & $\sim$75 & 103 & 141 & 0.81 \\ 
W21-033 & 88 & 42 & 16 & 300 & 171 & $\sim$22 & 87 & 65 & 0.51 \\ 
W21-034 & 9 & ... & 396 & 300 & 752 & $\sim$110 & 279 & 223 & 1.67 \\ 
W21-035 & 126 & 52 & 70 & 300 & 183 & $\sim$9 & 80 & 62 & 0.48 \\ 
W21-037 & 39 & 56 & 20 & 300 & 239 & 45 & 127 & 95 & 0.74 \\ 
W21-038 & 6 & $\sim$225 & 410 & 300 & 446 & ... & 232 & 174 & 1.35 \\ 
W21-040 & 98 & 77 & 132 & 300 & 256 & 24 & 73 & 66 & 0.47 \\ 
W21-042 & 17 & 70 & 272 & 300 & 505 & $\sim$54 & 176 & 112 & 0.96 \\ 
W21-044 & 20 & $\sim$35 & $\sim$61 & 300 & 286 & 120 & 194 & 170 & 1.21 \\ 
W21-051 & 24 & 17 & 108 & 300 & 830 & 116 & 310 & 256 & 1.89 \\ 
W21-052 & 24 & 46 & 68 & 300 & 505 & 92 & 208 & 173 & 1.27 \\ 
W21-058 & 51 & 58 & 123 & 300 & 299 & 26 & 116 & 88 & 0.68 \\ 
W21-060 & 14 & 91 & 426 & 300 & 372 & ... & 147 & 133 & 0.93 \\ 
W21-062 & 76 & 68 & ... & 300 & 123 & ... & 60 & 38 & 0.33 \\ 
W21-064 & 10 & ... & 414 & 300 & 953 & 91 & 188 & 226 & 1.38 \\ 
W21-068 & 7 & 69 & 413 & 300 & 767 & 66 & 309 & 220 & 1.76 \\ 
W21-069 & 18 & $\sim$60 & 112 & 300 & 560 & $\sim$95 & 240 & 177 & 1.39 \\ 
W21-070 & 31 & 90 & 271 & 300 & 690 & 79 & 160 & 165 & 1.08 \\ 
W21-074 & 42 & 61 & 156 & 300 & 870 & 88 & 268 & 230 & 1.66 \\ 
W21-080 & 13 & 45 & 189 & 300 & 771 & 107 & 206 & 164 & 1.23 \\ 
W21-084 & 8 & $\sim$18 & 256 & 300 & 523 & 82 & 177 & 141 & 1.06 \\ 
W21-085 & 103 & 52 & 97 & 300 & 853 & 84 & 184 & 203 & 1.29 \\ 
W21-088 & 14 & $\sim$43 & 188 & 300 & 1420 & $\sim$77 & 293 & 251 & 1.81 \\ 
W21-089 & 41 & 51 & 98 & 300 & 411 & 90 & 226 & 187 & 1.38 \\ 
W21-094 & 36 & 104 & 154 & 300 & 317 & $\sim$27 & 98 & 76 & 0.58 \\ 
W21-095 & 58 & 49 & 60 & 300 & 215 & ... & 102 & 76 & 0.59 \\ 
W21-098 & 33 & ... & 31 & 300 & 460 & 39 & 126 & 103 & 0.76 \\ 
W21-101 & 67 & 26 & 70 & 300 & 438 & 51 & 87 & 99 & 0.62 \\ 
W21-103 & 59 & 37 & 157 & 300 & 777 & 45 & 143 & 144 & 0.96 \\ 
W21-104 & 56 & 70 & 48 & 300 & 253 & 60 & 150 & 120 & 0.90 \\ 
W21-106 & 3 & $\sim$145 & 315 & 300 & 835 & 436 & 253 & 188 & 1.47 \\ 
W21-108 & 38 & 49 & 134 & 300 & 413 & $\sim$99 & 196 & 182 & 1.26 \\ 
W21-118 & 40 & 85 & 20 & 300 & 202 & 41 & 118 & 89 & 0.69 \\ 
W21-123 & 38 & 21 & 54 & 300 & 729 & 127 & 226 & 264 & 1.63 \\ 
W21-124 & 276 & 56 & 49 & 300 & 163 & $\sim$7 & 60 & 41 & 0.34 \\ 
W21-125 & 15 & 167 & 192 & 300 & 480 & 27 & 195 & 133 & 1.09 \\ 
W21-128 & 14 & 67 & 225 & 300 & 617 & 211 & 178 & 234 & 1.38 \\ 
W21-132 & 17 & 42 & ... & 300 & 314 & 50 & 128 & 112 & 0.80 \\ 
W21-134 & 134 & 71 & 173 & 300 & 308 & 35 & 100 & 93 & 0.64 \\ 
W21-135 & 10 & 51 & 156 & 300 & 483 & 101 & 195 & 146 & 1.14 \\ 
W21-139 & 19 & 86 & 57 & 300 & 228 & ... & 144 & 101 & 0.82 \\ 
W21-142 & 20 & 78 & 286 & 300 & 643 & 61 & 242 & 190 & 1.44 \\ 
W21-147 & 37 & 92 & 156 & 300 & 384 & 54 & 174 & 131 & 1.02 \\ 
W21-148 & 149 & 73 & 28 & 300 & 139 & $\sim$16 & 76 & 59 & 0.45 \\ 
W21-152 & 14 & 70 & 180 & 300 & 514 & $\sim$109 & 172 & 148 & 1.07 \\ 
W21-155 & 14 & $\sim$60 & 263 & 300 & 647 & $\sim$118 & 137 & 146 & 0.94 \\ 
W21-161 & 71 & 97 & 133 & 300 & 211 & 38 & 130 & 115 & 0.82 \\ 
W21-162 & 28 & 91 & 469 & 300 & 371 & $\sim$21 & 168 & 120 & 0.96 \\ 
W21-164 & 124 & 64 & 101 & 300 & 294 & 46 & 136 & 103 & 0.80 \\ 
W21-167 & 26 & 101 & 128 & 300 & 295 & 70 & 89 & 113 & 0.67 \\ 
W21-169 & 19 & 77 & 383 & 300 & 311 & 27 & 111 & 105 & 0.72 \\ 
W21-172 & 24 & 68 & 176 & 300 & 403 & 87 & 95 & 92 & 0.62 \\ 
W21-173 & 109 & 95 & 53 & 300 & 195 & 24 & 92 & 72 & 0.55 \\ 
W21-177 & 31 & 58 & 114 & 300 & 254 & $\sim$45 & 158 & 120 & 0.93 \\ 
W21-179 & 185 & 33 & 14 & 300 & 109 & ... & 52 & 38 & 0.30 \\ 
\enddata
\tablenotetext{a}{Flux in units of 10$^{-17}$ ergs cm$^{-2}$ s$^{-1}$.}
\end{deluxetable}
\end{longrotatetable}

\begin{longrotatetable}
\begin{deluxetable}{rrrrrrrrrr}
\tablecaption{Spectra of \hii\ Regions \label{table7_h2_spectra}}
\tablehead{
\colhead{Name} &
\colhead{H$\alpha$ flux$^{a}$} &
\colhead{H$\beta$} &
\colhead{\oiii\,$\lambda$5007} &
\colhead{H$\alpha$} &
\colhead{\nii\,$\lambda$6584} &
\colhead{\oi\,$\lambda$6300} &
\colhead{\sii\,$\lambda$6716} &
\colhead{\sii\,$\lambda$6731} &
\colhead{\sii:H$\alpha$}
}
\startdata
HII-01 & 1709 & 49 & 39 & 300 & 107 & ... & 29 & 21 & 0.17 \\ 
HII-02 & 598 & 46 & 45 & 300 & 108 & ... & 28 & 20 & 0.16 \\ 
HII-03 & 68 & 79 & ... & 300 & 147 & ... & 41 & 25 & 0.22 \\ 
HII-04 & 2014 & 70 & 14 & 300 & 119 & ... & 27 & 19 & 0.15 \\ 
HII-05 & 145 & 63 & $\sim$6 & 300 & 107 & ... & 36 & 26 & 0.21 \\ 
HII-06 & 44 & 36 & ... & 300 & 98 & ... & 20 & 21 & 0.14 \\ 
HII-07 & 299 & 60 & 55 & 300 & 142 & ... & 18 & 15 & 0.11 \\ 
HII-08 & 57 & 34 & ... & 300 & 134 & ... & 19 & 13 & 0.10 \\ 
HII-09 & 4945 & 68 & 5 & 300 & 98 & 2 & 30 & 22 & 0.18 \\ 
HII-10 & 80 & 64 & ... & 300 & 79 & ... & 24 & 19 & 0.14 \\ 
HII-11 & 11 & 55 & ... & 300 & 91 & ... & 28 & 35 & 0.21 \\ 
HII-12 & 125 & 34 & ... & 300 & 89 & ... & 36 & 24 & 0.20 \\ 
HII-13 & 179 & 71 & ... & 300 & 70 & ... & 16 & 12 & 0.09 \\ 
HII-14 & 209 & 69 & 33 & 300 & 135 & ... & 56 & 38 & 0.31 \\ 
HII-15 & 223 & 53 & 42 & 300 & 110 & ... & 27 & 19 & 0.16 \\ 
HII-16 & 20 & 45 & $\sim$16 & 300 & 109 & ... & 21 & 16 & 0.12 \\ 
HII-17 & 61 & 49 & ... & 300 & 82 & ... & 26 & 18 & 0.15 \\ 
HII-18 & 67 & 40 & ... & 300 & 132 & $\sim$6 & 24 & 18 & 0.14 \\ 
HII-19 & 90 & 39 & $\sim$9 & 300 & 129 & $\sim$15 & 58 & 41 & 0.33 \\ 
HII-20 & 143 & 77 & 25 & 300 & 119 & ... & 28 & 19 & 0.16 \\ 
HII-21 & 27 & 59 & ... & 300 & 107 & ... & 24 & 18 & 0.14 \\ 
HII-22 & 32 & 60 & ... & 300 & 98 & ... & 27 & 17 & 0.15 \\ 
HII-23 & 1065 & 41 & 6 & 300 & 87 & 5 & 27 & 21 & 0.16 \\ 
HII-24 & 327 & 75 & $\sim$10 & 300 & 58 & ... & 19 & 14 & 0.11 \\ 
HII-25 & 130 & 65 & 24 & 300 & 52 & ... & 20 & 18 & 0.13 \\ 
HII-26 & 2152 & 65 & 4 & 300 & 78 & 3 & 24 & 18 & 0.14 \\ 
HII-27 & 188 & 49 & ... & 300 & 65 & ... & 19 & 13 & 0.11 \\ 
HII-28 & 277 & 76 & 22 & 300 & 123 & ... & 26 & 18 & 0.15 \\ 
HII-29 & 157 & 68 & ... & 300 & 102 & ... & 32 & 23 & 0.18 \\ 
HII-30 & 151 & 52 & $\sim$9 & 300 & 135 & ... & 38 & 25 & 0.21 \\ 
HII-31 & 24 & 56 & ... & 300 & 106 & ... & 54 & 38 & 0.31 \\ 
HII-32 & 347 & 69 & $\sim$3 & 300 & 87 & ... & 37 & 28 & 0.22 \\ 
HII-33 & 57 & 63 & ... & 300 & 126 & ... & 33 & 25 & 0.19 \\ 
HII-34 & 182 & 56 & ... & 300 & 116 & ... & 35 & 25 & 0.20 \\ 
HII-35 & 1522 & 104 & 21 & 300 & 189 & 8 & 34 & 24 & 0.19 \\ 
HII-36 & 1759 & 74 & 14 & 300 & 124 & 1 & 21 & 15 & 0.12 \\ 
HII-37 & 190 & 73 & ... & 300 & 113 & ... & 34 & 22 & 0.19 \\ 
HII-38 & 6424 & 57 & 8 & 300 & 114 & ... & 30 & 22 & 0.18 \\ 
HII-39 & 283 & 76 & 17 & 300 & 127 & 8 & 29 & 22 & 0.17 \\ 
HII-40 & 871 & 66 & 28 & 300 & 130 & 3 & 51 & 27 & 0.26 \\ 
HII-41 & 194 & 64 & 10 & 300 & 129 & $\sim$4 & 22 & 16 & 0.13 \\ 
HII-42 & 724 & 26 & 16 & 300 & 145 & ... & 37 & 27 & 0.21 \\ 
HII-43 & 117 & 46 & 41 & 300 & 127 & ... & 37 & 25 & 0.20 \\ 
HII-44 & 1157 & 60 & 38 & 300 & 133 & 2 & 27 & 22 & 0.16 \\ 
HII-45 & 899 & 34 & 29 & 300 & 124 & ... & 51 & 36 & 0.29 \\ 
\enddata
\tablenotetext{a}{Flux in units of 10$^{-17}$ ergs cm$^{-2}$ s$^{-1}$.}
\end{deluxetable}
\end{longrotatetable}

\begin{deluxetable}{cccrrl}
\tablewidth{0pt}
\tablecaption{Additional M51 Soft X-ray or Radio Source Matches\tablenotemark{a}}
\tablehead{
\colhead {X-ray/Radio ID\tablenotemark{b}} &  
\colhead {R.A.} &  
\colhead {Decl.} &  
\colhead {Diam.} &
\colhead {R} &
\colhead {Comments}
\\
\colhead{} &
\colhead{(J2000)} &
\colhead{(J2000)} &
\colhead{(pc)} &
\colhead{(kpc)} &
\colhead{} 
}
\startdata
\hline\\[-9pt]
& & Soft X-ray Sources & & & \\
\hline
X-145   & 13:29:46.34 & +47:11:15.1 &  19 & 3.1  & Likely SNR  \\
X-164   & 13:29:50.08 & +47:11:39.5 &  6  & 1.2  & Modest \sii:\ha, possible SNR  \\
X-211   & 13:29:52.76 & +47:11:40.0 &  49 & 0.1 & Blowout, S of nucleus \\
X-224   & 13:29:53.54 & +47:11:26.5 &   5  & 0.8 & Modest \sii:\ha, possible SNR   \\
X-269   & 13:29:55.44 & +47:11:43.5 &   6  & 1.2 & Modest \sii:\ha, possible SNR  \\
X-317   & 13:29:59.57 & +47:11:11.6 &  24  & 3.4 & Ill-defined; possible SNR  \\
X-336   & 13:30:01.10 & +47:13:32.9 &   7  & 5.9 & Likely SNR   \\
X-359   & 13:30:04.32 & +47:08:41.3 &  65  & 9.3 & Large shell, GMOS, likely SNR   \\
X-368   & 13:30:05.06 & +47:10:35.9 &  24  & 6.3 & Likely SNR   \\
\hline\\[-9pt]
& & Radio Sources & & & \\
\hline
M07-073  & 13:29:55.41 & +47:14:01.9 & 17.5 & 5.9  & X271; modest \sii:\ha, likely SNR  \\
M07-076  & 13:29:55.60 & +47:12:02.9 & 18.7 & 1.5  & Modest \sii:\ha, possible SNR  \\
\enddata
\tablenotetext{a}{Positions and sizes are for the associated optical counterpart to the X-ray or radio source.}
\tablenotetext{b}{X-ray IDs from \cite{kuntz16}; radio IDs from \citet{maddox07}.}
\label{softx_tab}
\end{deluxetable}


\clearpage


\end{document}